\input harvmac
% S-Tables Macro
%\tableofcontents

%
\message{S-Tables Macro v1.0, ACS, TAMU (RANHELP@VENUS.TAMU.EDU)}
%
% Help Text
%
\newhelp\stablestylehelp{You must choose a style between 0 and 3.}%
\newhelp\stablelinehelp{You should not use special hrules when
stretching
a table.}%
\newhelp\stablesmultiplehelp{You have tried to place an S-Table
inside another S-Table.  I would recommend not going on.}%
%
% Line Thicknesses (Values)
%
\newdimen\stablesthinline
\stablesthinline=0.4pt
\newdimen\stablesthickline
\stablesthickline=1pt
%
% Border and Internal Line Thicknesses
%
\newif\ifstablesborderthin
\stablesborderthinfalse
\newif\ifstablesinternalthin
\stablesinternalthintrue
\newif\ifstablesomit
\newif\ifstablemode
\newif\ifstablesright
\stablesrightfalse
%
% Save Registers
%
\newdimen\stablesbaselineskip
\newdimen\stableslineskip
\newdimen\stableslineskiplimit
%
% Counters
%
\newcount\stablesmode
\newcount\stableslines
\newcount\stablestemp
\stablestemp=3
\newcount\stablescount
\stablescount=0
\newcount\stableslinet
\stableslinet=0
%
% Table Style Selection
%
% 0 - Centered
% 1 - Left Justified
% 2 - Right Justified
% 3 - Not Justified
%
\newcount\stablestyle
\stablestyle=0
%
% Element Buffering Definitions
%
\def\stablesleft{\quad\hfil}%
\def\stablesright{\hfil\quad}%
%
% Vertical Bar Activation
%
\catcode`\|=\active%
%
% Strut Control
%
\newcount\stablestrutsize
\newbox\stablestrutbox
\setbox\stablestrutbox=\hbox{\vrule height10pt depth5pt width0pt}
\def\stablestrut{\relax\ifmmode%
                         \copy\stablestrutbox%
                       \else%
                         \unhcopy\stablestrutbox%
                       \fi}%
%
% Misc. Internal Stuff
%
\newdimen\stablesborderwidth
\newdimen\stablesinternalwidth
\newdimen\stablesdummy
\newcount\stablesdummyc
\newif\ifstablesin
\stablesinfalse
%
% Table Macros
%
%
%
%
\def\stablesadj{%
  \ifcase\stablestyle%
    \hbox to \hsize\bgroup\hss\vbox\bgroup%
  \or%
    \hbox to \hsize\bgroup\vbox\bgroup%
  \or%
    \hbox to \hsize\bgroup\hss\vbox\bgroup%
  \or%
    \hbox\bgroup\vbox\bgroup%
  \else%
    \errhelp=\stablestylehelp%
    \errmessage{Invalid style selected, using default}%
    \hbox to \hsize\bgroup\hss\vbox\bgroup%
  \fi}%
\def\stablesend{\egroup%
  \ifcase\stablestyle%
    \hss\egroup%
  \or%
    \hss\egroup%
  \or%
    \egroup%
  \or%
    \egroup%
  \else%
    \hss\egroup%
  \fi}%
\def\stablestart{%
  \ifstablesin%
    \errhelp=\stablesmultiplehelp%
    \errmessage{An S-Table cannot be placed within an S-Table!}%
  \fi
  \global\stablesintrue%
  \global\advance\stablescount by 1%
  \message{<S-Tables Generating Table \number\stablescount}%
  \begingroup%
  \stablestrutsize=\ht\stablestrutbox%
  \advance\stablestrutsize by \dp\stablestrutbox%
  \ifstablesborderthin%
    \stablesborderwidth=\stablesthinline%
  \else%
    \stablesborderwidth=\stablesthickline%
  \fi%
  \ifstablesinternalthin%
    \stablesinternalwidth=\stablesthinline%
  \else%
    \stablesinternalwidth=\stablesthickline%
  \fi%
  \tabskip=0pt%
  \stablesbaselineskip=\baselineskip%
  \stableslineskip=\lineskip%
  \stableslineskiplimit=\lineskiplimit%
  \offinterlineskip%
  \def\borderrule{\vrule width \stablesborderwidth}%
  \def\internalrule{\vrule width \stablesinternalwidth}%
  \def\thinline{\noalign{\hrule height \stablesthinline}}%
  \def\thickline{\noalign{\hrule height \stablesthickline}}%
  \def\trule{\omit\leaders\hrule height \stablesthinline\hfill}%
  \def\ttrule{\omit\leaders\hrule height \stablesthickline\hfill}%
  \def\tttrule##1{\omit\leaders\hrule height ##1\hfill}%
  \def\stablesel{&\omit\global\stablesmode=0%
    \global\advance\stableslines by 1\borderrule\hfil\cr}%
  \def\el{\stablesel&}%
  \def\elt{\stablesel\thinline&}%
  \def\eltt{\stablesel\thickline&}%
  \def\elttt##1{\stablesel\noalign{\hrule height ##1}&}%
  \def\elspec{&\omit\hfil\borderrule\cr\omit\borderrule&%
              \ifstablemode%
              \else%
                \errhelp=\stablelinehelp%
                \errmessage{Special ruling will not display properly}%
              \fi}%
  \def\stmultispan##1{\mscount=##1 \loop\ifnum\mscount>3
\stspan\repeat}%
  \def\stspan{\span\omit \advance\mscount by -1}%
  \def\multicolumn##1{\omit\multiply\stablestemp by ##1%
     \stmultispan{\stablestemp}%
     \advance\stablesmode by ##1%
     \advance\stablesmode by -1%
     \stablestemp=3}%
  \def\multirow##1{\stablesdummyc=##1\parindent=0pt\setbox0\hbox\bgroup%
    \aftergroup\emultirow\let\temp=}
  \def\emultirow{\setbox1\vbox to\stablesdummyc\stablestrutsize%
    {\hsize\wd0\vfil\box0\vfil}%
    \ht1=\ht\stablestrutbox%
    \dp1=\dp\stablestrutbox%
    \box1}%
%
%  \def\stvcen##1{\vtop{\vfill\hbox{##1}\vfill}}
% Currently does not work!
  \def\stpar##1{\vtop\bgroup\hsize ##1%
     \baselineskip=\stablesbaselineskip%
     \lineskip=\stableslineskip%

\lineskiplimit=\stableslineskiplimit\bgroup\aftergroup\estpar\let\temp=}%
  \def\estpar{\vskip 6pt\egroup}%
  \def\stparrow##1##2{\stablesdummy=##2%
     \setbox0=\vtop to ##1\stablestrutsize\bgroup%
     \hsize\stablesdummy%
     \baselineskip=\stablesbaselineskip%
     \lineskip=\stableslineskip%
     \lineskiplimit=\stableslineskiplimit%
     \bgroup\vfil\aftergroup\estparrow%
     \let\temp=}%
  \def\estparrow{\vfil\egroup%
     \ht0=\ht\stablestrutbox%
     \dp0=\dp\stablestrutbox%
     \wd0=\stablesdummy%
     \box0}%
  \def|{\global\advance\stablesmode by 1&&&}%
  \def\|{\global\advance\stablesmode by 1&\omit\vrule width 0pt%
         \hfil&&}%
\def\vt{\global\advance\stablesmode
by 1&\omit\vrule width \stablesthinline%
          \hfil&&}%
  \def\vtt{\global\advance\stablesmode by 1&\omit\vrule width
\stablesthickline%
          \hfil&&}%
  \def\vttt##1{\global\advance\stablesmode by 1&\omit\vrule width ##1%
          \hfil&&}%
  \def\vtr{\global\advance\stablesmode by 1&\omit\hfil\vrule width%
           \stablesthinline&&}%
  \def\vttr{\global\advance\stablesmode by 1&\omit\hfil\vrule width%
            \stablesthickline&&}%
\def\vtttr##1{\global\advance\stablesmode
 by 1&\omit\hfil\vrule width ##1&&}%
  \stableslines=0%
  \stablesomitfalse}
\def\stablesdef{\bgroup\stablestrut\borderrule##\tabskip=0pt plus 1fil%
  &\stablesleft##\stablesright%
  &##\ifstablesright\hfill\fi\internalrule\ifstablesright\else\hfill\fi%
  \tabskip 0pt&&##\hfil\tabskip=0pt plus 1fil%
  &\stablesleft##\stablesright%
  &##\ifstablesright\hfill\fi\internalrule\ifstablesright\else\hfill\fi%
  \tabskip=0pt\cr%
  \ifstablesborderthin%
    \thinline%
  \else%
    \thickline%
  \fi&%
}%
\def\endtable{\advance\stableslines by 1\advance\stablesmode by 1%
   \message{- Rows: \number\stableslines, Columns:
\number\stablesmode>}%
   \stablesel%
   \ifstablesborderthin%
     \thinline%
   \else%
     \thickline%
   \fi%
   \egroup\stablesend%
\endgroup%
\global\stablesinfalse}
%
% end of STABLES.TEX

\overfullrule=0pt \abovedisplayskip=12pt plus 3pt minus 3pt
\belowdisplayskip=12pt plus 3pt minus 3pt
%macros

\noblackbox
\input epsf
\newcount\figno
\figno=0
\def\fig#1#2#3{
\par\begingroup\parindent=0pt\leftskip=1cm\rightskip=1cm\parindent=0pt
\baselineskip=11pt \global\advance\figno by 1 \midinsert
\epsfxsize=#3 \centerline{\epsfbox{#2}} \vskip 12pt
\centerline{{\bf Figure \the\figno:} #1}\par
\endinsert\endgroup\par}
\def\figlabel#1{\xdef#1{\the\figno}}

\def\IR{\relax{\rm I\kern-.18em R}}

%%% Paragraphs

%%% special math symbols
\font\cmss=cmss10 \font\cmsss=cmss10 at 7pt
\def\rlx{\relax\leavevmode}
\def\inbar{\vrule height1.5ex width.4pt depth0pt}
\def\IC{\relax\,\hbox{$\inbar\kern-.3em{\rm C}$}}
\def\IN{\relax{\rm I\kern-.18em N}}
\def\IP{\relax{\rm I\kern-.18em P}}
\def\ZZ{\rlx\leavevmode\ifmmode\mathchoice{\hbox{\cmss Z\kern-.4em Z}}
 {\hbox{\cmss Z\kern-.4em Z}}{\lower.9pt\hbox{\cmsss Z\kern-.36em Z}}
 {\lower1.2pt\hbox{\cmsss Z\kern-.36em Z}}\else{\cmss Z\kern-.4em
 Z}\fi}
%%% misc.
\def\IZ{\relax\ifmmode\mathchoice
{\hbox{\cmss Z\kern-.4em Z}}{\hbox{\cmss Z\kern-.4em Z}}
{\lower.9pt\hbox{\cmsss Z\kern-.4em Z}} {\lower1.2pt\hbox{\cmsss
Z\kern-.4em Z}}\else{\cmss Z\kern-.4em Z}\fi}

\def\narrowplus{\kern -.04truein + \kern -.03truein}
\def\narrowminus{- \kern -.04truein}
\def\narrowminussub{\kern -.02truein - \kern -.01truein}

\def\frac#1#2{{#1\over #2}}

\def\IZ{\relax\ifmmode\mathchoice
{\hbox{\cmss Z\kern-.4em Z}}{\hbox{\cmss Z\kern-.4em Z}}
{\lower.9pt\hbox{\cmsss Z\kern-.4em Z}} {\lower1.2pt\hbox{\cmsss
Z\kern-.4em Z}}\else{\cmss Z\kern-.4em Z}\fi}
\def\IB{\relax{\rm I\kern-.18em B}}
\def\IC{{\relax\hbox{$\inbar\kern-.3em{\rm C}$}}}
\def\ID{\relax{\rm I\kern-.18em D}}
\def\IE{\relax{\rm I\kern-.18em E}}
\def\IF{\relax{\rm I\kern-.18em F}}
\def\IG{\relax\hbox{$\inbar\kern-.3em{\rm G}$}}
\def\IGa{\relax\hbox{${\rm I}\kern-.18em\Gamma$}}
\def\IH{\relax{\rm I\kern-.18em H}}
\def\II{\relax{\rm I\kern-.18em I}}
\def\IK{\relax{\rm I\kern-.18em K}}
\def\IP{\relax{\rm I\kern-.18em P}}

\font\cmss=cmss10 \font\cmsss=cmss10 at 7pt
\def\IR{\relax{\rm I\kern-.18em R}}

\def\1{{\bf 1}}
\def\3{{\bf 3}}
\def\7{{\bf 7}}
\def\2{{\bf 2}}
\def\8{{\bf 8}}

\def\hat{\widehat}
\def\quabla{{\sqcap}\!\!\!\!{\sqcup}}

\def\o{\over}
%

%
%       \eqn\label{a+b=c} gives displayed equation, numbered
%    consecutively within sections.
%     \eqnn and \eqna define labels in advance (of eqalign?)
%
\def\eqnn#1{\xdef #1{(\secsym\the\meqno)}\writedef{#1\leftbracket#1}%
\global\advance\meqno by1\wrlabeL#1}
\def\eqna#1{\xdef #1##1{\hbox{$(\secsym\the\meqno##1)$}}
\writedef{#1\numbersign1\leftbracket#1{\numbersign1}}%
\global\advance\meqno by1\wrlabeL{#1$\{\}$}}
\def\eqn#1#2{\xdef #1{(\secsym\the\meqno)}\writedef{#1\leftbracket#1}%
\global\advance\meqno by1$$#2\eqno#1\eqlabeL#1$$}

%%

%\DuffWD

\lref\vafai{C.~Vafa,
``Superstrings and topological strings at large N,''
J.\ Math.\ Phys.\  {\bf 42}, 2798 (2001), hep-th/0008142.}

\lref\civ{F.~Cachazo, K.~A.~Intriligator and C.~Vafa,
``A large N duality via a geometric transition,''
Nucl.\ Phys.\ B {\bf 603}, 3 (2001), hep-th/0103067.}

\lref\cveticone{M.~Cvetic, G.~W.~Gibbons, H.~Lu and C.~N.~Pope,
 ``Cohomogeneity one manifolds of Spin(7) and G(2) holonomy,''
 Phys.\ Rev.\ D {\bf 65}, 106004 (2002), hep-th/0108245;
``M-theory conifolds,''
 Phys.\ Rev.\ Lett.\  {\bf 88}, 121602 (2002), hep-th/0112098;
``A G(2) unification of the deformed and resolved conifolds,''
Phys.\ Lett.\ B {\bf 534}, 172 (2002), hep-th/0112138.}

\lref\cvetictwo{M.~Cvetic, G.~W.~Gibbons, H.~Lu and C.~N.~Pope,
``New complete non-compact Spin(7) manifolds,''
 Nucl.\ Phys.\ B {\bf 620}, 29 (2002), hep-th/0103155.}

\lref\brand{A.~Brandhuber, J.~Gomis, S.~S.~Gubser and S.~Gukov,
``Gauge theory at large N and new G(2) holonomy metrics,''
Nucl.\ Phys.\ B {\bf 611}, 179 (2001), hep-th/0106034.}

\lref\fawad{S.~F.~Hassan,
``T-duality, space-time spinors and R-R fields in curved backgrounds,''
Nucl.\ Phys.\ B {\bf 568}, 145 (2000), hep-th/9907152.}

\lref\brandtwo{A.~Brandhuber,
``G(2) holonomy spaces from invariant three-forms,''
Nucl.\ Phys.\ B {\bf 629}, 393 (2002), hep-th/0112113.}

\lref\salamontwo{R. ~Bryant, S.~Salamon,
``On the construction of some complete metrics with exceptional
holonomy,'' Duke Math. J. {\bf 58} (1989) 829;
G.~W.~Gibbons, D.~N.~Page and C.~N.~Pope,
``Einstein Metrics On S**3 R**3 And R**4 Bundles,''
Commun.\ Math.\ Phys.\  {\bf 127}, 529 (1990).}

\lref\kovalev{A. Kovalev,
``Twisted connected sums and special Riemannian holonomy,''
math-DG/0012189.}

\lref\joyce{D. ~Joyce,
``Compact Riemannian 7 manifolds with holonomy $G_2$ I, J. Diff. Geom.
{\bf 43} (1996) 291;
II: J. Diff. Geom.{\bf 43} (1996) 329.}

\lref\grayone{A. Gray, L. Hervella,
``The sixteen classes of almost Hermitian manifolds and their linear 
invariants,'' Ann. Mat. Pura Appl.(4) {\bf 123} (1980) 35.}

\lref\salamon{ S. Chiossi, S. Salamon,
``The intrinsic torsion of $SU(3)$ and $G_2$ structures,''
 Proc. conf. Differential Geometry Valencia 2001.}

\lref\gauntlett{J.~P.~Gauntlett, D.~Martelli and D.~Waldram,
``Superstrings with intrinsic torsion,'' hep-th/0302158.}

\lref\gukov{S.~Gukov,
``Solitons, superpotentials and calibrations,''
Nucl.\ Phys.\ B {\bf 574}, 169 (2000), hep-th/9911011.}

\lref\ach{B.~S.~Acharya and B.~Spence,
``Flux, supersymmetry and M theory on 7-manifolds,''
arXiv:hep-th/0007213.}

\lref\bw{C.~Beasley and E.~Witten,
 ``A note on fluxes and superpotentials in M-theory compactifications on
manifolds of G(2) holonomy,'' JHEP {\bf 0207}, 046 (2002),
hep-th/0203061.}

\lref\behr{K.~Behrndt and C.~Jeschek,
``Fluxes in M-theory on 7-manifolds and G structures,''
JHEP {\bf 0304}, 002 (2003), hep-th/0302047;
``Fluxes in M-theory on 7-manifolds: G-structures and 
superpotential,'' hep-th/0311119.}

\lref\bertwo{K.~Behrndt and C.~Jeschek,
``Superpotentials from flux compactifications of M-theory,'', hep-th/0401019.}

\lref\gray{M. Fernandez and A. Gray, ``Riemannian manifolds with
structure group $G_2$,'' Ann. Mat. Pura. Appl. {\bf 32} (1982),
19-45.}

\lref\graytwo{M. Fernandez and L. Ugarte,
``Dolbeault cohomology for $G_2$ manifolds,''
Geom. Dedicata, {\bf 70} (1998) 57.}

\lref\ivan{T.~Friedrich and S.~Ivanov,
 ``Parallel spinors and connections with skew-symmetric torsion
  in string theory,'' math.dg/0102142;
T.~Friedrich and S.~Ivanov,
``Killing spinor equations in dimension 7 and geometry of integrable
 $G_2$-manifolds,'' math.dg/0112201.
P.~Ivanov and S.~Ivanov,
``SU(3)-instantons and $G_2$, Spin(7)-heterotic string solitons,'' math.dg/0312094.}

\lref\tp{G.~Papadopoulos and A.~A.~Tseytlin, ``Complex geometry of conifolds
and 5-brane wrapped on 2-sphere,''
Class.\ Quant.\ Grav.\  {\bf 18}, 1333 (2001).hep-th/0012034.}

\lref\lust{G.~L.~Cardoso, G.~Curio, G.~Dall'Agata, D.~Lust, P.~Manousselis and G.~Zoupanos,
``Non-Kaehler string backgrounds and their five torsion classes,''
Nucl.\ Phys.\ B {\bf 652}, 5 (2003), hep-th/0211118.}

\lref\louis{S.~Gurrieri, J.~Louis, A.~Micu and D.~Waldram,
``Mirror symmetry in generalized Calabi-Yau compactifications,''
Nucl.\ Phys.\ B {\bf 654}, 61 (2003), hep-th/0211102.}

\lref\rstrom{A.~Strominger, ``Superstrings with torsion,'' Nucl.\
Phys.\ B {\bf 274}, 253 (1986).}

\lref\mal{J.~M.~Maldacena,
``The large N limit of superconformal field theories and supergravity,''
Adv.\ Theor.\ Math.\ Phys.\  {\bf 2}, 231 (1998)
[Int.\ J.\ Theor.\ Phys.\  {\bf 38}, 1113 (1999), hep-th/9711200.}

\lref\ks{I.~R.~Klebanov and M.~J.~Strassler,
``Supergravity and a confining gauge theory: Duality cascades and
chiSB-resolution of naked singularities,'' JHEP {\bf 0008}, 052 (2000), hep-th/0007191.}

\lref\mn{J.~M.~Maldacena and C.~Nunez,
``Towards the large N limit of pure N = 1 super Yang Mills,''
Phys.\ Rev.\ Lett.\  {\bf 86}, 588 (2001), hep-th/0008001.}

\lref\vafai{C.~Vafa,
``Superstrings and topological strings at large N,''
J.\ Math.\ Phys.\  {\bf 42}, 2798 (2001), hep-th/0008142.}

\lref\civ{F.~Cachazo, K.~A.~Intriligator and C.~Vafa,
``A large N duality via a geometric transition,''
Nucl.\ Phys.\ B {\bf 603}, 3 (2001), hep-th/0103067.}

\lref\syz{A.~Strominger, S.~T.~Yau and E.~Zaslow,
``Mirror symmetry is T-duality,''
Nucl.\ Phys.\ B {\bf 479}, 243 (1996), hep-th/9606040.}

\lref\tduality{E.~Bergshoeff, C.~M.~Hull and T.~Ortin,
``Duality in the type II superstring effective action,''
Nucl.\ Phys.\ B {\bf 451}, 547 (1995), hep-th/9504081;
P.~Meessen and T.~Ortin,
``An Sl(2,Z) multiplet of nine-dimensional type II supergravity theories,''
Nucl.\ Phys.\ B {\bf 541}, 195 (1999), hep-th/9806120.}

\lref\eot{J.~D.~Edelstein, K.~Oh and R.~Tatar,
``Orientifold,
 geometric transition and large N duality for SO/Sp gauge  theories,''
JHEP {\bf 0105}, 009 (2001), hep-th/0104037.}

\lref\dotu{K.~Dasgupta, K.~Oh and R.~Tatar, {``Geometric
transition, large N dualities and MQCD dynamics,''} Nucl.\ Phys.\
B {\bf 610}, 331 (2001), hep-th/0105066; {``Open/closed string
dualities and Seiberg duality from geometric transitions in
M-theory,''} JHEP {\bf 0208}, 026 (2002), hep-th/0106040.}

\lref\dotd{K.~Dasgupta, K.~h.~Oh, J.~Park and R.~Tatar, ``Geometric
transition versus cascading solution,'' JHEP {\bf 0201}, 031
(2002), hep-th/0110050.}

\lref\ohta{K.~Ohta and T.~Yokono,
``Deformation of conifold and intersecting branes,''
JHEP {\bf 0002}, 023 (2000), hep-th/9912266.}

\lref\dott{K.~h.~Oh and R.~Tatar,
``Duality and confinement
in N = 1 supersymmetric theories from geometric  transitions,''
Adv.\ Theor.\ Math.\ Phys.\  {\bf 6}, 141 (2003), hep-th/0112040.}

\lref\edelstein{J.~D.~Edelstein and C.~Nunez,
``D6 branes and M-theory geometrical transitions from gauged  supergravity,''
JHEP {\bf 0104}, 028 (2001), hep-th/0103167.}

\lref\candelas{P.~Candelas and X.~C.~de la Ossa, ``Comments on
conifolds,'' Nucl.\ Phys.\ B {\bf 342}, 246 (1990).}

\lref\minasianone{R.~Minasian and D.~Tsimpis,
``Hopf reductions, fluxes and branes,''
Nucl.\ Phys.\ B {\bf 613}, 127 (2001), hep-th/0106266.}

\lref\imamura{Y.~Imamura,
``Born-Infeld action and Chern-Simons term from Kaluza-Klein monopole in
M-theory,''
Phys.\ Lett.\ B {\bf 414}, 242 (1997), hep-th/9706144;
A.~Sen,
``Dynamics of multiple Kaluza-Klein monopoles in M and string theory,''
Adv.\ Theor.\ Math.\ Phys.\  {\bf 1}, 115 (1998), hep-th/9707042;
``A note on enhanced gauge symmetries in M- and string theory,''
JHEP {\bf 9709}, 001 (1997), hep-th/9707123.}

\lref\robbins{K.~Dasgupta, G.~Rajesh, D.~Robbins and S.~Sethi,
``Time-dependent warping, fluxes, and NCYM,''
JHEP {\bf 0303}, 041 (2003), hep-th/0302049;
K.~Dasgupta and M.~Shmakova,
``On branes and oriented B-fields,''
Nucl.\ Phys.\ B {\bf 675}, 205 (2003), hep-th/0306030.}

\lref\svw{S.~Sethi, C.~Vafa and E.~Witten,
``Constraints on low-dimensional string compactifications,''
Nucl.\ Phys.\ B {\bf 480}, 213 (1996), hep-th/9606122;
K.~Dasgupta and S.~Mukhi,
``A note on low-dimensional string compactifications,''
Phys.\ Lett.\ B {\bf 398}, 285 (1997), hep-th/9612188.}

\lref\kachruone{S.~Kachru, M.~B.~Schulz, P.~K.~Tripathy and S.~P.~Trivedi,
``New supersymmetric string compactifications,''
JHEP {\bf 0303}, 061 (2003), hep-th/0211182;
S.~Kachru, M.~B.~Schulz and S.~Trivedi,
``Moduli stabilization from fluxes in a simple IIB orientifold,''
JHEP {\bf 0310}, 007 (2003), hep-th/0201028.}

\lref\hitchin{N. ~Hitchin,
``Stable forms and special metrics'',
Contemp. Math., {\bf 288}, Amer. Math. Soc. (2000).}

\lref\giveon{S.~S.~Gubser,
 ``Supersymmetry and F-theory realization of the deformed conifold with
three-form flux,'' hep-th/0010010;
A.~Giveon, A.~Kehagias and H.~Partouche,
``Geometric transitions, brane dynamics and gauge theories,''
JHEP {\bf 0112}, 021 (2001), hep-th/0110115.}

\lref\bonan{E. Bonan,
``Sur le varietes remanniennes a groupe d'holonomie $G_2$ ou Spin(7),''
C. R. Acad. Sci. paris {\bf 262} (1966) 127.}

%\lref\bismut{J. M. Bismut,
%``A local index theorem for non-K\"ahler manifolds,''
%Math. Ann. {\bf 284} (1989) 681.}

%\lref\monar{F. Cabrera, M. Monar, A. Swann,
%``Classification of $G_2$ structures,''
%J. London Math. Soc. {\bf 53} (1996) 98;
%F. Cabrera,
%``On Riemannian manifolds with $G_2$ structure,''
%Bolletino UMI A {\bf 10} (1996) 98.}

\lref\amv{M.~Atiyah, J.~M.~Maldacena and C.~Vafa,
``An M-theory flop as a large N duality,''
J.\ Math.\ Phys.\  {\bf 42}, 3209 (2001), hep-th/0011256.}

\lref\syz{A.~Strominger, S.~T.~Yau and E.~Zaslow,
``Mirror symmetry is T-duality,''
Nucl.\ Phys.\ B {\bf 479}, 243 (1996), hep-th/9606040.}

\lref\tduality{E.~Bergshoeff, C.~M.~Hull and T.~Ortin,
``Duality in the type II superstring effective action,''
Nucl.\ Phys.\ B {\bf 451}, 547 (1995), hep-th/9504081;
P.~Meessen and T.~Ortin,
``An Sl(2,Z) multiplet of nine-dimensional type II supergravity theories,''
Nucl.\ Phys.\ B {\bf 541}, 195 (1999), hep-th/9806120.}

\lref\adoptone{J.~D.~Edelstein, K.~Oh and R.~Tatar,
``Orientifold, geometric transition and large N duality for SO/Sp
gauge theories,'' JHEP {\bf 0105}, 009 (2001), hep-th/0104037.}

\lref\ohta{K.~Ohta and T.~Yokono,
``Deformation of conifold and intersecting branes,''
JHEP {\bf 0002}, 023 (2000), hep-th/9912266.}

\lref\adoptfo{K.~h.~Oh and R.~Tatar, ``Duality and confinement in
N = 1 supersymmetric theories from geometric transitions,'' Adv.\
Theor.\ Math.\ Phys.\  {\bf 6}, 141 (2003), hep-th/0112040.}

\lref\edelstein{J.~D.~Edelstein and C.~Nunez, ``D6 branes and
M-theory geometrical transitions from gauged supergravity,'' JHEP
{\bf 0104}, 028 (2001), hep-th/0103167.}

\lref\candelas{P.~Candelas and X.~C.~de la Ossa,
``Comments On Conifolds,''
Nucl.\ Phys.\ B {\bf 342}, 246 (1990).}

\lref\minasianone{R.~Minasian and D.~Tsimpis,
``Hopf reductions, fluxes and branes,''
Nucl.\ Phys.\ B {\bf 613}, 127 (2001), hep-th/0106266.}

\lref\pandoz{L.~A.~Pando Zayas and A.~A.~Tseytlin,
``3-branes on resolved conifold,''
JHEP {\bf 0011}, 028 (2000), hep-th/0010088.}

\lref\rBB{K.~Becker and M.~Becker, ``M-Theory on
eight-manifolds,'' Nucl.\ Phys.\ B {\bf 477}, 155 (1996),
hep-th/9605053.}

\lref\bbdgs{K.~Becker, M.~Becker, P.~S.~Green, K.~Dasgupta and
E.~Sharpe, ``Compactifications of heterotic strings on
non-K\"ahler complex manifolds. II,'' Nucl.\ Phys.\ B {\bf 678},
19 (2004), hep-th/0310058.}

\lref\bbdg{K.~Becker, M.~Becker, K.~Dasgupta and P.~S.~Green,
``Compactifications of heterotic theory on non-K\"ahler complex
manifolds. I,'' JHEP {\bf 0304}, 007 (2003), hep-th/0301161.}

\lref\GP{E.~Goldstein and S.~Prokushkin, ``Geometric model for
complex non-K\"ahler manifolds with SU(3) structure,''
hep-th/0212307.}

\lref\townsend{P.~K.~Townsend,
``D-branes from M-branes,''
Phys.\ Lett.\ B {\bf 373}, 68 (1996), hep-th/9512062.}

\lref\sav{K.~Dasgupta, G.~Rajesh and S.~Sethi,
``M theory, orientifolds and G-flux,''
JHEP {\bf 9908}, 023 (1999), hep-th/9908088.}

\lref\beckerD{K.~Becker and K.~Dasgupta,
``Heterotic strings with torsion,''
JHEP {\bf 0211}, 006 (2002), hep-th/0209077.}

\lref\ks{I.~R.~Klebanov and M.~J.~Strassler,
 ``Supergravity and a confining gauge theory: Duality cascades
  and $\chi_{SB}$-resolution of naked singularities,''
JHEP {\bf 0008}, 052 (2000), hep-th/0007191.}

\lref\gv{R.~Gopakumar and C.~Vafa,
``On the gauge theory/geometry correspondence,''
Adv.\ Theor.\ Math.\ Phys.\  {\bf 3}, 1415 (1999),hep-th/9811131.}

\lref\dvu{R.~Dijkgraaf and C.~Vafa,
``Matrix models, topological strings, and supersymmetric gauge theories,''
Nucl.\ Phys.\ B {\bf 644}, 3 (2002), hep-th/0206255.}
%%CITATION = HEP-TH 0206255;%%

\lref\ps{J.~Polchinski and M.~J.~Strassler, ``The String Dual of a
Confining Four-Dimensional Gauge Theory ,'' hep-th/0003136.}
%%CITATION = HEP-TH 0206255;%%

\lref\susskind{L.~Susskind,
``The anthropic landscape of string theory,'' hep-th/0302219.}

\lref\banks{T.~Banks, M.~Dine and E.~Gorbatov,
``Is there a string theory landscape?,'' hep-th/0309170.}

\lref\hori{K.~Hori and C.~Vafa,
``Mirror symmetry,'' hep-th/0002222;
M.~Aganagic, A.~Klemm and C.~Vafa,
``Disk instantons, mirror symmetry and the duality web,''
Z.\ Naturforsch.\ A {\bf 57}, 1 (2002), hep-th/0105045;
M.~Aganagic, A.~Klemm, M.~Marino and C.~Vafa,
``Matrix model as a mirror of Chern-Simons theory,''
JHEP {\bf 0402}, 010 (2004), hep-th/0211098.}

\lref\karch{M.~Aganagic, A.~Karch, D.~Lust and A.~Miemiec,
``Mirror symmetries for brane configurations and branes at singularities,''
Nucl.\ Phys.\ B {\bf 569}, 277 (2000), hep-th/9903093.}

\lref\dmconi{K.~Dasgupta and S.~Mukhi,
``Brane constructions, conifolds and M-theory,''
Nucl.\ Phys.\ B {\bf 551}, 204 (1999), hep-th/9811139.}

\lref\toappear{S.~Alexander, K.~Becker, M.~Becker, K.~Dasgupta, 
A.~Knauf, R.~Tatar,
``In the realm of the geometric transitions,'' hep-th/0408192.}

\lref\dvd{R.~Dijkgraaf and C.~Vafa,
``A perturbative window into non-perturbative physics,'' hep-th/0208048.}

\lref\civu{F.~Cachazo, S.~Katz and C.~Vafa,
``Geometric transitions and N = 1 quiver theories,'' hep-th/0108120.}

\lref\civd{F.~Cachazo, B.~Fiol, K.~A.~Intriligator, S.~Katz and C.~Vafa,
``A geometric unification of dualities,'' Nucl.\ Phys.\ B {\bf 628}, 3 (2002),
hep-th/0110028.}

\lref\radu{R.~Roiban, R.~Tatar and J.~Walcher,
``Massless flavor in geometry and matrix models,''
Nucl.\ Phys.\ B {\bf 665}, 211 (2003), hep-th/0301217.}

\lref\radd{K.~Landsteiner, C.~I.~Lazaroiu and R.~Tatar,
``(Anti)symmetric matter and superpotentials from IIB orientifolds,''
JHEP {\bf 0311}, 044 (2003), hep-th/0306236.}

\lref\radt{K.~Landsteiner, C.~I.~Lazaroiu and R.~Tatar,
``Chiral field theories from conifolds,''
JHEP {\bf 0311}, 057 (2003), hep-th/0310052.}

\lref\radp{K.~Landsteiner, C.~I.~Lazaroiu and R.~Tatar,
``Puzzles for matrix models of chiral field theories,'' hep-th/0311103.}

\lref\ber{M.~Bershadsky, S.~Cecotti, H.~Ooguri and C.~Vafa,
 ``Kodaira-Spencer theory of gravity and exact results for quantum string amplitudes,''
Commun.\ Math.\ Phys.\  {\bf 165}, 311 (1994), hep-th/9309140.}

\lref\bismut{J. M. Bismut,
``A local index theorem for non-K\"ahler manifolds,''
Math. Ann. {\bf 284} (1989) 681.}

\lref\monar{F. Cabrera, M. Monar and A. Swann,
``Classification of $G_2$ structures,''
J. London Math. Soc. {\bf 53} (1996) 98;
F. Cabrera,
``On Riemannian manifolds with $G_2$ structure,''
Bolletino UMI A {\bf 10} (1996) 98.}

\lref\kath{Th. Friedrich and I. Kath,
``7-dimensional compact Riemannian manifolds with killing spinors,''
Comm. Math. Phys. {\bf 133} (1990) 543;
Th. Friedrich, I. Kath, A. Moroianu and U. Semmelmann,
``On nearly parallel $G_2$ structures,'' J. geom. Phys. {\bf 23} (1997) 259;
S. Salamon,
``Riemannian geometry and holonomy groups,''
Pitman Res. Notes Math. Ser., {\bf 201} (1989);
V. Apostolov and S. Salamon,
``K\"ahler reduction of metrics with holonomy $G_2$,''
math-DG/0303197.} 

\lref\ooguri{H.~Ooguri and C.~Vafa,
``The C-deformation of gluino and non-planar diagrams,''
Adv.\ Theor.\ Math.\ Phys.\  {\bf 7}, 53 (2003), hep-th/0302109;
``Gravity induced C-deformation,''
Adv.\ Theor.\ Math.\ Phys.\  {\bf 7}, 405 (2004), hep-th/0303063.}

\lref\tv{T.~R.~Taylor and C.~Vafa,
``RR flux on Calabi-Yau and partial supersymmetry breaking,''
Phys.\ Lett.\ B {\bf 474}, 130 (2000), hep-th/9912152.}

\lref\cvetic{M.~Cvetic, G.~W.~Gibbons, H.~Lu and C.~N.~Pope,
  ``Ricci-flat metrics, harmonic forms and brane resolutions,''
  Commun.\ Math.\ Phys.\  {\bf 232}, 457 (2003), hep-th/0012011.}

\lref\bdkkt{M.~Becker, K.~Dasgupta, S.~Katz, A.~Knauf and R.~Tatar,
  ``Geometric transitions, flops and non-K\"ahler manifolds. II,'' hep-th/0511099.}

\lref\nekra{N.~Nekrasov,~H.~Ooguri and C.~Vafa, `S-duality and
Topological Strings,'' hep-th/0403167.}

\lref\lustu{G.~L.~Cardoso, G.~Curio, G.~Dall'Agata and D.~Lust,
``BPS action and superpotential for heterotic string compactifications  with
fluxes,'' JHEP {\bf 0310}, 004 (2003),hep-th/0306088.}

\lref\lustd{G.~L.~Cardoso, G.~Curio, G.~Dall'Agata and D.~Lust,
``Heterotic string theory on non-K\"ahler manifolds with H-flux
and gaugino condensate,'' hep-th/0310021.}

\lref\bd{M.~Becker and K.~Dasgupta, ``K\"ahler versus non-K\"ahler
compactifications,'' hep-th/0312221.}

\lref\micu{S.~Gurrieri and A.~Micu,
``Type IIB theory on half-flat manifolds,''
Class.\ Quant.\ Grav.\  {\bf 20}, 2181 (2003), hep-th/0212278.}

\lref\dal{G.~Dall'Agata and N.~Prezas,
``N = 1 geometries for M-theory and type IIA strings with fluxes,''
hep-th/0311146.}

\lref\douo{M.~R.~Douglas,``The statistics of string / M theory vacua,''
JHEP {\bf 0305}, 046 (2003), hep-th/0303194.}

\lref\wittenchern{E.~Witten,``Chern-Simons gauge theory as a
string theory,'' hep-th/9207094.}

\lref\ms{D.~Martelli and J.~Sparks,``G Structures, fluxes and
calibrations in M Theory,'' hep-th/0306225.}

\lref\fg{A.~F.~Frey and A.~Grana,``Type IIB solutions with
interpolating supersymetries ,'' hep-th/0307142.}

\lref\bbs{K.~Becker, M.~Becker and R.~Sriharsha,``PP-waves,
M-theory and fluxes,'' hep-th/0308014.}

\lref\minu{P. Kaste,~R. Minasian,~A. Tomasiello,''Supersymmetric M theory
Compactifications with
Fluxes on Seven-Manifolds and G Structures'', JHEP {\bf 0307} 004, 2003,
hep-th/0303127.}

 \lref\mind{S. Fidanza,~R. Minasian,~A. Tomasiello,''Mirror Symmetric
SU(3) Structure Manifolds with NS fluxes'',
hep-th/0311122.}

\lref\beru{K.~Behrndt and M.~Cvetic, ``General N=1 Supersymmetric Flux
Vacua of
(Massive) Type IIA String Theory'', hep-th/0403049.}

\lref\dalu{G.~Dall'Agata, ``On Supersymmetric Solutions of Type IIB
Supergravity with General Fluxes'', hep-th/0403220.}

\Title{\vbox{\hbox{hep-th/0403288}
\hbox{SU-ITP-04/10}\hbox{UCB-PTH-04/09}\hbox{LBNL-54768}
\hbox{UMD-PP-04/035}}} {\vbox{ \vskip-4.5in
\hbox{\centerline{Geometric Transitions,}}\vskip.1in \hbox{\centerline{
Flops and Non-K\"ahler
Manifolds: I}}}}

\vskip-.2in \centerline{\bf Melanie Becker$^1$,~ Keshav
Dasgupta$^2$, ~Anke Knauf$^{1,3}$, ~Radu Tatar$^4$} \vskip.2in
\centerline{\it ${}^1$~Department of Physics, University of
Maryland, College Park, MD 20742}
%\vskip.02in
\centerline{\tt melanieb@physics.umd.edu, anke@umd.edu}\vskip.02in
\centerline{\it ${}^2$~Department of Physics, Stanford
University, Stanford CA 94305}
%\vskip.02in
\centerline{\tt keshav@itp.stanford.edu} \vskip.02in
\centerline{\it ${}^3$~II. Institut f\"ur Theoretische Physik,
Universit\"at Hamburg} \centerline{\it ~~~~~ Luruper Chaussee
149, 22761 Hamburg, Germany} \vskip.02in \centerline{\it
${}^4$~Theoretical Physics Group, LBL Berkeley, CA 94720}
%\vskip.02in
\centerline{\tt rtatar@socrates.Berkeley.EDU}

\vskip.2in

\centerline{\bf Abstract}

\noindent We construct a duality cycle which provides a complete
supergravity description of geometric transitions in type II
theories via a flop in M-theory. This cycle
connects the different supergravity descriptions before and after
the geometric transitions. Our construction
reproduces many of the known phenomena studied earlier in the
literature and allows us to describe some new and interesting
aspects in a simple and elegant fashion. A precise supergravity
description of new torsional manifolds that appear on the type IIA
side with branes and fluxes and the corresponding geometric
transition are obtained. A local description of new $G_2$
manifolds that are circle fibrations over non-K\"ahler manifolds
is presented.

\Date{}

\centerline{\bf Contents}\nobreak\medskip{\baselineskip=12pt
\parskip=0pt\catcode`\@=11

\noindent {1.} {Introduction} \leaderfill{1} \par 
\noindent {2.} {Geometric Transitions, Fluxes and Gauge Theories} \leaderfill{6} \par 
\noindent {3.} {The Type IIB Background From M-Theory Dual} \leaderfill{9} \par 
\noindent {4.} {Mirror Formulas using Three T-Dualities} \leaderfill{13} \par
\noindent \quad{4.1.} {Metric Components} \leaderfill{13} \par 
\noindent \quad{4.2.} {$B_{NS}$ Components} \leaderfill{14} \par 
\noindent \quad{4.3.} {Background Simplications} \leaderfill{16} \par  
\noindent \quad{4.4.} {Rewriting the Deformed Conifold Background} \leaderfill{18} \par  
\noindent {5.} {Chain 1: The Type IIA Mirror Background} \leaderfill{22} \par 
\noindent \quad{5.1.} {Searching for the $d\theta_1 d\theta_2$ term} \leaderfill{28} \par 
\noindent \quad{5.2.} {Physical meaning of $f_1$ and $f_2$} \leaderfill{38} \par 
\noindent \quad{5.3.} {$B$-fields in the Mirror set-up} \leaderfill{39} \par 
\noindent \quad{5.4.} {The Mirror Manifold} \leaderfill{41} \par 
\noindent {6.} {Chain 2: The M-Theory Description of the Mirror} \leaderfill{44} \par 
\noindent \quad{6.1.} {One-Forms in M-Theory} \leaderfill{44} \par 
\noindent \quad{6.2.} {M-Theory lift of the Mirror IIA Background} \leaderfill{46} \par 
\noindent {7.} {Chain 3: M-Theory Flop and Type IIA Reduction} \leaderfill{55} \par 
\noindent \quad{7.1.} {The Type IIA Background} \leaderfill{60} \par 
\noindent \quad{7.2.} {Analysis of Type IIA Background and Superpotential} \leaderfill{61} \par 
\noindent {8.} {Discussion and Future Directions} \leaderfill{66} \par
\noindent \quad{8.1.} {Future Directions} \leaderfill{67} \par 
\noindent Appendix {1.} {Algebra of $\alpha$} \leaderfill{70} \par 
\noindent Appendix {2.} {Details on $G_2$ Structures} \leaderfill{71} \par 
\catcode`\@=12 \bigbreak\bigskip}

\newsec{Introduction}

During the last years there has been tremendous progress toward
constructing the string theory dual descriptions of large N gauge
theories. The first steps in this direction were made by
considering the conformal ${\cal N} = 4$, $D=4$ super Yang-Mills
theory \mal\ and later on more realistic field theories with
${\cal N} = 1$ supersymmetry and confinement were described in
\ks,\ps,\mn\ and \vafai. In a slightly different context, the
connection between gauge theories and topological string theory
was discussed in \gv\ for type IIA strings and in \dvu\ for type
IIB strings, the latter leading to the powerful Dijkgraaf-Vafa
conjecture by which non-perturbative computations in field
theories can be performed using perturbative expansions in matrix
models.

The seminal work of Vafa \vafai\ very clearly showed how to embed
the topological duality of \gv\ into the framework of the AdS/CFT
correspondence. The basic idea of \vafai\ was to consider the
${\cal N} = 1$ theory resulting from type IIA superstring theory
in the deformed conifold background $T^{*}S^3$ in the presence of
$N$ D6 branes wrapped around the Lagrangian $S^3$ cycle and
filling the external space and compute the corresponding
superpotential of this theory. On the other hand it was known that
the superpotential for the field theory living in the four
noncompact directions of the D6 branes is described by a
Chern-Simons gauge theory on $S^3$ \wittenchern, whose
superpotential can be computed in terms of topological field
theory amplitudes. Therefore, a connection between large $N$
Chern-Simons theory/topological string duality to ordinary
superstring theory and the AdS/CFT correspondence was
established. The idea was soon extended to many more rather
interesting models in \civ,\eot,\civu,\civd,\radu,\radd,\radt\
and \radp.

In all the above mentioned models the superpotentials computed by
topological strings are mapped by the AdS/CFT like correspondences to
geometries generated by fluxes on the dual closed string side.
Since we start with D6 branes, we expect to have  RR two-form
fluxes in the dual closed string solution. These fluxes thread
through a holomorphic $P^1$ cycle inside a blow up of a conifold.
The superpotential is a product between the RR two-form fluxes and
the K\"ahler form associated with the 
four-cycle that is Hodge dual to $P^1$. It turns out \vafai\
that open/closed string duality requires another term in the
superpotential. This term originates from the field theory gluino
condensate, as topological open string computations imply the
existence of a term linear in the gluino condensate \ber, which
gets mapped into the size of the holomorphic two-cycle. Therefore
one has to have a term linear in the size of the holomorphic
two-cycle which can be obtained if this is multiplied by a
four-form. Furthermore, this four-form should be of NS type.

The quest for this four-form has been the subject of intense
scrutiny in the last years. As there is no D-brane which can
create it, the flux originates from changes in the geometry. The
best way to understand its appearance is to go to the mirror type
IIB picture and consider the superpotential \tv  \eqn\tve{W =
\int \Omega \wedge (H_{RR} + \varphi H_{NS}),~~\varphi = \chi +
i~e^{-\Phi},} where $\chi$ is the axion, $\Phi$ is the dilaton,
$H_{RR}$ is the RR three-form flux and $H_{NS}$ is the NS
three-form flux. From here we can go to the mirror type IIA
picture
 and the fluxes map into the RR two-form flux and an NS four-form
 flux respectively. Of course, the main question is why would the
 NS fluxes appear when one
starts with brane configurations involving only D branes? A
partial answer to this question was given in \louis, where the
origin of the NS four-form $F_4^{NS}$ was related to the fact that
the (3,0) form $\Omega=\Omega^++i\Omega^-$ is not closed for the
type IIA compactification, and therefore $d\Omega^+\sim
F_4^{NS}$. The fact that $d \Omega \ne 0$ allows an extra term in
the superpotential $\int d \Omega \wedge J$ , $J$ being the
fundamental two-form\foot{Notice that this term also plays a crucial role 
in the S-duality conjecture of \vafai.}. 
Manifolds with the property that the real
part of the (3,0) form is not closed, while the imaginary part
satisfies $d\Omega^-=0=d(J\wedge J)$ are called half flat
manifolds. Half flat manifolds are examples of non-K\"ahler
manifolds\foot{The manifold that we will eventually get in type
IIA side later in this paper, will however be more general than
the half-flat manifold in the sense that both $d\Omega^\pm \ne 0$.}.

At this point one might wonder, {{why would a non-K\"ahler
manifold appear as a result of a geometric transition from a
brane configuration with D6 branes wrapped on a cycle of a Calabi
Yau manifold?}} One of the goals of this paper is to give an
answer to this question. In short the answer is this: {\it{the
fluxes live on a non-K\"ahler manifold because the D6 branes
actually are themselves wrapped on a cycle inside a non-K\"ahler
manifold and the geometric transition is a flop inside a $G_2$
manifold with torsion.}}

To arrive to the non-K\"ahler geometry where the D branes are
wrapped we start with the type IIB solution corresponding to D5
branes wrapped on the resolution cycle $P^1$ of the resolved
conifold \pandoz. The supergravity solution involves, besides the
RR three-form, a NS three-form and a RR five-form. These fields
are required by the string equations of motion. Similarly as in
\civ\ the presence of the NS three-form before the transition is a
signal that a NS three-form should also exist after the geometric
transition has taken place.

As the resolved conifold is a toric manifold we can easily
identify three $S^1$ coordinates and take a T-duality in these
directions.\foot{The outcome of these T-dualities is different
from the ones of \dotu,\dotd\ and  \dott\ where one T-duality takes
a type IIB picture to a type IIA brane configuration.}
 \foot{There
have been previous attempts to relate the resolved conifold and
the deformed conifold by starting with the deformed conifold
\minasianone. As the deformed conifold does not admit a $T^3$
fibration using this manifold as a starting point may seem more
problematic, although, we have been informed that there are 
some papers that overcome this problem \hori.} By applying Buscher's T-duality formulas we observe
that the NS three-form transforms into the type IIA metric in such
a way that the resulting manifold is not K\"ahler. \foot{This is a
generalization of the notion of the mirror symmetry, where the B field
in the type IIB picture is traded for a non  K\"ahlerity in type IIA.
A similar observation has been made in \mind\ where the generalized mirror
symmetry exchange was between a non closed $J + i B$ and a nonclosed
$\Omega$.} The
D5 branes get mapped into D6 branes which are wrapped on a three-cycle
inside a {\it{ non-K\"ahler deformation of a deformed conifold}}.

The non-K\"ahler deformation of a deformed conifold is then
locally lifted to M theory. The lift leads to a $G_2$
manifold\foot{By this we mean a manifold endowed with an almost $G_2$
structure. The holonomy however is contained in $G_2$. For more details on the
almost $G_2$ structure the reader may want to refer to the work of \salamon\ and the references given 
in Appendix 2.} 
which is
a deformation of the manifold constructed in \brand, \cveticone, 
and as such is
described in terms of some left invariant one forms\foot{For an earlier
discussion of special holonomy spaces like $spin(7)$ and the corresponding
one-forms, the reader may want to see \cvetictwo.}. 
These
one forms reduce to the ones of \brand, \cveticone\ in the absence
of B fields and they can be exchanged giving rise to a flop
transformation. The resulting type IIA geometry describes a
{\it{non-K\"ahler deformation of a resolved conifold}}. The
non-K\"ahler deformation has a non-closed (3,0) form whose
derivative can be used to construct the additional contribution
to the superpotential $\int d \Omega \wedge J$. These non-K\"ahler
deformations of resolved and deformed conifolds, and the $G_2$ manifolds
resulting from their lift to M-theory are, to the best of our knowledge,
first concrete examples. In earlier literature these manifolds were
anticipated as solutions of type II and M-theories although no concrete
examples were presented.

To summarize, we propose {\it{a new geometric transition in the
type IIA theory}} which relates D-branes wrapped on cycles of
non-K\"ahler manifolds and fluxes on other non-K\"ahler manifolds.
When lifted to M theory this transition describes a flop inside a
$G_2$ manifold with torsion. \foot{This is related to the recent
results of \nekra. This paper discusses a topological string model
concluding that the non-integrability of the complex structure is
related to the existence of Lagrangian NS branes called ``NS
two-branes'' in \vafai. In our language, the non-K\"ahlerity
condition will be related the existence of the ``NS two-brane''.
It would be extremely interesting to relate this non-K\"ahlerity
appearing in the supergravity description to a corresponding
effect in the Chern-Simons theory.} This would lead to a deeper
understanding of non-K\"ahler geometries. These geometries have
only recently been discussed in some detail in the context of
heterotic strings in \beckerD, \lust, \bbdg, \lustu, \bbdgs,
\lustd\ and \bd\ and also of type II and M theories in \micu, \minu, \dal,
\mind, \beru, \dalu.

Our new geometric transition would enrich the ``landscape
picture'' advocated in \susskind,\douo, \banks\ in the sense of extra 
identifications between branes on cycles of non-K\"ahler
geometries and fluxes on cycles of related non-K\"ahler
geometries.

This paper is organized as follows: In section 2 we give a very
brief review on the subject of geometric transitions and outline
of the calculation that we will perform in this paper. Our
starting point is the type IIB metric describing $D5$ branes
wrapping a $P^1$ of a resolved conifold and through a series of
T-duality transformations and a flop we shall be able to describe
the geometric transition taking place in the type IIA mirror in
great detail. In section 3 we give an alternative way to derive
the metric using fourfold compactifications in M-theory in the
presence of fluxes. Section 4 discusses the mirror formulas that
we will use to get the full background in the type IIA theory. In
the absence of fluxes, it is known that the mirror type IIA
picture involves $D6$ branes wrapping an $S^3$ of a deformed
conifold \vafai.
In section 4.4 we write the metric of the deformed
conifold in a simpler way by making a coordinate transformation.
We will discuss the reason why a deformed conifold may not
have a $T^3$ fibration. Section 5 begins the study of the mirror
manifold. We will present an explicit way to get the mirror
manifold in the type IIA theory. We will show that the naive
$T^3$ direction of the resolved conifold does not lead to the
right mirror metric, which can nevertheless be determined by a
set of restricted coordinate transformations. These aspects will
be discussed in sections 5.1 and 5.2. 
In section 5.3 we will determine the
$B$ field background and the metric for the mirror manifold
will appear as eqn. (5.64) (and later as a final metric in eqn. (6.23)). 
In section 6 we begin
our ascent to M-theory. In the absence of any fluxes in the type
IIB picture, we expect a manifold with $G_2$ holonomy after
lifting to M-theory. In the presence of fluxes, we will also get
a seven dimensional manifold which now has a $G_2$ structure and
torsion\foot{In this paper we will interchangeably use $G_2$ structure
and $G_2$ holonomy. The seven dimensional manifold that we construct
will have an almost $G_2$ structure, although we will not check the
holonomy here. More detailed discussions will be relegated to part
II of this paper. We thank K. Behrndt and G. Dall'Agata for correspondences
on this issue. See also \behr, \monar.}.
The generic study of $G_2$ holonomy manifolds has been
done earlier using left invariant one-forms \amv.
For our case we will
also have one forms that are appropriately shifted by the
background $B$ fields of the type IIA theory. Locally these one
forms look exactly like the ones without fluxes. However,
globally the system is much more involved, as we do not have any
underlying $SU(2)$ symmetry. The M-theory lift of the mirror type
IIA manifold is given as eqn (6.24), (6.25).

After lifting to M-theory we shall discuss the flop taking place
in the resulting $G_2$ manifold. This will be studied in
section 7 using the one forms that we devised earlier. We shall
show that for torsional $G_2$ manifolds the flop is a little
subtle. We discuss this in detail and compute the form of the
metric after the flop. The result for the metric is given as eqn.
(7.17) of section 7.1. Knowing the M-theory metric after the flop,
helps us to get the corresponding type IIA metric easily by
dimensional reduction. The resulting manifold in the type IIA
theory is non-K\"ahler and the metric is given in eqn. (7.18). We
show that the metric is basically a non-K\"ahler deformation of
the resolved conifold. Interestingly, this is a rather similar
situation as in the type IIA manifold {\it before} the geometric
transition has taken place, whose
metric is a non-K\"ahler deformation of the deformed
conifold. In this section we further study some properties of
these manifolds like non-K\"ahlerity and the underlying
superpotential. We leave a detailed discussion on various aspects
of the whole duality chain for part II of this paper. We end with
a discussion in section 8.

\newsec{Geometric Transitions, Fluxes and Gauge Theories}

We will summarize here some useful facts about geometric
transitions, fluxes and field theory results.

Geometric transitions are examples of generalised AdS/CFT correspondence
which relate D-branes in the open string picture and fluxes in the
closed string picture. There are several types of geometric
transitions depending on the framework in which we formulate them.
The type IIB geometric transition, which starts with $D5$ branes
wrapping a $P^1$ of a resolved conifold, has a parallel counterpart
in which $D5$ branes wrap a vanishing two cycle of a conifold. This is
the Klebanov-Strassler model \ks. In fact, these two models have identical
behaviors in the IR of the corrresponding four-dimensional ${\cal N} =1$
gauge theories. However the UV behaviors are different. In the UV the
geometric transition models give rise to six-dimensional gauge theories,
whereas the Klebanov-Strassler model remains four dimensional. Both these
models show cascading behavior. The cascading behavior in the geometric
transition models manifest as an infinite sequence of flop transitions \civd.
The corresponding brane constructions for these models have been developed
earlier in \dotu. The precise equivalences between the two models
were shown in \dotd\ using (a) T-dual brane constructions, and (b) M-theory
four-fold compactifications. We will not go into the details of this in the
present paper but instead delve directly to the supergravity aspects
of geometric transitions in both type II and M-theories, with the starting
point being $D5$ wrapped on resolution $P^1$ cycle of a resolved conifold.
(For the Klebanov-Strassler model, the duality chain is not obvious).
Readers interested in the details of the
equivalence should look up the
above mentioned references. Some related work has also been done in
\giveon.

The figure below gives an overview of the geometric transitions
and flops that will be discussed in this paper:

\vskip.3in

\centerline{\epsfbox{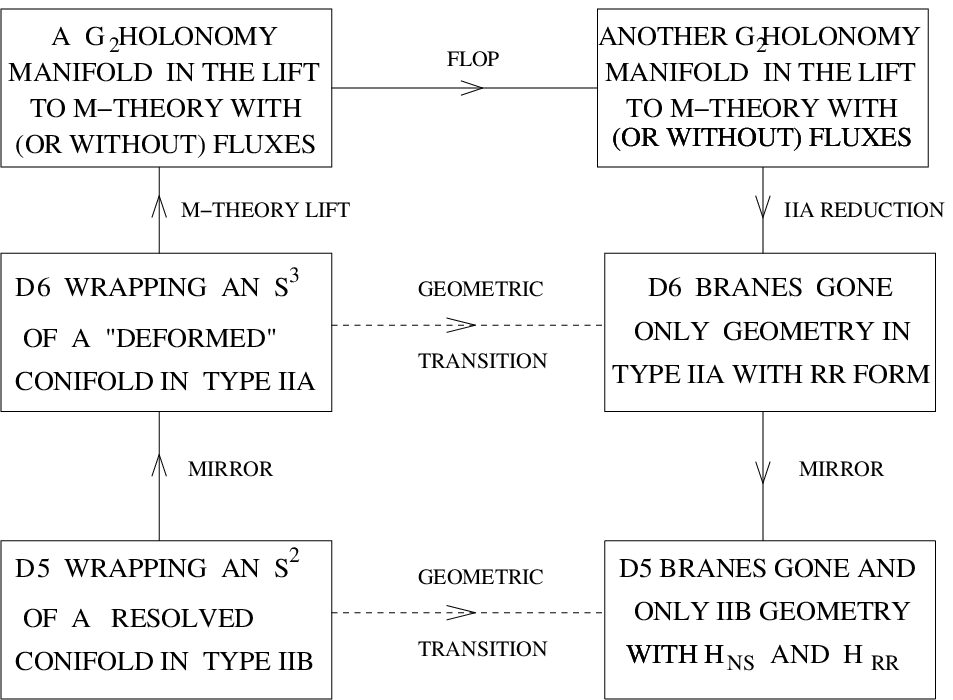}}\nobreak

\vskip.3in

\noindent Let us elaborate this figure. The type IIB theory,
depicted at the bottom left level, is the most studied one in
different approaches \mn, \ks\ and \vafai. Here one starts with
the field theory living on D5 branes wrapped on the resolution
$P^1$ cycle of a resolved conifold. In the strong coupling limit
of the field theory the $P^1$ cycle shrinks but the theory avoids
the singularity by opening up an $S^3$ cycle inside a deformed
conifold. In the figure, this is given by a dotted line pointing
to the box on the lower right side of the picture, where a
geometric transition has taken place. The deformed geometry
encodes the information about the strongly coupled field theory
as the size of the $S^3$ cycle is identified with the gluino
condensate in the field theory \vafai.

There is an analog version of this process for the type IIA string
which is depicted in the middle line of the figure. This time one
starts with the field theory living on D6 branes wrapped on the
$S^3$ cycle inside a deformed conifold. In the strong coupling
limit of the field theory the $S^3$ cycle shrinks but the theory
avoids the singularity by opening up a $P^1$ cycle inside a
resolved conifold (given by the next dotted line). As before, the
complexified volume of the $P^1$ cycle is identified with the
gluino condensate appearing in the field theory.

The transition looks mysterious if seen from ten dimensions but
its understanding can be simplified by going up to eleven
dimensions where it appears as a flop transition inside a $G_2$
manifold \amv\ as seen in the upper line of the figure. For the
above mentioned case of D6 branes wrapped on an $S^3$ cycle, the
$G_2$ manifold appears as a cone over a $Z_N$ quotient of $S^3
\times \tilde{S}^3$ and the flop switches the two $S^3$ cycles.
The type IIB transition has also been lifted to an M theory
picture involving a warped fourfold compactification \dotd.
Alternatively, by using one T-duality, the geometric transition
has also been discussed in the brane configuration language in
\dotu\ and \dott.

Even though there is a compelling evidence in favor of using
geometric transitions to describe strongly coupled field
theories, there are still some unanswered questions. One of them
is related to the form of the superpotential (1.1). Intuitively
one would think that if we start with D branes, the supergravity
solution should involve only RR fluxes;  but, as it turns out,
 we need NS fluxes, too. In the language of the Klebanov-Strassler model \ks,
both RR and NS fluxes appear naturally as we have fractional D branes but this
is not obvious in
the language of \vafai.

One goal of this paper is to fill this gap and clarify the
presence of the NS fluxes. We shall start with a known solution
for the type IIB  configuration with wrapped D5 branes on the
resolved conifold by including both NS and RR fluxes. This is
depicted in the lower left box of the figure. We first go up one
step by using three T-dualities and as a result we get a
non-K\"ahler type IIA geometry located in the middle left box of
the figure. Then we go up to M theory, obtaining a $G_2$ manifold
with torsion. We follow the upper arrow by performing a flop
inside the  $G_2$ manifold and then descend to obtain another
non-K\"ahler type IIA geometry. We shall leave the last step and
a detailed discussion on various aspects of the duality chain,
 for a future publication as we are
mostly concerned here with the type IIA geometric transition.

\noindent In the figure the dark arrows represent the directions
that we will be following in this paper. The dotted arrows
represent the connection between two geometries that are related
by a geometric transition. The duality cycle will therefore be
powerful enough to give the precise supergravity background for
{\it all} the examples studied in the literature so far. Notice
also the fact that the key difference between the work presented
here and some of the related work \minasianone\
 is that our starting point involves $D5$ branes
wrapping a resolved conifold and {\it not} a deformed conifold
with fluxes. We believe that a deformed conifold with fluxes {\it
does not} have any obvious $T^3$ fibration and so a simple
Strominger, Yau and Zaslow (SYZ) \syz\ analysis may not be easy to
perform (see however \hori). We will elaborate more on this as we go along.

\noindent We will begin by describing the first box in this
picture (located on the lower left part of the figure): the type
IIB background.

\newsec{The Type IIB Background From M-Theory Dual}

The type IIB background with D5 wrapping an $S^2$ of a resolved
conifold has been discussed in detail in \pandoz. The metric of
the system is shown to follow from the standard D3 brane metric
at a point on the non-compact manifold. In this section we will
give an {\it alternative} derivation from M-theory
with fluxes that at least reproduces {\it locally} some aspects of \pandoz. 
One of the major advantages of using M-theory as
opposed to the type IIB theory is the drastic reduction of the
field content. The bosonic field content of M-theory consists of
only the metric and four-form fluxes. Furthermore preserving
supersymmetry in lower dimensions puts some constraints on the
fluxes. The constraint equations are generically {\it linear} and
therefore one can avoid the complicated second order equations
that we would get by solving equations of motion.

To be more specific, we shall consider a {\it non-compact}
fourfold in  M-theory with G-fluxes. The non-compact fourfold is a
$T^2$ fibration over a resolved conifold base. The $T^2$
fibration will be trivial (for the time being) and therefore the
manifold is almost a product manifold. At this point one might get a
little worried by the fact that the trivial fibration may force
the Euler characteristic to vanish and therefore may disallow
fluxes. Of course by having a 
non-compact manifold we may still have the possibility of
non-zero fluxes, but we can get non-zero Euler number in this framework
simply by allowing the $T^2$ fiber to degenerate far away, i.e not in the 
local neighborhood.
There is one immediate advantage of having such a
fibration. The $T^2$ torus doesn't degenerate locally
and therefore when we shrink the fiber torus to zero size
to go to the type IIB framework, there will be no seven branes
(and possibly orientifold seven planes) in our local neighborhood.
 This will simplify the
subsequent analysis. Also, having a non-compact manifold allows
us to put as many branes in the setup as we like. For the compact
case, the number of branes (and fluxes) is constrained by an
anomaly cancellation rule \rBB, \svw.

Let us therefore consider a fourfold that is locally 
of the form ${\cal M}_6 \times
T^2$, where ${\cal M}_6$ denotes the resolved conifold which is
oriented along $x^{4,5,6,7,8,9}$, with one of the $S^2$ along
$x^{4,5}$ and the other $S^2$ along $x^{8,9}$. The second $S^2$
degenerates at the radial distance $x^7 = 0$, whereas the other
$S^2$ has a finite size. The coordinate $x^6$ is the usual $U(1)$
fibration. Using angular coordinates in terms of which the metric
is generically written, the two $S^2$'s have coordinates
$\theta_1, \phi_1$ and $\theta_2, \phi_2$. The radial coordinate
is $r \equiv x^7$ and the $U(1)$ coordinate is $\psi \equiv x^6$,
the latter being non-trivially fibered over the two $S^2$'s. The
product torus $T^2$ is oriented along $x^3$ and $x^{11}$, with
$x^{11} \equiv x^a$ being the M-theory direction\foot{As we shall soon see, the
coordinates ($\psi, \theta_i, \phi_i, r$) will not be the right coordinates
to express the local metric. The correct set of coordinate system will be 
provided in sec. 5.}.

In the presence of G-fluxes with components $G_{457a}$ and
$G_{3689}$, the  backreaction on the metric has been worked out in
\rBB. The metric picks up a warp factor $\Delta$ which is a
function of the internal (radial) coordinate only. The generic
form of the metric is \eqn\mthmetric{ds^2 = \Delta^{-1}
ds^2_{012} + \Delta^{1/2} ds^2_{{\cal M}_6 \times T^2},} where
$ds^2_{012}$ denotes the Minkowski directions. In case that two
covariantly constant spinors of definite chirality on the
internal space can be found, supersymmetry requires  the internal
G-fluxes to be primitive and hence self-dual in the eight
dimensional sense\foot{This is the case we shall be interested
in. The generalization to non-chiral spinors on the internal
space was worked out in \ms, \fg\ and \bbs. In this case the
primitivity condition is replaced by a more general equation.}.
Finally, there is also a spacetime component $G_{012m}$, where
$x^m$ is one of the internal space directions. This component of
the flux is given in terms of warp factor as $G_{012m} = \del_m
\Delta^{-3/2}$. The warp factor, $\Delta$, in turn satisfies the
equation $\quabla \Delta^{3/2} =$ sources.

Let us {\it replace} all the fluxes with M2 branes. This
situation has been considered earlier in section 4 of \bbdgs. The
metric of the system is the metric of $N$ M2 branes at a point on
the fourfold ${\cal M}_6 \times T^2$  \eqn\metofmtwo{ds^2 =
H_2^{-2/3} ~ds^2_{012} + H_2^{1/3}~ ds^2_{{\cal M}_6 \times T^2},}
where $H_2$ is the harmonic function of the M2 branes. As
discussed in \bbdgs, there is a one-to-one connection between the
two pictures: the harmonic function in the M2 brane framework is
related to the warp factor describing the flux via the relation
$H_2 = \Delta^{3/2}$. One can easily show that the source
equations work out correctly using the above identification
\bbdgs.

Imagine now that instead of M2 branes we have $M$ M5 branes in
our framework. These M5 branes wrap three-cycles inside the
fourfold. The three-cycles are an $S^1$ product over an $S^2$
base. We will assume that the M5 branes wrap the directions
$x^{a,4,5}$ inside the fourfold. The metric ansatz for the
wrapped M5 brane is \eqn\mfivemet{ds^2 = H_5^\alpha~ds^2_{012} +
H_5^\beta~ds^2_{S^2 \times S^1} + H_5^\gamma~ds^2_{36789},} where
$H_5$ is the harmonic function of the wrapped M5 branes, and
$\alpha,\beta,\gamma$ are constants. Observe that generically
$\beta \ne \gamma$ and therefore the metric of wrapped M5 branes
is warped differently along the $S^2 \times S^1$ and the
remaining $x^{3,6,7,8,9}$ directions.

The above discussion was in the absence of any fluxes. Let us
switch on three-form potentials $C_{a45}$ and $C_{378}$. In the
presence of these potentials the M5 branes will contain the
following world-volume term \townsend \eqn\sourceterm{S = -{1\o 4}
\int \Gamma^{il}\Gamma^{jm} \Gamma^{kn} (F_{ijk} -
C_{ijk})(F_{lmn} - C_{lmn}),} in addition to the usual source
term that contributes a {\it five} dimensional delta function in
the supergravity equations of motion. We have also denoted the
field strength of the self dual two form propagating on the
$M5$ branes as $F_{ijk}$.
In the
absence of three-form sources this term would be absent and the
equations of motion will only contain a five dimensional delta
function source related to the M5 branes. In the presence of $C$
fields, the above action will induce an M2 brane charge and
therefore the delta function contribution to the action will
become eight dimensional. Therefore the background configuration
will be the usual M2 brane background, implying that $\beta =
\gamma$. When reduced to the type IIB theory by shrinking the
fiber torus to zero size, the metric will locally resemble the D3
brane metric obtained in \pandoz. Here we have provided a
derivation of this metric from M-theory. For the M2 brane
background the warp factor will satisfy the usual equation
\rBB,\svw\ \eqn\wareqsa{\ast \quabla H_5 - 4\pi^2 X_8 + {1\o 2} G
\wedge G = -4\pi^2 \sum_{i=1}^n \delta^8(y-y_i),} where the Hodge
duality is over the unwarped metric, $n$ is the number of
fractional M2 branes situated at points $y_i$ in the fourfold and
$X_8$ is a polynomial in powers of the curvature that contains 
information about the seven-branes.

{}From the M-theory point of view, our choice of G-fluxes
immediately reduces  to the $H_{NS}$ and $H_{RR}$ fluxes in the
type IIB theory. The fact that supersymmetry requires the G-fluxes
to be primitive, implies that the NS and the RR fluxes should be
dual to each other. This duality gives rise to {\it linear}
equations in the type IIB theory. 
As expected, notice that the primitivity of G-fluxes in
M-theory implies that the form should be of type (2,2) \rBB. This
means that the NS and the RR fluxes when combined to form a
three-form ${\cal G} \equiv H_{NS} + \varphi H_{RR}$ with
$\varphi$ being the usual axion-dilaton scalar, will be a (2,1) form.
This is where we face a problem. The (2,1) nature of the fluxes
is in agreement with the analysis of \ks\ but unfortunately not with \pandoz\
(see also \cvetic). This means that the global geometry of \pandoz\ breaks 
susy, and therefore our background derived from M-theory should be the 
right global description and {\it not} \pandoz. From the way we derived the 
background, the metric would {\it locally} resemble the metric of 
\pandoz\ but globally there will be extra seven branes (see \bdkkt\ for the 
full story).

To summarize, starting with M-theory we have rederived the form of
the metric describing wrapped D5 metric in the type IIB theory.
The choice of fluxes fixes the complex structure to some
particular value $\tau$. Using this we define a one form $dz =
dx^3 + \tau dx^a$. The final background can be presented in a
compact form using the notation of \pandoz\ (with $\alpha = -2/3,
\beta = \gamma = 1/3$) \eqn\wrapmfivebg{\eqalign{& ds^2 =
H_5^{-2/3}~ds^2_{012} + H_5^{1/3}~ ds^2_{{\cal M}_6 \times T^2,}
\cr & G = e_{\theta_1} \wedge e_{\phi_1} \wedge G_1 +
e_{\theta_2} \wedge e_{\phi_2} \wedge G_2,}} where $e_{\theta_i}$
and $e_{\phi_i}$ with $i = 1,2$ are defined in \pandoz.
The forms
$G_1$ and $G_2$ also have an explicit representation. They can be
defined in terms of the remaining forms $dz, dr$ and $e_\psi$
(defined in \pandoz) as \eqn\defg{\eqalign{& G_1 = {\bar \tau} f'
~dz \wedge dr + {\bar \tau} ~dz \wedge e_\psi - {\rm c.c,} \cr &
G_2 = {\bar \tau} g' ~dz \wedge dr - {\bar \tau} ~dz \wedge
e_\psi - {\rm c.c,}}} where $f$ and $g$ are solutions to the
linear equations following from supersymmetry. Observe also the
fact that we haven't yet fixed the value of $\tau$, the complex
structure of the $x^{3,a}$ torus\foot{The complex structure of the 
base will be discussed later.}. 
It is not too difficult to see
that the complex structure can be fixed to $\tau = i$ so that we
end up with a square torus. The analysis follows closely to the
one discussed in \beckerD, so we will not repeat it here. The fact
that fluxes are not constant here (as opposed to the constant
fluxes in \beckerD) does not alter the result for the complex
structure.

%(1) Give the fourfold derivation of the exact type IIB background.

%(2) Identify the $T^3$ fibration.

\newsec{Mirror Formulas using Three T-dualities}

In this section we will determine the formulas for the metric and
fluxes $B_{NS}$ and $B_{RR}$ of a general type IIA manifold that
is the mirror of a six dimensional type IIB manifold that is a
$T^3$-fibration over a three dimensional base. According to
Strominger, Yau and Zaslow \syz\ the mirror manifold can be
determined by performing three T-dualities on the fiber. We shall
be using the T-duality formulas of \tduality. Later on we will
use our general result for the particular case of the resolved
conifold.

\subsec{Metric Components}

We will start by determining the metric components of the mirror
manifold. Let us call the $T^3$ directions of the lagrangian $T^3$ fibered
manifold with which we start as $x, y$ and $z$. We will be
performing three T-dualities \tduality\ along these directions in
the order $x, y, z$. The starting metric in the type IIB theory
has the following components \eqn\meetcok{\eqalign{ ds^2 = &~
j_{\mu \nu}dx^\mu ~dx^\nu + j_{x\mu} dx ~dx^\mu +  j_{y \mu} dy~
dx^\mu +  j_{z\mu} dz ~ dx^\mu +  j_{xy} dx ~dy \cr
 & ~ + j_{xz} dx ~dz +  j_{zy} dz ~ dy +  j_{xx} dx^2 +  j_{yy}dy^2 +  j_{zz}~dz^2}}
where $\mu, \nu \ne x, y, z$, and the $j's$ are for now arbitrary.
After a straightforward calculation we obtain the form of the
metric of the mirror manifold \eqn\finmetccc{\eqalign{ds^2 = &
\left( G_{\mu\nu} - {G_{z\mu}G_{z\nu} - {\cal B}_{z\mu} {\cal
B}_{z\nu} \over G_{zz}} \right) dx^\mu~dx^\nu +2 \left( G_{x\nu} -
{G_{zx}G_{z\nu} - {\cal B}_{zx} {\cal B}_{z\nu}
 \over G_{zz}} \right) dx~dx^\mu  \cr
& ~ + 2\left( G_{y\nu} - {G_{zy}G_{z\nu} - {\cal B}_{zy} {\cal B}_{z\nu}
 \over G_{zz}}\right) dy~dx^\nu +
2\left( G_{xy} - {G_{zx}G_{zy} - {\cal B}_{zx} {\cal B}_{zy} \over
G_{zz}}\right) dx~dy  \cr & ~ + {dz^2\over G_{zz}} + 2{{\cal
B}_{\mu z} \over G_{zz}} dx^\mu~dz + 2{{\cal B}_{xz} \over G_{zz}}
dx~dz + 2{{\cal B}_{yz} \over G_{zz}} dy~dz \cr &~+  \left( G_{xx}
- {G^2_{zx} - {\cal B}^2_{zx} \over G_{zz}} \right) dx^2 + \left(
G_{yy} - {G^2_{zy} - {\cal B}^2_{zy} \over G_{zz}} \right)
dy^2.}} The various components of the metric can be written as
\eqn\gmunu{\eqalign{G_{\mu\nu} = &~~ {j_{\mu\nu}j_{xx} -
j_{x\mu}j_{x\nu} + b_{x\mu}b_{x\nu} \over j_{xx}} -
{(j_{y\mu}j_{xx} - j_{xy} j_{x \mu} + b_{xy} b_{x\mu})
(j_{y\nu}j_{xx}
 - j_{xy} j_{x \nu} + b_{xy} b_{x\nu}) \over
j_{xx}(j_{yy}j_{xx}- j_{xy}^2 + b_{xy}^2)} \cr & ~~ +
{(b_{y\mu}j_{xx} - j_{xy} b_{x \mu} + b_{xy}
j_{x\mu})(b_{y\nu}j_{xx}
 - j_{xy} b_{x \nu} + b_{xy} j_{x\nu})\over
j_{xx}(j_{yy}j_{xx}- j_{xy}^2 + b_{xy}^2)},}}

\eqn\gmuz{\eqalign{G_{\mu z} = &~~ {j_{\mu z}j_{xx} -
j_{x\mu}j_{xz} + b_{x \mu}b_{xz} \over j_{xx}} - {(j_{y\mu}j_{xx}
- j_{xy} j_{x \mu} + b_{xy} b_{x\mu}) (j_{yz}j_{xx} - j_{xy} j_{x
z} + b_{xy} b_{xz}) \over j_{xx}(j_{yy}j_{xx}- j_{xy}^2 +
b_{xy}^2)} \cr & ~~ + {(b_{y\mu}j_{xx} - j_{xy} b_{x \mu} +
b_{xy} j_{x\mu})(b_{yz}j_{xx} - j_{xy} b_{x z} + b_{xy}
j_{xz})\over j_{xx}(j_{yy}j_{xx}- j_{xy}^2 + b_{xy}^2)},}}

\eqn\gzz{\eqalign{G_{zz} = &~~ {j_{zz}j_{xx} - j^2_{xz} +
b^2_{xz}\over j_{xx}} - {(j_{yz}j_{xx} - j_{xy} j_{xz} + b_{xy}
b_{xz})^2 \over j_{xx}(j_{yy}j_{xx}- j_{xy}^2 + b_{xy}^2)} \cr &
~~ + {(b_{yz}j_{xx} - j_{xy} b_{x z} + b_{xy} j_{xz})^2 \over
j_{xx}(j_{yy}j_{xx}- j_{xy}^2 + b_{xy}^2)},}}

\eqn\gymu{ G_{y \mu} = -{b_{y \mu} j_{xx} - b_{x \mu} j_{xy} + b_{xy}
 j_{\mu x} \over j_{yy}j_{xx}- j_{xy}^2 + b_{xy}^2},
~ G_{y z} = -{b_{y z} j_{xx} - b_{x z} j_{xy} + b_{xy} j_{z x}
\over j_{yy}j_{xx}- j_{xy}^2 + b_{xy}^2},}

\eqn\gyy{G_{yy} = {j_{xx} \over j_{yy}j_{xx}- j_{xy}^2 +
b_{xy}^2},~ G_{xx} = {j_{yy} \over j_{yy}j_{xx}- j_{xy}^2 +
b_{xy}^2}, ~G_{xy} = {-j_{xy} \over j_{yy}j_{xx}- j_{xy}^2 +
b_{xy}^2},}

\eqn\gmux{G_{\mu x} = {b_{\mu x} \over j_{xx}} + {(j_{\mu y} j_{xx} -
 j_{xy} j_{x \mu} + b_{xy} b_{x \mu}) b_{xy} \over
j_{xx}(j_{yy}j_{xx}- j_{xy}^2 + b_{xy}^2)}
+ {(b_{y \mu} j_{xx} - j_{xy} b_{x \mu} + b_{xy} j_{x \mu}) j_{xy}
 \over j_{xx}(j_{yy}j_{xx}- j_{xy}^2 + b_{xy}^2)},}

\eqn\gzx{ G_{z x} = {b_{z x} \over j_{xx}} + {(j_{z y} j_{xx} -
j_{xy} j_{x z} + b_{xy} b_{x z}) b_{xy} \over
j_{xx}(j_{yy}j_{xx}- j_{xy}^2 + b_{xy}^2)}
 + {(b_{y z} j_{xx} - j_{xy} b_{x z} + b_{xy} j_{xz}) j_{xy}
  \over j_{xx}(j_{yy}j_{xx}- j_{xy}^2 + b_{xy}^2)}.}
In the above formulae we have denoted the type IIB
$B$ fields as $b_{mn}$, whose explicit
form will be computed in the next section.
We will use this more general formula for the particular case of
the resolved conifold a little later. Our next goal is
to determine the NS fluxes on the mirror manifold for the most
general case.

\subsec{$B_{NS}$ Components}

\noindent For the generic case we will switch on all the
components of the $B$ field \eqn\bcompolk{\eqalign{ b = & ~~
b_{\mu\nu} ~ dx^\mu \wedge dx^\nu + b_{x \mu} dx \wedge dx^\mu +  b_{y \mu}
~ dy~\wedge dx^\mu + b_{z \mu} ~ dz \wedge dx^\mu \cr & ~~ + ~ b_{xy} 
~ dx \wedge dy +
 b_{xz} ~ dx  \wedge dz +  b_{zy}~ dz  \wedge dy.}}

In the later sections we will concentrate on the special
components that describe a resolved conifold with branes.
\noindent After applying again the T-dualities, the NS component
of the $B$ field in the mirror set-up will take the form
\eqn\bbfielc{\eqalign{ {\tilde B} = & ~~ \left( {\cal B}_{\mu\nu}
+ {2 {\cal B}_{z[\mu} G_{\nu]z} \over G_{zz}} \right) dx^\mu
\wedge dx^\nu + \left( {\cal B}_{\mu x} + {2 {\cal B}_{z[\mu}
G_{x]z} \over G_{zz}}\right)
 dx^\mu \wedge dx  \cr
& ~~ \left( {\cal B}_{\mu y} + {2 {\cal B}_{z[\mu} G_{y]z} \over G_{zz}}
 \right) dx^\mu \wedge dy
+ \left( {\cal B}_{xy} + {2 {\cal B}_{z[x} G_{y]z} \over G_{zz}}
\right) dx \wedge dy \cr & ~~ + {G_{z \mu} \over G_{zz}} dx^\mu
\wedge dz + {G_{z x} \over G_{zz}} dx \wedge dz + {G_{z y} \over
G_{zz}} dy \wedge dz.}} Here the $G_{mn}$ components have been
given above, and the various ${\cal B}$  components can now be
written as \eqn\bmunu{\eqalign{ {\cal B}_{\mu\nu} = & ~~
{b_{\mu\nu} j_{xx} + b_{x \mu} j_{\nu x} - b_{x \nu} j_{\mu x}
\over j_{xx}} \cr & + ~~ {2 (j_{y[\mu}j_{xx} - j_{xy}j_{x[\mu} +
b_{xy} b_{x[\mu}) (b_{\nu]y}j_{xx} - b_{\nu]x}j_{xy} - b_{xy}
j_{\nu]x}) \over j_{xx}(j_{yy}j_{xx}- j_{xy}^2 + b_{xy}^2)},}}

\eqn\bmuz{\eqalign{ {\cal B}_{\mu z} = & ~~ {b_{\mu z} j_{xx} +
b_{x \mu} j_{z x} - b_{x z} j_{\mu x} \over j_{xx}} \cr & + ~~ {2
(j_{y[\mu}j_{xx} - j_{xy}j_{x[\mu} + b_{xy} b_{x[\mu})
(b_{z]y}j_{xx} - b_{z]x}j_{xy} - b_{xy} j_{z]x}) \over
j_{xx}(j_{yy}j_{xx}- j_{xy}^2 + b_{xy}^2)},}}

\eqn\bmuy{{\cal B}_{\mu y} = {j_{\mu y} j_{xx} - j_{xy} j_{x \mu} + b_{xy} b_{x \mu}
 \over j_{yy}j_{xx}- j_{xy}^2 + b_{xy}^2}, ~~~
{\cal B}_{z y} = {j_{z y} j_{xx} - j_{xy} j_{x z} + b_{xy} b_{x z} \over j_{yy}j_{xx}-
 j_{xy}^2 + b_{xy}^2},}

\eqn\bmux{ {\cal B}_{\mu x} = {j_{\mu x} \over j_{xx}} - {j_{xy} (j_{\mu y} j_{xx} -
j_{xy} j_{x \mu} + b_{xy} b_{x \mu}) \over
j_{xx}(j_{yy}j_{xx}- j_{xy}^2 + b_{xy}^2)} + {b_{xy} (b_{x\mu}j_{xy} - b_{y\mu}j_{xx} -
b_{xy}j_{xz})
 \over j_{xx}(j_{yy}j_{xx}- j_{xy}^2 + b_{xy}^2)},}

\eqn\bzx{ {\cal B}_{z x} = {j_{z x} \over j_{xx}} - {j_{xy} (j_{z y} j_{xx} -
j_{xy} j_{xz} + b_{xy} b_{x z}) \over
j_{xx}(j_{yy}j_{xx}- j_{xy}^2 + b_{xy}^2)} + {b_{xy} (b_{xz}j_{xy} - b_{yz}j_{xx} -
 b_{xy}j_{xz})
\over j_{xx}(j_{yy}j_{xx}- j_{xy}^2 + b_{xy}^2)},}

\eqn\bxy{{\cal B}_{xy} = {-b_{xy} \over j_{yy}j_{xx}- j_{xy}^2 + b_{xy}^2}.}

In the above analysis, there is one subtlety related to the
compactness of the $x,y,z$ directions. The type IIB $B$ fields
defined wholly along these directions, i.e. ${\cal B}_{yz}, {\cal
B}_{zx}$ and ${\cal B}_{xy}$, should be {\it periodic}. This would
mean, for example, if we specify a value of ${\cal B}_{yz}$  as
(say) $\alpha_{yz}$ then this should also be equal to
$-\alpha_{yz}$ because of periodicity. This implies that the
values of ${\cal B}_{yz}, {\cal B}_{zx}$ and ${\cal B}_{xy}$
found in \bmuy, \bzx\ and \bxy\ are {\it ambiguous} up to a
possible sign. Later on we shall use consistency conditions to fix
the sign.

Furthermore, observe that we haven't yet discussed how the RR $B$
fields look like in the mirror set-up.  We will eventually
compute the form of these fields when we perform the M-theory lift
of the type IIA mirror. We also need to see how the
fermions transform under mirror symmetry. This will be important
in order to understand the complex structure of the mirror
manifold and to check whether it is integrable or not. This will
be discussed in the sequel to this paper. The string coupling
constant in the type IIB theory and the one in the type IIA mirror
are related in the following way \eqn\ccconss{g_A = {g_B \o
\sqrt{(j_{xx}j_{yy} - j_{xy}^2 + b_{xy}^2)~G_{zz}}},} where we
have defined $G_{zz}$ in \gzz. This coupling constant is in
general a function of the internal coordinates.

\subsec{Background Simplifications}

The background given in the above set of formulas can be written
in a {\it compact} form which will be helpful to see the fibration
structure more clearly
\eqn\metcomp{\eqalign{ds^2 = & {1\over G_{zz}} (dz + {\cal B}_{\mu z}
dx^\mu + {\cal B}_{xz} dx
+ {\cal B}_{yz} dy)^2 -{1\over G_{zz}}
(G_{z\mu} dx^\mu + G_{zx} dx + G_{zy} dy )^2 \cr
& + G_{\mu\nu} dx^\mu dx^\nu + 2 G_{x \nu} dx dx^\nu + 2 G_{y \nu} dy dx^\nu +
2 G_{xy} dx dy + G_{xx} dx^2 + G_{yy} dy^2.}}
The above compact form can be simplified even further for the particular
example we are interested in. More concretely the $G_{mn}$
components ($m,n = \mu, x, y, z$) become rather simple if one
assumes the following choices of $j_{mn}, b_{mn}$
\eqn\chofjb{j_{\mu x} = j_{\mu y} = j_{\mu z} = 0; ~~~ b_{xy} =
b_{zx} = b_{zy} = 0; ~~~ b_{\mu\nu} = 0.}
In this case the negative components in the metric vanish. The
above assumption implies that the type IIB metric of a D5
wrapping an $S^2$ of the resolved conifold has no off-diagonal
components. This can be easily checked and we shall elaborate this
further later on. The type IIB $b$ field choice tells us that
off-diagonal components are allowed but the cross terms vanish. This
can also be verified easily. With this choice of type IIB metric
the metric components of the mirror take the following form

\eqn\nowgmunu{G_{\mu\nu} = j_{\mu\nu} + {b_{x\mu} b_{x \nu} \over j_{xx}} +
 {(b_{y\mu}j_{xx} - b_{x\mu} j_{xy})(b_{y\nu} j_{xx} - b_{x\nu} j_{xy}) \over
j_{xx}(j_{yy} j_{xx} - j_{xy}^2)},}

\eqn\nowgzxy{G_{\mu z} ~=~ G_{zx} ~= ~G_{zy} ~= ~0,}

\eqn\nowgyyzzxx{G_{xx} = {j_{yy} \over j_{yy} j_{xx} - j^2_{xy}},
~~ G_{yy} = {j_{xx}\over j_{yy} j_{xx} - j_{xy}^2}, ~~G_{zz} =
j_{zz}- {j_{xz}^2 \over j_{xx}} - {(j_{yz}j_{xx} - j_{xy}
j_{xz})^2 \over j_{xx}(j_{yy} j_{xx} - j_{xy}^2)},}

\eqn\nowgxy{G_{xy} = {-j_{xy} \over j_{yy}j_{xx} - j_{xy}^2},}

\eqn\nowgmuxz{G_{\mu x} = {b_{\mu x} \over j_{xx}} + {(b_{y
\mu}j_{xx} - b_{x \mu} j_{xy})j_{xy} \over j_{xx}(j_{yy}j_{xx} -
j_{xy}^2)}, ~~ G_{y \mu} = -{b_{y \mu}j_{xx} - b_{x \mu} j_{xy}
\over j_{yy}j_{xx} - j_{xy}^2}.}
On the other hand, the ${\cal B}$ fields appearing in the metric
and fluxes (not to be confused with the ${\tilde B}$ fields in
the type IIA picture) take the form

\eqn\nowbmunu{{\cal B}_{\mu \nu}~ = ~ {\cal B}_{y \mu} ~ = ~
{\cal B}_{x \mu} ~ = ~ {\cal B}_{xy} ~ = 0,}

\eqn\nowbmuzxy{ {\cal B}_{\mu z} = b_{\mu z} + {b_{x\mu} j_{xz}
\over j_{xx}} - {(j_{yz}j_{xx} - j_{xy}j_{xz})(b_{\mu y} j_{xx} -
b_{\mu x} j_{xy}) \o j_{xx}(j_{yy}j_{xx} - j_{xy}^2)},}

\eqn\nowbzx{{\cal B}_{zx} = {j_{zx} \over j_{xx}} -
{j_{xy}(j_{zy}j_{xx} - j_{xy}j_{xz}) \over j_{xx}(j_{yy}j_{xx} -
j_{xy}^2)}, ~~~~ {\cal B}_{y z} = -{j_{yz}j_{xx} - j_{xy} j_{zx}
\over j_{yy} j_{xx} - j_{xy}^2}.}
Using the above choices of $G_{mn}$ and ${\cal B}_{mn}$ one can
easily show that all components of the mirror NS flux vanish,
${\tilde B}_{mn} = 0$.

Naively one would expect that the background of the mirror
manifold that we just derived corresponds to a deformed conifold
of \ks\ in the presence of fluxes. In order to see the relation to
the deformed conifold  our mirror background can be simplified
further. But before doing so, let us rewrite the deformed conifold
background of \ks\ in a suggestive way so that a comparison can be
made.

\subsec{Rewriting the Deformed Conifold Background}

The metric of a D6 brane wrapping a three cycle of a deformed
conifold has been discussed earlier in \edelstein. Let us
recapitulate the result. To obtain the metric, one defines a
K\"ahler potential ${\cal F}$ as a function of $\rho^2 \equiv
{\rm tr} (W^\dagger W)$ with $W$ being a complex $2 \times 2$
matrix satisfying ${\rm det}~W = -\epsilon^2/2$. The quantity
$\rho$ is basically the radial parameter and $\epsilon$ is a real
number. The generic form of the Ricci flat K\"ahler background is
determined from (see \candelas\ for details) \eqn\conimet{ds^2 =
{\cal F}'~{\rm tr} (dW^\dagger dW) + {\cal F}''~\vert {\rm
tr}(W^\dagger dW)\vert^2,} where the primes are defined as ${\cal
F}' = d{\cal F}/d\rho^2$ and the determinant of the deformed
conifold metric is given by $\epsilon^{-8}(\rho^4 -
\epsilon^4)^2$. For a Ricci flat metric ${\cal F}'$ becomes equal
to \eqn\feqto{ {(\sqrt{2}\epsilon)^{-{2\over 3}} (2\epsilon^2
\rho^2 \sqrt{\rho^4 - \epsilon^4} - 2 \epsilon^6~ {\rm
ch}^{-1}(\rho^2/\epsilon^2))^{1/3} \over
\sqrt{\rho^4-\epsilon^4}}.} In this form it is not too difficult
to write the metric of a bunch of D6 branes wrapping the three
cycle of a deformed conifold. If we take the limit $\rho \to
\epsilon$ the metric becomes the metric of an $S^3$ space. The
generic form of the wrapped D6 metric is \eqn\gendsixmet{ds^2 =
{\cal A}_0~ ds^2_{0123} + {\cal A}_1~ d\rho^2 + {\cal A}_2
~ds_1^2 + {\cal A}_3 ~ds_2^2 + {\cal A}_4~ ds_4^2 + {\cal A}_5~
ds^2_3,} where ${\cal A}_i$ are some specific functions of the
radial coordinate $\rho^2$ and the metric components $ds_i$ are
given by \eqn\metcomptwo{\eqalign{ & ds_1^2 = (d\psi + {\rm cos}
\theta_1~d\phi_1 + {\rm cos} \theta_2~d\phi_2)^2, \cr & ds_2^2 =
d\theta_1^2 + {\rm sin}^2\theta_1~d\phi_1^2, ~~~ ds_4^2 =
d\theta_2^2 + {\rm sin}^2\theta_2~d\phi_2^2, \cr & ds_3^2 = 2~{\rm
sin} \psi~ (d\phi_1 d \theta_2~{\rm sin} \theta_1 + d\phi_2 d
\theta_1~{\rm sin} \theta_2) + 2~{\rm cos} \psi ~( d\theta_1
d\theta_2 - d\phi_1 d\phi_2 ~{\rm sin} \theta_1 {\rm sin}
\theta_2).}} The appearance of ${\rm sin} \psi$ and ${\rm cos}
\psi$ in the above metric is a little disconcerting as the expected
$U(1)$ symmetry acting on $\psi$ as $\psi \to \psi + c$ is not
present \minasianone. This means that the deformed conifold {\it
cannot} be written as a simple $T^3$ fibration over a three
dimensional base. An immediate consequence of this is that the
usual SYZ technique cannot be applied. On the other hand, D5
branes wrapped on a resolved conifold do have the required $U(1)$
isometries related to constant shifts in $\psi, \phi_1, \phi_2$,
so that we can perform three T-dualities and obtain the mirror
manifold, as we are doing in this paper. The $T^3$ fibration
corresponds to the $\psi, \phi_1, \phi_2$ torus. In the notations
of the previous subsection, ($\psi, \phi_1, \phi_2$)~$\to$~($z, x, y$)
for the type IIB resolved conifold (we will give a more precise 
mapping soon).

In order to compare the deformed conifold metric with the metric
obtained for our mirror manifold, let us assume that we fix the
value of $\psi$ in $ds_3$ in \metcomptwo\ as $\psi = \psi_0$. 
It is then convenient to perform the following
change of coordinates
 $\theta_2, \phi_2$: \eqn\tranthe{
\pmatrix{{\rm sin}~\theta_2~ d\phi_2 \cr d\theta_2} \to
\pmatrix{{\rm cos}~ \psi_0 & {\rm sin}~ \psi_0 \cr -{\rm sin}~ \psi_0 &
{\rm cos}~ \psi_0} \pmatrix{{\rm sin}~\theta_2 ~d\phi_2 \cr
d\theta_2},} with the other coordinates $\theta_1$ and $\phi_1$
remaining unchanged. Although identical in spirit, the above
transformation is {\it different} from equation (2.2) of
\minasianone. Under the transformation \tranthe, the metric
component $ds_3^2$ changes to \eqn\dsthree{ds_3^2 ~ \to~ 2
d\theta_1~d\theta_2 - 2{\rm sin}~\theta_1~{\rm
sin}~\theta_2~d\phi_1 ~d\phi_2,} so that the $\psi_0$ dependence is
completely removed. Although this may mean that we regain the
$U(1)$ isometry but this is only because of the obvious delocalisation procedure. 
In general, for non-constant $\psi$ in \tranthe, the $\psi$ dependence would enter into $ds_1^2$ in
such a way that a shift in the other coordinates would fail to remove
it. Observe that the metric components $ds_2^2$ and $ds_4^2$
remain unaltered.

To summarize: the above observation tells us that a simple $T^3$
fibration of a deformed  conifold does not exist. In a new
coordinate system (which is discussed in \minasianone) it might
be possible to regain some of the $U(1)$ isometries, although a
SYZ description of the corresponding mirror appears futile (see \hori\ 
for some proposals to reconcile this were given).
Nevertheless, the above change of coordinates will be useful to
understand the type IIA mirror background and compare the
metric with the expected deformed conifold metric.

In the following we will rewrite the metric \gendsixmet\ in such a
way that it can be mapped to the mirror manifold obtained by using
three T-dualities. The readers interested in the type IIA mirror
background may want to skip this part and go directly to the next
section. If we define a warp factor $h \equiv h(\rho)$ then the
generic metric of D6 branes wrapping a three cycle of a deformed
conifold can be written as \eqn\nordi{ds^2 = h^\alpha~
ds^2_{0123} + h^\beta~ dr^2 + h^\gamma~ ds_1^2 + h^\delta~(ds_2^2
+ ds_4^2) + h^\rho~ds_3^2,} where ${\alpha,\beta,\gamma,\rho}$
are the various numerical powers for the wrapped D6 branes
(compare with the previous formula \gendsixmet)  and $ds_i$ ($i =
1, 2, 3$) are defined earlier. We have also taken a slightly
simplified case where corresponding to ${\cal A}_3 = {\cal A}_4$
in our earlier notation, as this is the case we are interested in.
The above metric can now be written in a more suggestive way \ohta
\eqn\metsugges{\eqalign{ds^2 = & h^\alpha~ds^2_{0123} +
(h^{\beta/2}dr)^2 + h^\gamma ~(d\psi + {\rm cos}~\theta_1~d\phi_1
+ {\rm cos}~\theta_2~d\phi_2)^2 + \cr & ~~~~ + (h^\delta -
h^\rho)~ ({\rm sin}~\psi~{\rm sin}~\theta_1~d\phi_1 + {\rm
cos}~\psi~d\theta_1 - d\theta_2)^2 ~ + \cr & ~~~~ + (h^\delta -
h^\rho)~ ({\rm cos}~\psi~{\rm sin}~\theta_1~d\phi_1 - {\rm
sin}~\psi~d\theta_1 + {\rm sin}~\theta_2~d\phi_2)^2 ~ + \cr &
~~~~ +  (h^\delta + h^\rho)~ ({\rm sin}~\psi~{\rm
sin}~\theta_1~d\phi_1 + {\rm cos}~\psi~d\theta_1 + d\theta_2)^2 ~
+ \cr & ~~~~ + (h^\delta + h^\rho)~ ({\rm cos}~\psi~{\rm
sin}~\theta_1~d\phi_1 - {\rm sin}~\psi~d\theta_1 - {\rm
sin}~\theta_2~d\phi_2)^2.}} 
The above metric cannot be the full global picture, as we know that 
the corresponding type IIB manifold has extra seven branes. Thus we
have to add {\it extra} six branes in this scenario to complete the picture.
However we can still study the {\it local} metric using different set of coordinates
that are more suited for this case.  
Our next goal therefore is to rewrite this
metric in terms of the T-dual coordinates $x, y, z$ that we used
earlier to get the fields of the mirror manifold, by using the
identification ($\psi, \phi_1, \phi_2$)~$\to$~($z, x, y$) and
$\mu,\nu =
\theta_1, \theta_2$. Omitting the $r$ and the $ds^2_{0123}$ term,
we might expect 
the metric \metsugges\ can be written as
\eqn\metwritxyz{\eqalign{ds^2 = & ~~~h^\gamma ~(dz + f^1_{zx}~dx
+ f^2_{zy}~dy)^2 + (h^\delta - h^\rho)~(f^3_{xx}~dx +
f^4_{x\mu}~dx^\mu)^2 + \cr &+ (h^\delta - h^\rho)~(f^5_{yy}~dy +
f^6_{xy}~dx + f^7_{x\mu}~dx^\mu)^2 + (h^\delta +
h^\rho)~(f^8_{xx}~dx + f^9_{x\mu}~dx^\mu)^2 + \cr &~~~~~ ~~~~ +
(h^\delta + h^\rho)~(f^{10}_{yy}~dy + f^{11}_{xy}~dx +
f^{12}_{x\mu}~dx^\mu)^2},} where $f^i_{mn}, ~i= 1, 2,...,12, ~~m,n
= x, y, z$ can be easily related to the coefficients in
\metsugges\ once the precise relation between ($\psi, \phi_1, \phi_2$) and 
($z, x, y$) is spelled out. 
Having written the metric in the form of \metsugges\
still does not tell us that D6 branes wrapped on
$S^3$ of a deformed conifold should have a $T^3$ fibration,
because of the appearance of ${\rm sin}~\psi$ and ${\rm
cos}~\psi$ in the product. On the other hand writing the metric in
the form \metwritxyz\ will help us to relate it to the mirror
metric that we derived in the previous section. We will do this
in the next section.

Let us comment a little more on the transformation \tranthe. As
mentioned earlier, this transformation with non-constant $\psi$ would
 remove the $\psi$
dependence in $ds_3$ and bring it to the form \dsthree. However
the $d\psi$ fibration structure will now change because ${\rm
cos}~\theta_2~d\phi_2$ changes under \tranthe. The change will
generically introduce some terms proportional to $d\theta_2$ in
the $d\psi$ fibration structure. The precise change will be
\eqn\psichange{{\rm cot}~\theta_2~dy ~\to ~ {\rm
cot}~{\bar\theta_2}~({\rm cos}~\psi~dy + {\rm
sin}~\psi~d\theta_2),} where $\bar\theta$ is the change in
$\theta$ under the transformation \tranthe. Now the change
\psichange\ explicitly introduces the $\psi$ dependence in the
fibration structure but removes it from the other parts of the
metric. In the {\it delocalized} limit, the $\psi$ values are
basically constant and therefore can be
approximated by constants. {\it This is the only assumption that
we will consider at this stage}. Under this assumption the
$d\theta$ dependent term appearing in the fibration structure
\psichange\ can be absorbed by a shift in $d\psi$ as $d\psi ~\to
~d(\psi - a~{\rm ln~ sin}~{\theta_2})$, where we have approximated
$\bar\theta$ by $\theta$ and $a$ is a constant. Under this
transformation and in the delocalization limit the $d\psi$
fibration structure does not change too much from its original
value. In this way we can recover a simplified form of the
deformed conifold metric. Observe that this doesn't mean that we
generate a $U(1)$ isometry in a theory that didn't have an
isometry before transformation. We can only get the metric with
$\psi$ isometry in the delocalization limit.

Thus to summarise, we can get rid of the $\psi$
dependences in $ds_3$ by restricting to a specific value of
$\psi$, i.e $\psi = \psi_0 \equiv \langle\psi\rangle$. This choice of $\psi$ can be easily
obtained from the mirror map (that we are going to discuss
in the next section). 
The mirror can be determined from performing
three T-dualities along the $z, x$ and $y$
directions. T-dualities require that we delocalize the  $z,
x$ and $y$ directions. Since the resolved conifold
metric is already independent of these directions, delocalizing
simply amounts to setting $\psi = \psi_0 \equiv \langle\psi\rangle$ in \tranthe\
when we do the transformation. A
somewhat related discussion on the transformation of $ds_3^2$ has
been given in \ohta. Therefore we will consider the specific
delocalized limit of the deformed conifold where the $\psi_0$
dependences in $ds_3$ will appear from \dsthree\ by applying \tranthe.
Later on we will argue how generic value of $\psi$ can appear in the metric.

For completeness and since we will need these expressions for
later comparison we will list the expressions for the vielbeins
describing $D6$ wrapped on an $S^3$ of a deformed conifold without any
additional six-branes as:
\eqn\vieiib{\eqalign{& e^1_{\phi_1} = \sqrt{h_+}~{\rm cos}~\psi ~
{\rm sin}~\theta_1, ~~~ e^1_{\theta_1} = -\sqrt{h_+}~{\rm
sin}~\psi, ~~~~e^1_{\phi_2} = \sqrt{h_+}~{\rm sin}~\theta_2 \cr &
e^2_{\phi_1} = \sqrt{h_+}~{\rm sin}~\psi ~ {\rm sin}~\theta_1,
~~~e^2_{\theta_1} = \sqrt{h_+}~{\rm cos}~\psi,
~~~~~~e^2_{\theta_2} = \sqrt{h_+}\cr & e^3_{\phi_1} =
\sqrt{h_-}~{\rm cos}~\psi ~ {\rm sin}~\theta_1, ~~~
e^3_{\theta_1} = -\sqrt{h_-}~{\rm sin}~\psi, ~~~~e^3_{\phi_2} =
\sqrt{h_-}~{\rm sin}~\theta_2 \cr & e^4_{\phi_1} =
\sqrt{h_-}~{\rm sin}~\psi ~ {\rm sin}~\theta_1, ~~~
e^4_{\theta_1} = \sqrt{h_-}~{\rm cos}~\psi, ~~~~~~e^4_{\theta_2} =
-\sqrt{h_-}\cr & e^5_\psi = \sqrt{h^\gamma}, ~~ e^5_{\phi_1} =
\sqrt{h^\gamma}~ {\rm cos}~\theta_1, ~~ e^5_{\phi_2} =
\sqrt{h^\gamma}~{\rm cos}~\theta_2, ~~e^6_r = \sqrt{h^\beta}}}
where $h_+ = h^\delta + h^\rho$ and $h_- = h^\delta - h^\rho$.
{}Once we express these vielbeins using our
local coordinates, many useful properties of the background such as the fundamental
form, holomorphic three form etc., can be easily extracted.

\newsec{Chain 1: The Type IIA Mirror Background}

In this section we will determine the exact form of the mirror
manifold and apply our generic formulas to the special case $D5$
branes wrapping a resolved conifold in the type IIB theory. We
will find that the manifold is not quite a deformed conifold in
the presence of fluxes as one would have naively expected, rather
it will turn out to be a non-K\"ahler manifold that could even be
non-complex.

To determine the precise form of the manifold let us first present
the metric for D5 branes wrapped on an $S^2$ of a resolved
conifold. It is given in \pandoz\ in the following form
\eqn\metresconi{\eqalign{ds^2 = & ~h^{-1/2} ds^2_{0123} + h^{1/2}
\Big[\gamma' dr^2 + {1\o 4} \gamma' r^2 (d\psi + {\rm
cos}~\theta_1 d\phi_1 + {\rm cos}~\theta_2 d\phi_2)^2 + \cr &
~~~+ {1\o 4}\gamma(d\theta_1^2 + {\rm sin}^2~\theta_1 d\phi_1^2)
+ {1\o 4}(\gamma + 4a^2) (d\theta_2^2 + {\rm sin}^2~\theta_2
d\phi_2^2)\Big],}} where we have used the notations of \pandoz\
and $\gamma$ is defined as a function of $r^2$ only\foot{We have
also defined the radial coordinate $r$ and $\gamma'$ in the
following way: $\gamma' \equiv {d\gamma \o dr^2} = {2\o 3}
{\sqrt{\gamma + 6a^2}\o \gamma + 4a^2}$ and the Ricci flatness
condition gives rise to the equality $r = \sqrt{\gamma
\sqrt{\gamma + 6a^2}}$.}. The presence of wrapped
$D5$ branes in the metric is
signalled by the harmonic function $h$ whose functional form can be 
extracted from \pandoz. 
Observe that the parameter $a$ creates
an asymmetry between the two spheres denoted by $\theta_1,
\phi_1$ and $\theta_2, \phi_2$. As discussed in \pandoz, for
small $r$ (the radial coordinate) the $S^3$ denoted by $\psi,
\theta_1, \phi_1$ shrinks to zero size whereas the other sphere
remains finite with radius $a$. This is the resolving parameter.
As can be easily seen, when the resolving parameter goes to zero
size, the manifold becomes a conifold, and the metric works out
correctly. Furthermore, the curvature remains regular all
through. Notice also that the metric \metresconi\ has three
isometries related to constant shifts in $\psi, \phi_1$ and
$\phi_2$ as $\psi \to \psi + c_1, \phi_1 \to \phi_1 + c_2, \phi_2
\to \phi_2 + c_3$. But there are no isometries along $r,
\theta_1$ and $\theta_2$ directions because of the warp factors
and the $d\psi$ fibration structure. Therefore there is a natural
$T^3$ structure associated with $\psi, \phi_1, \phi_2$
directions. This $T^3$ could be a special lagrangian submanifold if the global metric of \pandoz\ 
preserve supersymmetry (see also \karch).  
A direct way to 
see this would be to evaluate the condition required for a cycle to be lagrangian. 
Alternatively, one can see that the $\phi_1, \phi_2$ directions lead to a brane-box
configuration after two T-dualities \karch. This configuration preserves susy when the size of the 
resolution circle is zero. Furthermore, there a
T-duality along $\psi$ direction has been shown earlier to lead to a susy 
preserving configuration \dmconi. Thus
$\psi, \phi_1, \phi_2$ lead to a lagrangian submanifold that preserves susy after three T-dualities in the 
conifold limit. Unfortunately the global metric of \pandoz\ do not preserve susy \cvetic, but our M-theory configuration
does. In sec. 3 we showed that the fourfold would preserve susy with primitive fluxes. We took a fourfold that locally 
looks like a product manifold of a $T^2$ fiber and a resolved conifold base. It is now clear \bdkkt\ that 
{\it globally} the manifold
will preserve susy when the fiber degenerates over the base. This means that we would require seven branes, and 
the base will be K\"ahler instead of a Calabi-Yau manifold. 
The product structure of sec. 3 should then be regarded as
though we have moved the seven branes far away. In fact this is exactly how we can study pure ${\cal N } =1$ 
$SU(N)$ gauge theory! This way one might expect
the three cycle ($\psi, \phi_1, \phi_2$) would form a Lagrangian submanifold
on which we can do T-dualities. 
However, as we will see below, the T-duality
directions are not the naively expected isometry directions. The
T-dualities in this scenario are a little subtle as we now
elaborate. 

To begin, we need to first convert ($\psi, \phi_1, \phi_2$) into suitable coordinate system by which we 
can express our local metric and also perform the mirror map, as the original coordinate choice do not
suffice
(see our recent work \bdkkt\ for details). Since the global
metric of \pandoz\ break susy, we can only trust the local metric; and then add seven branes to make the system
supersymmetric. The full global picture is now understood in \bdkkt. Thus to write the local metric,
we can use the T-duality coordinates ($z, x, y$)
that we referred to in the previous section. Therefore we shall use the following
mapping to relate the above metric to the one 
presented earlier
\eqn\miapmet{\eqalign{& (x, y, z) ~\to ~(\phi_1, \phi_2, \psi),
\cr & (dx, dy, dz) ~ = ~ \left({1\o 2}{\sqrt{h^{1/2}\gamma}}~{\rm
sin}~\langle\theta_1\rangle~d\phi_1, {1\o 2} {\sqrt{h^{1/2}(\gamma+
4a^2)}}~{\rm sin}~\langle\theta_2\rangle~d\phi_2, {1\o 2} {r_0\sqrt{\gamma'
h^{1/2}}}~ d\psi \right).}} 
where we have picked a point ($r_0, \langle\theta_i\rangle, \langle\phi_i\rangle, \psi_0$) to define 
\miapmet. To avoid clutter $\gamma \equiv \gamma(r_0), \gamma' \equiv \gamma'(r_0), h \equiv h(r_0)$ 
henceforth unless mentioned 
otherwise. 
The physical meaning of
$x,y,z$ can be given as follows: under a single T-duality along $\psi$,
the system maps to an intersecting brane configuration \dotu, \dotd, 
\ohta; $x, y$ and $z$ form the coordinates of the branes. More precisely,
we are in fact converting the two spheres with coordinates 
($\phi_1, \theta_1$) and ($\phi_2, \theta_2$) to tori with coordinates
($x, \theta_1$) and ($y, \theta_2$) respectively. Recall that a sphere 
is topologically the same as a tori with a {\it degenerating} cycle 
(i.e. if we shrink one of the cycles of the $T^2$ to zero size then this would 
be topologically the same as a sphere) and 
therefore this mapping would be locally indistinguishable (but will not have the 
full [0, $2\pi$] isometries of $\phi_i$ in \metresconi). 
Furthermore, this mapping will 
be particularly useful to perform many simplifying manipulations later in the
paper which are otherwise difficult in the absence of the full global metric.
The local metric will become \bdkkt:
\eqn\locmet{\eqalign{ds^2 ~=~ & dr^2 + \Bigg(dz + \sqrt{\gamma' \o \gamma}~r_0 ~{\rm
cot}~\langle\theta_1\rangle~ dx  + \sqrt{\gamma' \o (\gamma+4a^2)~}~r_0~ {\rm
cot}~\langle\theta_2\rangle~dy\Bigg)^2~ + \cr
& ~~~~~~~~~~~ + \Bigg[{\gamma\sqrt{h} \o 4}~d\theta_1^2 + dx^2\Bigg] +
\Bigg[{(\gamma + a^2)\sqrt{h} \o 4}~d\theta_2^2 + dy^2\Bigg] + ....}}
where we see that the two tori are square tori. We have to soon modify this further, but before that let us 
find the effect of the
seven branes in this scenario. 
Locally, if we keep the seven branes very far away, then the metric will be \locmet. When
the seven branes are somewhat nearby, but still far away so that we can study pure ${\cal N} =1$ SYM, 
we can keep the radial direction delocalized, but $\theta_i$ arbitrary, i.e ($r_0, \langle\theta_i\rangle$) $~\to~$
($r_0, \theta_i$) in \locmet. 
With this map,
we can now write the
various components of the wrapped D5 metric:
\eqn\comedfi{\eqalign{& j_{zz} = 1 , ~~j_{xx} =  1 +  {\gamma' \o
\gamma}~ r_0^2 ~{\rm cot}^2~\theta_1, \cr & j_{yy} =  1 +  {\gamma'
\o \gamma + 4a^2}~ r_0^2~ {\rm cot}^2~\theta_2, \cr & j_{zx} =
{\sqrt{\gamma'\o \gamma}}~ r_0~ {\rm cot}~\theta_1, ~~ j_{zy} =
\sqrt{\gamma' \o \gamma + 4a^2} ~r_0 ~{\rm cot}~\theta_2, \cr &
j_{xy} =
 {\gamma' \o \sqrt{\gamma(\gamma + 4a^2)}}~ r_0^2~
 {\rm cot}~\theta_1 ~{\rm cot}~\theta_2, ~~j_{rr} = \gamma' h^{1/2}, \cr
& j_{\theta_1\theta_1} =  {1\o 4}\gamma h^{1/2}, ~~
j_{\theta_2\theta_2} = {1\o 4} h^{1/2} (\gamma + 4a^2)}} with the
rest of the components zero. The $B_{NS}$ fields on the other
hand have the following components (see also sec. 4 of \pandoz):
\eqn\bfico{b = {\cal J}_1 ~d\theta_1 \wedge dx + {\cal J}_2~
d\theta_2 \wedge dy} with the rest of the components zero and
${\cal J}_i$ are now functions of the radial and the angular coordinates {\it globally},
i.e ($r, \theta_1, \theta_2$) although we have used local coordinate differentials to write 
$b$ over a given coordinate patch. 
In \pandoz\ the $B$ field was only function
of the radial coordinates. Here since we converted all the spheres in the 
metric to tori, we will keep $B$ as a generic function of ($r, \theta_1, 
\theta_2$) to preserve supersymmetry globally. 
Notice also that the choice of the $B$ field and the metric is 
consistent with the assumptions that we made in
the previous section, namely $b_{xy} = b_{yz} = b_{zx} = 0$ and
the cross term $j_{(x,y,z)\mu} = 0$. The small
and the large radius behavior of $\gamma$ is \pandoz:
\eqn\smlarbe{\gamma_{r_0\to 0} = {1\o \sqrt{6} a} r_0^2 - {1\o 72
a^4} r_0^4 + {\cal O}(r_0^6), ~~~~ \gamma_{r_0 \to \infty} = r_0^{4/3} -
2 a^2 + {\cal O}(r_0^{-4/3})} and so we can use the first relation to 
define $\gamma(r_0)$. 
In the above set of components
\comedfi, if we ignore the overall $h^{1/2}$ dependences, observe
that for small $r$ (which we will concentrate on mostly) $\gamma$
is a small quantity, and thus terms like $j^{-1}$ (which we will
encounter soon) could be expanded in powers of $\gamma$ (or $r$),
because the $\theta_i$ dependences can be made generically small.
We will however try to avoid making approximations and
concentrate on the exact values as far as possible.

Before moving ahead one comment is in order. The metric of $D5$
wrapping an $S^2$ of a resolved conifold has  no $j_{\theta_1
\theta_2}$ component, i.e. no $d\theta_1 d\theta_2$ cross term.
However, our anticipation will be to have such a cross term in
the mirror, see e.g. \metcomptwo. We know that T-dualities {\it
cannot} generate such terms (in the absence of $B$ fields). In
the presence of $B$ fields, as we show below, cross term of the
form $d\theta_1 d\theta_2$ do get generated. However these cross
terms combine together with $dx$ and $dy$ terms (as will become
obvious soon) and therefore do not generate the {\it single}
$d\theta_1 d\theta_2$ term. We will discuss a way to generate
this later.

The expected mirror manifold will have the following form of the
metric \metcomp: \eqn\metcomthree{\eqalign{ds^2 = & {1\over
G_{zz}} (dz + {\cal B}_{\mu z} dx^\mu + {\cal B}_{xz} dx + {\cal
B}_{yz} dy)^2 + \cr & + G_{\mu\nu} dx^\mu dx^\nu + 2G_{x \nu} dx
dx^\nu + 2G_{y \nu} dy dx^\nu + 2G_{xy} dx dy + G_{xx} dx^2 +
G_{yy} dy^2.}} The $dz$ fibration structure is more or less
consistent in form, so lets check whether the components work out
fine. By denoting \eqn\defalpha{\alpha^{-1} = j_{xx} j_{yy} -
j^2_{xy} + b^2_{xy} = j_{xx} j_{yy} - j^2_{xy},} we write:
\eqn\bfibg{\eqalign{& {\cal B}_{xz} = -\alpha ~j_{xz} =
-\sqrt{\gamma'\o \gamma}~ \alpha~r_0~ {\rm cot}~\theta_1 \cr &
{\cal B}_{yz} = -\alpha ~j_{yz}  = -\sqrt{\gamma' \o \gamma +
4a^2} ~\alpha~r_0 ~{\rm cot}~\theta_2 \cr & {\cal B}_{\mu z} =
b_{\mu z} + \alpha(b_{x\mu} j_{xz} + b_{y \mu} j_{yz}).}} This
can combine together with \bfibg\ to give the following fibration
structure: \eqn\actfibstr{(dz - b_{z\mu}~dx^\mu) - \alpha~j_{xz}
(dx - b_{x\theta_1}~d\theta_1) - \alpha~j_{yz} (dy -
b_{y\theta_2}~d\theta_2)} where we have kept the $B$ field
component $b_{\mu z}$. The above form of the fibration is highly
encouraging because it looks similar to (4.36). And since $\alpha
= 1 +$ higher orders, up to those terms we seem to be getting the
fibration structure in somewhat expected form.
%The above fibration includes all
%the higher T-duality terms and is therefore exact (up to higher order $\alpha'$ corrections)

\noindent Let us now look at other terms. \eqn\othtern{\eqalign{&
G_{xx} =  \alpha~j_{yy}, ~~~~ G_{yy} =  \alpha~j_{xx} \cr &
G_{\mu\nu} = j_{\mu\nu} + \alpha(j_{yy}~b_{x\mu}~b_{x\nu} +
j_{xx}~b_{y\mu} ~b_{y\nu} - j_{xy}(b_{y\mu} ~b_{x\nu} + b_{x\mu}
~b_{y\nu}))}} The existence of cross terms in the above formula
is very important. This tells us that we can have components like
$G_{\theta_1 \theta_2}$. Such terms do exist in the usual
deformed conifold metric and are {\it absent} in the  resolved
conifold metric. The other terms will be:
\eqn\ottretop{\eqalign{& G_{xy} = {-j_{xy} \o j_{yy}j_{xx} -
j^2_{xy} + b^2_{xy}} = -\alpha~j_{xy} \cr & G_{x\mu} =
\alpha(j_{yy}~b_{\mu x} + j_{xy} b_{y\mu}), ~~~~ G_{y \mu} =
\alpha(j_{xx}~b_{\mu y} + j_{xy} b_{x \mu})}} Again, the
existence of cross term is important, they will give rise to
components like $G_{x \theta_2}$ and $G_{y \theta_1}$.
Such terms are {\it not} present in the resolved conifold
setting, but do exist in the deformed conifold metric! Finally
there is the $zz$ component \eqn\gzzc{G_{zz} = \alpha} The above
term is again of order one. Let us furthermore introduce the
shorthand notation \eqn\defAandB{j_{xz} = A = \Delta_1~{\rm
cot}~\theta_1, ~~~~ j_{yz} = B = \Delta_2~{\rm cot}~\theta_2,}
with $\Delta_1(r_0)$ and $\Delta_2(r_0)$ being warp factors. Now combining
everything together we get the following mirror manifold: 
\eqn\mirman{\eqalign{ds^2 = &~~ g_1~\left[(dz - b_{z\mu}~dx^\mu)
- \alpha~\Delta_1~{\rm cot}~\theta_1~ (dx - b_{x\theta_1}~d\theta_1) -
\alpha~\Delta_2~{\rm cot}~\theta_2~(dy - b_{y\theta_2}~d\theta_2)+
..\right]^2 + \cr &~~~~~ +  g_2~ d\theta_1^2 + g_3~d\theta_2^2   +
g_4~(dx - b_{x\theta_1}~d\theta_1)^2 +\cr & ~~~~~ + g_5~(dy -
b_{y\theta_2}~d\theta_2)^2 -  g_7~(dx -
b_{x\theta_1}~d\theta_1)(dy - b_{y\theta_2}~d\theta_2)}} where
$g_i \equiv g_i(r=r_0, \theta_1, \theta_2)$ are some functions of $r_0,
\theta_1, \theta_2$ coordinates and can be easily determined from
our analysis above. They are given as \eqn\gis{\eqalign{&
g_1=\alpha^{-1}, ~~~~~g_2={\gamma \sqrt{h} \o 4}, ~~~~~~ g_3=
{(\gamma+4a^2) \sqrt{h} \o 4}, \cr & g_4=\alpha~j_{yy}, ~~~~
g_5=\alpha j_{xx}, ~~~~~~ g_7=2\alpha~j_{xy}.}}

At this point let us compare our metric \mirman\ to the metric
of  the wrapped D6-branes on $S^3$ of a deformed conifold. The
generic form of that metric is given by (we take the {\it
delocalized} metric in \metcomptwo) \eqn\dsixdco{\eqalign{ds^2 =
& ~~ {\tilde g}_1~(dz + \tilde\Delta_1~{\rm cot}~\theta_1~dx +
\tilde\Delta_2~{\rm cot}~\theta_2~dy + ..)^2 + \cr & ~~ {\tilde
g}_2~ [d\theta_1^2 + dx^2] + {\tilde g}_3~[d\theta_2^2 + dy^2] +
{\tilde g}_4~[d\theta_1 ~d\theta_2 - dx~dy]}} where ${\tilde
g}_i$ are again some functions of $r_0, \theta_1, \theta_2$ that
could be easily evaluated. Let us now compare the two metrics:

\vskip.1in

\noindent $\bullet$ As a general rule, everywhere where we would expect
$dx$ or $dy$, they are replaced in \mirman\ by the appropriate
${\cal B}$--dependent fibration stucture
$(dx-b_{x\theta_1}d\theta_1)$ or $(dy-b_{y\theta_2}d\theta_2)$,
respectively. In fact, this non--trivial
fibration will be responsible for making the manifold \mirman\ a
non-K\"ahler space, as we will show later. If we define $d\hat
x=dx-b_{x\theta_1}d\theta_1$ and $d\hat y=dy-b_{y\theta_2}d\theta_2$
for {\it constant} $b_{x\theta_1}$ and $b_{y\theta_2}$,
we find agreement between \mirman\ and \dsixdco\ in all terms involving
$dx$ and $dy$, up to differing warp factors.

\noindent $\bullet$ We also see that the $d\theta_1~d\theta_2$ cross
term is now entirely absorbed in the fibration structure and there is
{\it no single} $d\theta_1~d\theta_2$ term in \mirman.
Whatever $d\theta_1 d\theta_2$
terms are generated actually combine with $dx$ and $dy$ to give
us the ${\cal B}$-dependent fibration, and therefore no {\it extra}
$d\theta_1 d\theta_2$ term appears.

\noindent $\bullet$ Apart from the $dx$-- and $dy$--fibration mentioned above
the $dz$ 
fibration structure of both the metrics have similar form modulo
some warp factors and relative signs.
The relative
signs between the $dx, dy$ terms in \mirman\ and \dsixdco\
can be fixed if we fix ${\cal B}_{yz},
{\cal B}_{zx}$ and ${\cal B}_{xy}$ in \bmuy, \bzx, and \bxy\ as
{\it minus} of themselves. In the following we will assume
that we have fixed the signs of the $B$ fields. This way the
fibration structures of \mirman\ and \dsixdco\ would tally.

\noindent $\bullet$ {}From the transformation \tranthe\ we expect
the  coefficients $g_3$ and $g_5$ to be the same. But a careful
analysis \gis\ shows that they are in fact different. The
coefficients $g_2$ and $g_4$ are also different, which could in
principle be because \tranthe\ do not act on them. But $g_{3,5}$
should be the same if we hope to recover the deformed conifold
scenario.

\noindent In the following we will try to argue a possible way
to  generate the $\theta$ cross terms. We will see that the
T-duality directions are slightly different from the naively
expected directions (which we called $x,y,z$). In the process we
will also fix the coefficients $g_i$. We begin with the search
for the $d\theta_1 ~d\theta_2$ term in the metric.

\subsec{Searching for the $d\theta_1 d\theta_2$ term}

The absence of the $d\theta_1 d\theta_2$ term in \mirman\ is a
near miss. As one can see that the metric that we get in \mirman\
is almost the metric of a deformed conifold (in the delocalized
limit) when we switch off the $B$ fields except that we are
missing the $d\theta_1 d\theta_2$ term. Furthermore this term
should come in the metric with the precise coefficient $\alpha
j_{xy}$.

The absence of this term however raises some doubts about the
directions that we made our T-dualities. But since we almost
reproduced the correct form of the metric, we cannot be too far
from the right choice of the isometry directions. Now whatever
new isometry directions we choose in the resolved
conifold\foot{We have been a little sloppy here. By resolved
conifold we will always mean $D5$ wrapped on $T^2$ of our local geometry
unless mentioned otherwise.}
 side should keep the present form of the mirror metric intact. This is a
 strong restriction because we cannot change the
$dz, dx$ and $dy$ fibration structure any more as they are
already in the expected format. The only things that we could
fiddle with are the $d\theta_i$ terms in the mirror side.
Therefore the question is: what changes in the resolved side are
we allowed to perform that would {\it only} affect the $d\theta_i$
parts of the mirror manifold?

An immediate guess would be to change the $\theta_i$ terms so as
to generate a $j_{\theta_1 \theta_2}$ directly in the resolved
conifold setup in type IIB theory \metresconi. A way to get this
would be to go to a new coordinate system given by:
\eqn\necosy{\theta_1 ~\to ~ \theta_1 + \gamma~\theta_2, ~~~~~
\theta_2 ~\to ~ \theta_2 + \beta~\theta_2,} where $\gamma, \beta$
are small integers. This will give us the necessary cross term
and will change the other terms to \eqn\otterto{{\rm
cot}~\theta_1~dx ~\to ~ (\gamma~\theta_2 + {\rm
cot}~\theta_1 + ...)~dx, ~~~ dx ~\to ~ (1 + \gamma~\theta_2~{\rm
cot}~\theta_1 + ...)~dx,} and similar changes to the $dy$ terms. In
other words the warp factors in front of the $dx,dy$ terms will
change and the metric will have a $j_{\theta_1 \theta_2}$ term.

Making a mirror transformation now to the resolved conifold
metric will generate the requisite $d\theta_1 d\theta_2$ term
term, but the coefficient of this is an arbitrary number.
Therefore will not explain the $\alpha j_{xy}$ coefficient that
we require. Instead of this, we can perform the following
infinitesimal rotation on the sphere coordinates of the type IIB
metric \metresconi: \eqn\rotonsp{ \pmatrix{dx \cr d\theta_1}~~\to
~~ \pmatrix{1 & \epsilon_1 \cr -\epsilon_1 & 1}\pmatrix{dx \cr
d\theta_1}, ~~~~~ \pmatrix{dy \cr d\theta_2}~~\to ~~ \pmatrix{1 &
\epsilon_2 \cr -\epsilon_2 & 1}\pmatrix{dy \cr d\theta_2}.} This
will keep the metric of the two tori ($\theta_1, x$) and
($\theta_2, y$) invariant, but will generate $j_{z\mu},
j_{x\mu}, j_{y\mu}$ and $j_{\mu\nu}$ components. Interestingly,
one can easily verify from the T-duality rules that this changes
only the  $G_{\mu\nu}$ and $G_{z\mu}$ components with all other
metric components $G_{mn}$ invariant. The change in $G_{\mu\nu}$
can be written as \eqn\metchgmunu{\eqalign{G_{\mu\nu} =
j_{\mu\nu} & ~+ \alpha~[j_{yy}~b_{x\mu}b_{x\nu} +
j_{xx}~b_{y\mu}b_{y\nu} - j_{xy}(b_{y\mu}b_{x\nu} +
b_{x\mu}b_{y\nu})] ~+ \cr & ~ -\alpha~[j_{yy}~j_{x\mu}j_{x\nu} +
j_{xx}~j_{y\mu}j_{y\nu} - j_{xy}(j_{y\mu}j_{x\nu} +
j_{x\mu}j_{y\nu})].}} The second line is by replacing $b
\leftrightarrow j$. Observe that only the second line in
\metchgmunu, which we shall call $G_{\mu\nu}^{\rm new}$,
introduces new components in the metric. Similarly, $G_{z\mu}$
can be written as: \eqn\gmuz{G_{\mu z} = j_{\mu z} -
\alpha~[j_{yy}~j_{x\mu}j_{xz} + j_{xx}~j_{y\mu}j_{yz} -
j_{xy}(j_{y\mu}j_{xz} + j_{x\mu}j_{yz})].} Both these components
will contribute to $d\theta_1^2, d\theta_2^2$ and $d\theta_1
d\theta_2$ (by keeping only the $j$ components) in addition to
the terms that we already have in \mirman, as: \eqn\condstwo{ds^2
\to ds^2 + G_{\mu\nu}^{\rm new}~dx^\mu dx^\nu - {G_{z\mu}
G_{z\nu} \o G_{zz}}~ dx^\mu dx^\nu.}

This is almost what we might require, but again the coefficient
is arbitrary. And there seems no compelling reason for a
particular coefficient to show up in the metric. Both the above
analysis have failed to provide a specific reason for the $\alpha
j_{xy}$ coefficient in the metric. Therefore it is time now to
look at other possibilities as we have exploited the
transformations on $x, y$ and $\theta_i$, but haven't yet
considered the possibilities of a $z$ transformation.
Can we change the $dz$ terms without spoiling the consistency of
the mirror manifold?

A little thought will tell us that the allowed changes should be
done directly to the line element, so that we can define
distances properly in both type IIB as well as the mirror type
IIA. We can start defining some transformation on $x, y, z$ and
$\theta_i$ using some (as yet) unknown functions. However, this
procedure is rather involved, because eventually we have to
determine $dx, dy, dz$ and $d\theta_i$ thereby giving rise to set
of PDE's. Therefore it will be easier if we make transformations
{\it directly} on $dx, dy, dz$ and $d\theta_i$. Transformations
on the infinitesimal shifts are, on the other hand, not always
integrable. Therefore to avoid this problem, let us start by
making transformations on {\it finite} shifts $\delta x, \delta
y, \delta z, \delta \theta_i$. We will later integrate these
expressions to get the transformations on the coordinates $x, y,
z$ and $\theta_i$. Furthermore as we are at a fixed 
point in the radial direction, i.e 
$r_0$, we
will consider shifts only for a fixed $r$. 

For {finite} shifts $\delta x, \delta y, \delta z, \delta
\theta_i$ of the coordinates of the resolved conifold a typical
distance $d$ on the resolved conifold can be written in terms of
distances $d_1, d_2$ on the two tori
with coordinates ($x,
\theta_1$) and ($y, \theta_2$) as: \eqn\lineel{\|~ d ~\| =
\sqrt{d_1^2 + d_2^2 + (\delta z + \Delta_1~{\rm
cot}~\theta_1~\delta x + \Delta_2~{\rm cot}~\theta_2~\delta y)^2}}
where we have ignored the contributions from the $dr^2$ term in \locmet\
just for simplicity. We have also defined the distances along
the two tori as: \eqn\linesp{d_1 = {1\o 2} \sqrt{\gamma
\sqrt{h} ~(\delta\theta_1)^2 + 4(\delta x)^2}, ~~~~ d_2 = {1\o 2}
\sqrt{(\gamma + 4 a^2) \sqrt{h} ~(\delta\theta_2)^2 + 4(\delta y)^2}.} Now
the  allowed change in the resolved side that would affect only
the $\delta\theta_i$ parts will be to change $\delta z$
to\foot{This can be motivated as follows. 
The generic shift between two
nearby points can be written as: $z_{(1)} - z_{(2)} = (\tilde
z_{(1)}-\tilde z_{(2)}) +
(\rho_{1(1)}~\theta_{1(1)}-\rho_{1(2)}~\theta_{1(2)}) +
 (\rho_{2(1)}~\rho_{2(1)}-\alpha_{2(2)}~\theta_{2(2)})$.
Now defining
$\rho_{j(1)}~\theta_{j(1)}-\rho_{j(2)}~\theta_{j(2)} =
\rho_j (\theta_{j(1)} - \theta_{j(2)})$, we get the
corresponding equation. Also since $\delta\theta_i = \theta_{i(1)} - \theta_{i(2)}$, 
there is no problem with the periodicity here. This will be clear later when we integrate these
shifts and write the final result in terms of ${\rm cos}~\theta_i$ and ${\rm sin}~\theta_i$, which are 
periodic variables.} 
\eqn\psipsi{\delta z ~\to~ \delta z +
\rho_1~\delta\theta_1 + \rho_2~\delta\theta_2.} This typically
means that we are slanting the $z$ direction along the $\theta_i$
directions. This would convert the line element  in \lineel\ to
\eqn\conichange{\|~ d ~\|^2 = d_1^2 + d_2^2 + \left[\delta\tilde
z + \sum_{i=1}^2 (\Delta_i~{\rm cot}~\theta_i~\delta x_i +
\rho_i~\delta\theta_i)\right]^2} with $\tilde z$ being the new
$z$ direction and $\rho_i$ are generic functions of ($r=r_0,
\theta_i$). We have also defined $x_1= x$ and $x_2 = y$. Now
without a loss of generality we can write the $\rho_i$'s as:
$\rho_1 = f_1~\Delta_1~{\rm cot}~\theta_1$ and $\rho_2 =
f_2~\Delta_2~{\rm cot}~\theta_2$ where $f_i$ are now generic
functions of ($r = r_0,\theta_i$) that we have to determine. Using
this, the line element will take the final form \eqn\fibcha{\|~ d
~\|^2 = d_1^2 + d_2^2  + \left[\delta\tilde z + \sum_{i=1}^2
(\Delta_i~{\rm cot}~\theta_1~\delta x_i + f_i~\Delta_i~{\rm
cot}~\theta_i~\delta\theta_i)\right]^2} The above changes could
also be viewed as having  generated the following new components
of the metric in the resolved side: $j_{\tilde z \theta_1},
j_{\tilde z \theta_2}, j_{x \theta_1}, j_{y\theta_2}$ and
$j_{\theta_1 \theta_2}$. As we discussed before these components
will only change the $\delta\theta_i$ part of the metric, i.e.
the $(\delta\theta_1)^2 + (\delta\theta_2)^2$ part and will keep all
the fibrations in the mirror picture intact. The changes in the
mirror metric can therefore be calculated from \condstwo.

Until now the arguments have been more or less parallel to the
arguments that we provided earlier for the changes in the $\delta
x, \delta y$ or $\delta\theta_i$ terms. However, as we show
below, the changes in the $\delta z$ part actually allows us to
fix the form of the functions $f_i$. To see how this is possible,
make the following changes in $\delta x, \delta y$ coordinates:
\eqn\dxdychange{\delta x ~\to~ \delta x - f_1~\delta\theta_1,
~~~~~ \delta y ~\to ~ \delta y - f_2~\delta \theta_2.} The effect
of this change is rather immediately obvious: it removes the
effect of the changes made earlier by the $\delta z$
transformation. However this change in $\delta x, \delta y$ can
also be assumed  as though $j_{x\theta_1}$ and $j_{y\theta_2}$
cross terms have been added in the metric. Taking into account
all the above changes, and also allowing a possible finite shift along the
radial direction $\delta r$, the line element on the resolved
conifold will take  the following form:
\eqn\metresconichange{\eqalign{\|~d~&\|^2  ~= ~ h^{1/2} ~\gamma'
(\delta r)^2 + (\delta \tilde z + \Delta_1~{\rm cot}~\theta_1~
\delta \tilde x + \Delta_2~{\rm cot}~\theta_2 ~\delta\tilde y)^2
+ (\delta\tilde x)^2 +  (\delta\tilde y)^2 +\cr & + {1\o
4}~(\gamma \sqrt{h}+4f_1^2)~(\delta\theta_1)^2  - 2f_1
\delta\tilde x~ \delta\theta_1 + {1\o 4}~(\gamma \sqrt{h} +
4a^2\sqrt{h} + 4f_2^2)~(\delta\theta_2)^2 - 2 f_2~\delta\tilde y~
\delta\theta_2.}} Observe that in the above line element cross
components have developed and the distances along the $\theta_i$
directions have changed by warp factors. Both these changes are
given in terms of the unknown functions $f_i$ (to be determined
soon)\foot{The physical meaning of $f_i$ will be discussed in the 
next subsection. For the time being we will view $f_i$ as being a consequence
of generic coordinate transformations, whose integral form will be presented
later in this section.}.

It is now important to consider some special limits of the
functions  $f_i$, as one can easily show that for small and
finite values of $f_1$ and $f_2$, a $\delta\theta_1
\delta\theta_2$ term does {\it not} get generated in the mirror
picture. Assuming that the $f_i$ are large, one could use a
special regularization scheme that would generate this term. The
coordinates $\tilde z, \tilde x_i, \theta_i$ are now the
coordinates in which the line element of the mirror manifold is
in a known format. Therefore the above transformation from ($z,
x, y$) ~$\to$~($\tilde z, \tilde x, \tilde y$) for the resolved
conifold means that we are writing the line element in the
coordinates of the mirror. To simplify the ensuing calculations,
let us use the notation introduced in \defAandB. With this
definition, we can re-express the line element \metresconichange\
as new components for the resolved conifold metric. The various
components can now be written as: \eqn\varcompo{\eqalign{&
j_{\tilde x \tilde x} = 1 + A^2, ~~ j_{\tilde y \tilde y} = 1 +
B^2, ~~ j_{\tilde x \theta_1} = - f_1, ~~ j_{\tilde y \theta_2} =
-f_2 \cr & j_{\theta_1 \theta_1} = {1\o 4}(4 f_1^2 + {\gamma
\sqrt{h}}), ~~ j_{\theta_2 \theta_2} = {1\o 4}(4f_2^2 + {\gamma
\sqrt{h}} + 4a^2\sqrt{h})  \cr &  j_{\tilde x \tilde z} = A, ~~
j_{\tilde y \tilde z} = B, ~~ j_{\tilde z \tilde z} = 1-
\epsilon, ~~ j_{\tilde x \tilde y} = AB}} with the radial and the
spacetime components remaining the same as earlier. Observe that
we have shifted the $j_{\tilde z \tilde z}$ component by a small
amount $\epsilon$ in \varcompo. Letting $\epsilon \to 0$ and
$f_i\to\infty$ will result in a finite $d\theta_1 d\theta_2$--term
in the mirror metric. This is our regularization scheme, so to
speak. Ergo, \varcompo\ is the correct IIB starting metric, which
we T--dualize along $\tilde x, \tilde y$ and $\tilde z$ to obtain
the mirror manifold. It can also be easily verified that both the
$B_{NS}$ and the $H_{RR}$ (given in the next section) remain
completely unchanged in forms because of their wedge structures and 
antisymmetrisations. The only change there is that now
everything is written by tilde-coordinates.

Let us now see the possible additional metric components that we
can  get after we make a mirror transformation. From the
T-duality rules we see that the new components are:
\eqn\newcom{\eqalign{& G_{\tilde z\theta_1} = -\alpha j_{\tilde
x\theta_1}[j_{\tilde x\tilde z} j_{\tilde y\tilde y} - j_{\tilde
x\tilde y} j_{\tilde y\tilde z}] = \alpha~ f_1~A \cr & G_{\tilde
z\theta_2} = -\alpha j_{\tilde y\theta_2}[j_{\tilde y\tilde z}
j_{\tilde x\tilde x} - j_{\tilde x\tilde y} j_{\tilde x\tilde z}]
= \alpha ~f_2~B \cr & G_{\theta_1 \theta_1} = j_{\theta_1
\theta_1} + \alpha~j_{\tilde y\tilde y}[b_{\tilde x\theta_1}^2 -
j_{\tilde x\theta_1}^2] = {\gamma\sqrt{h}\o 4} +
\alpha~(1+B^2)~b_{\tilde x\theta_1}^2 + \alpha~f_1^2 A^2 \cr &
G_{\theta_2 \theta_2} = j_{\theta_2 \theta_2} + \alpha~j_{\tilde
x\tilde x}[b_{\tilde y\theta_2}^2 - j_{\tilde y\theta_2}^2] =
{(\gamma + 4a^2)\sqrt{h} \o 4} + \alpha~(1+A^2)~b_{\tilde
y\theta_2}^2 + \alpha~f_2^2 B^2 \cr & G_{\theta_1 \theta_2} =
-\alpha j_{\tilde x\tilde y} [ b_{\tilde x\theta_1} b_{\tilde
y\theta_2} - j_{\tilde x\theta_1} j_{\tilde y\theta_2}] =
-\alpha~AB~b_{\tilde x\theta_1}~ b_{\tilde y\theta_2} +
\alpha~f_1~f_2~AB \cr & G_{\tilde z\tilde z} = \alpha(j_{\tilde
x\tilde x}j_{\tilde y\tilde y}j_{\tilde z\tilde z} -j_{\tilde
x\tilde x} j^2_{\tilde y\tilde z} - j_{\tilde y\tilde
y}j^2_{\tilde x\tilde z} - j_{\tilde z\tilde z} j^2_{\tilde
x\tilde y} + 2 j_{\tilde x\tilde y}j_{\tilde y\tilde z}j_{\tilde
z\tilde x}) = {\alpha - \epsilon}.}} The $B$ field dependent
terms in \newcom\ would reorganize themselves according to the
fibration structure that we discussed earlier. What remains now
is to see whether the additional terms (which depend on $f_i$)
can be used effectively. The distance along the
 new $\theta_1 \theta_2$ directions will be \condstwo:
\eqn\metonetwo{\eqalign{ds^2_{\theta_1 \theta_2} & = 2
\left(G_{\theta_1\theta_2}^{\rm new} - {G_{\tilde z\theta_1}
G_{\tilde z\theta_2} \o G_{\tilde z\tilde z}}
 \right)~\delta\theta_1~\delta\theta_2 \cr
& = - 2\alpha ~ f_1 f_2~j_{\tilde x\tilde y}\left[{\epsilon \o
\alpha - \epsilon}\right] ~\delta\theta_1~\delta\theta_2=
 - 2f_1f_2 ~j_{\tilde x\tilde y}
  ~\epsilon~\delta\theta_1~\delta\theta_2.}}
At this point we can use our freedom to define   the functions
$f_1$ and $f_2$. As discussed earlier, we see that for finite
values of the functions $f_i$ the above metric component is
identically zero in the limit $\epsilon\to 0$. However if we
define $f_i \equiv \epsilon^{-1/2} \beta_i$ such that $\beta_1
\beta_2 = -\alpha$, then \metonetwo\ implies
\eqn\metfin{ds^2_{\theta_1 \theta_2} = 2\alpha~j_{\tilde x\tilde
y}~\delta\theta_1 \delta\theta_2} which is what we require. In
this way we recover the elusive $\theta_1\theta_2$ component of
the metric.

At this point one might ask whether different choices for $f_1
f_2$ could be entertained. From the mirror metric \mirman\ we
observed that $-$ in the absence of $B$-fields $-$ the metric
resembles the deformed conifold in the delocalized limit (of
course with the $d\theta_1 d\theta_2$ absent). When we restore
back this term (via the above analysis) we should recover the
exact deformed conifold setup. This is possible, if in the
absence of $B$ fields, the product $f_1 f_2$ is proportional to
$\alpha$. Therefore, in the presence of fluxes, we believe that
this will continue to hold.

To show that the choice of $\beta_1 \beta_2 = -\alpha$ is
consistent, we have to determine $f_i$ individually. To see this
let us first bring the metric for the $\theta^2$ terms in
\newcom\ into a more canonical form:
\eqn\scaletheta{\delta\theta_1 ~\to~ {2 \o
\sqrt{h^{1/2}\gamma}}~\delta\tilde\theta_1, ~~~~~~~
\delta\theta_2 ~\to~ {2 \o
\sqrt{h^{1/2}(\gamma+4a^2)}}~\delta\tilde\theta_2.} Assuming now
that the above equation can be integrated to give a relation
between $\theta_i$ and $\tilde\theta_i$, the change in all terms
with a $b$--fibration can easily be absorbed in \bfico\ as:
\eqn\tildeb{\eqalign{b  & = ~~{2 {\cal J}_1 \o
\sqrt{h^{1/2}\gamma}}~d\tilde\theta_1 \wedge d\tilde x + {2 {\cal
J}_2 \o \sqrt{h^{1/2}(\gamma+4a^2)}}~d\tilde\theta_2 \wedge
d\tilde y \cr &= ~~ \tilde{b}_{\theta_1 x}~ d\tilde\theta_1
\wedge d\tilde x + \tilde{b}_{\theta_2 y}~ d\tilde\theta_2 \wedge
d\tilde y,}} where we have taken the infinitesimal limit to write
$b$ with one forms $d\theta_i$. The relation \tildeb\  implies
for the $\theta$ dependent metric components:
\eqn\gththco{\eqalign{G_{\tilde\theta_1 \tilde\theta_2} & =
-\alpha ~AB~\tilde{b}_{x\theta_1} \tilde{b}_{y\theta_2} +
\alpha~{4AB~f_1 f_2\o h^{1/2}\sqrt{\gamma(\gamma+4a^2)}} \cr
G_{\tilde\theta_1 \tilde\theta_1} & = 1 + \alpha (1+B^2)~
\tilde{b}_{x\theta_1}^2 + \alpha~{4\o h^{1/2}\gamma}~f_1^2 A^2,
\cr G_{\tilde\theta_2 \tilde\theta_2} & = 1 + \alpha (1+A^2)
~\tilde{b}_{y\theta_2}^2+ \alpha~{4\o h^{1/2}
(\gamma+4a^2)}~f_2^2 B^2.}} Note, that this changes \metonetwo\
to \eqn\metonetilde{\eqalign{ds^2_{\theta_1 \theta_2} & = -
2~{4f_1f_2 ~AB\o h^{1/2}\sqrt{\gamma (\gamma+4a^2)}}
~\epsilon~\delta\tilde\theta_1~\delta\tilde\theta_2,}} so we now
want to require \eqn\betaonetwo{\beta_1 \beta_2 = - {\alpha\o
4}~h^{1/2}\sqrt{\gamma(\gamma+4a^2)}.} To find out if this is
consistent with the other metric components, take a look at
\eqn\thetatwotwo{\eqalign{ds^2_{\tilde\theta_2 \tilde\theta_2}~
&~ = ~\left(G_{\tilde\theta_2 \tilde\theta_2}~
 - {G_{\tilde z\tilde \theta_2}
 G_{\tilde z\tilde\theta_2} \o G_{\tilde z\tilde z}}\right)
~(\delta\tilde\theta_2)^2 \cr &~ = ~\left(1 + \alpha (1+ A^2)
\tilde{b}_{y\theta_2}^2 - {4B^2~\beta_2^2 \o
h^{1/2}(\gamma+4a^2)}\right)~(\delta\tilde\theta_2)^2.}} {}From
the above analysis we see that the $\tilde b_{y\theta_2}$ term
will join $\delta\tilde y$ in the fibration as shown in \mirman,
and the rest of the term (which depends on $\beta_2$) will act as
a warp factor. The $\delta\tilde y$ term has warp factor $g_4$.
Can we use this to determine $\beta_2$?

At first this may seem impossible as we can have any coefficients
in front of $(\delta\tilde\theta_2)^2$ and $(\delta\tilde y -
\tilde{b}_{y\theta_2}~\delta\tilde\theta_2)^2$ or
$(\delta\tilde\theta_1)^2$ and the corresponding $(\delta\tilde x
- \tilde{b}_{x\theta_1}~\delta\tilde\theta_1)^2$. A little
thought will tell us that this is not quite true. Of course we
are allowed to have {\it any} coefficients in front of
$(\delta\tilde\theta_1)^2$ and $(\delta\tilde x - \tilde
b_{x\theta_1}~ \delta\tilde\theta_1)^2$, but the case for
$(\delta\tilde\theta_2)^2$ and
 $(\delta\tilde y - \tilde b_{y\theta_2}
~\delta\theta_2)^2$ is different. This is because of the
transformation \tranthe.  Under this transformation (in the
absence of fluxes) the coefficients of $\delta\theta_2$ and
$\delta y$ should be same so that under \tranthe\ the line
element does not change. Now we expect similar thing when we
switch on fluxes if we denote $\delta\hat y = \delta\tilde y -
\tilde b_{y\theta_2}~\delta\tilde\theta_2$, and define an
equivalent transformation between ($\delta\hat y,
\delta\tilde\theta_2$). Notice that there is no such constraint
on the $\delta\tilde x, \delta\tilde\theta_1$ term. Therefore
from the above argument and \gis\ we have the following equality:
\eqn\ftwos{ 1 - {4\o h^{1/2}(\gamma+4a^2)}~\beta_2^2 B^2 = \alpha
(1+A^2) ~~~\Rightarrow ~~ \beta_2 = \pm {1\o
2}\sqrt{\alpha~h^{1/2}(\gamma+4a^2)}.} Thus we fix $\beta_2$ or
equivalently $f_2$. As expected the line element for the
$\theta_1^2$ component can now be written in terms of $\beta_1$
as \eqn\metoneone{ds^2_{\theta_1 \theta_1} = \left(1 + \alpha~(1
+ B^2) \tilde{b}_{x\theta_1}^2 - {4A^2~\beta_1^2\o
h^{1/2}\gamma}\right)(\delta\tilde\theta_1)^2.} The $b$ dependent
term can be equivalently absorbed in the $\delta\tilde x$
fibration structure as $\delta{\hat x} \equiv \delta\tilde x +
\tilde b_{x\theta_1}~\delta\tilde\theta_1$. The rest of the
remaining term serve as warp factor for the
$(\delta\tilde\theta_1)^2$ term. What happens now if we argue an equality 
between the coefficients of the
$\delta\tilde\theta_1$ and $\delta{\hat x}$ terms?
This would imply:
\eqn\foneequa{1 - \beta_1^2~{4\o h^{1/2}\gamma}~ A^2 = \alpha
(1+B^2) ~~ \Rightarrow ~~ \beta_1 = \pm {1\o
2}\sqrt{\alpha~h^{1/2}\gamma}.} {}From \ftwos\ and \foneequa\ we
see that we can indeed choose the signs in a way that fulfills
\betaonetwo ! This is consistent with our assumption, implying
that in this setup we will see the tori metrics appear with one
unique warp factor. This is again expected in the case without
fluxes. What we see here is that, this remains true for the case
with fluxes also.

To summarize, we see that the T-duality directions  are $\tilde
z, \tilde x, \tilde y$ on the resolved side, and the line element
is more or less the same as \lineel\ with additional cross
components \metresconichange. To go from \lineel\ to
\metresconichange\ we have performed some transformations on the
finite shifts $\delta z, \delta x, \delta y$ and
$\delta\theta_i$. Defining three vectors $V_i$ as
\eqn\vectors{V_1 = \pmatrix{\delta z\cr \delta\theta_1\cr
\delta\theta_2}, ~~~ V_2 = \pmatrix{\delta x\cr \delta\theta_1\cr
\delta\theta_2}, ~~~ V_3 = \pmatrix{\delta y\cr \delta\theta_1\cr
\delta\theta_2}} and three matrices as: \eqn\matdef{\eqalign{ &
M_1 = \pmatrix{1& \Delta_1~f_1~{\rm cot}~\theta_1&
\Delta_2~f_2~{\rm cot}~\theta_2\cr 0& 1& 0\cr 0& 0& 1}, ~~~ M_2 =
\pmatrix{1& -f_1& 0\cr 0& 1& 0\cr 0& 0& 1},\cr & M_3 =
\pmatrix{{\rm cos}~\psi_0 & 0 & {\rm sin}~\psi_0 - f_2~{\rm cos}~\psi_0
\cr 0 & 1& 0\cr -{\rm sin}~\psi_0 & 0 & {\rm cos}~\psi_0 + f_2~{\rm
sin}~\psi_0}}} we can convert the line element \lineel\ to
\metresconichange\ via the transformations: \eqn\transf{V_1~\to
~M_1~V_1, ~~~~ V_2~\to ~M_2~V_2, ~~~~ V_3~\to~M_3~V_3.} This will
generate the T-duality directions 
$\tilde z, \tilde x, \tilde y$, as mentioned
above.

We can now try to go to the infinitesimal shifts given by the
forms $dx_i, d\theta_i$. At this point we therefore assume  that
we can replace \eqn\relshif{(\delta\tilde
x_i,~\delta\tilde\theta_i,~ \delta\tilde z) ~ \to ~ (d\tilde
x_i,~ d\tilde\theta_i,~ d\tilde z)} in the final line element.
This would imply that the line element with finite shifts can
serve as the metric for the mirror manifold. This is somewhat
strong assumption as the transformations that we made on the
finite shifts $\delta x_i, \delta z, \delta \theta_i$ may not
always be extrapolated to transformations on infinitesimal
shifts. It turns out that the transformation that we made can be
extrapolated if we assume that these transformations are
restricted on the plane ($x, y, z, \theta_i$) at $r = r_0$ which is,
of course, consistent with all our earlier assumptions. 

To see the effects of the finite shifts, first view the $z$ coordinate 
to be determined in terms of $\psi_1$ and $\psi_2$ as
$z \equiv \psi_1 - \psi_2$. In fact, as we will soon encounter in
the M-theory section, the coordinate $z$ which will eventually be
the $z$ coordinate of the type IIA mirror manifold, and the
M-theory eleventh direction $x_{11}$ can be written in terms of
$\psi_1$ and $\psi_2$ as: \eqn\psionetwom{dz ~\equiv~ d\psi_1 -
d\psi_2, ~~~~~~ dx_{11}~ \equiv ~ d\psi_1 + d\psi_2.} This means that in
the type IIB picture we can distribute the coordinates on two
different $S^3$'s parametrized by: ($\psi_1, x, \theta_1$) and
($\psi_2, y, \theta_2$). Existence of two $S^3$ here is just for
book keeping, and will eventually be related to the real $S^3$'s
of the mirror manifold\foot{In fact as will be clear from sec. 6, M-theory
will allow metric fibrations that are no longer constrained at $\psi = \psi_0$.}.

Let us now consider one $S^3$ with coordinates ($\psi_1, x,
\theta_1$). This $S^3$ is already at a fixed $r$ and now also at fixed ($\psi_2, y,
\theta_2$). From the analysis done above, we see that the
transformations generically lead to an integral of the form
(compare e.g. to \fibcha): \eqn\intg{\int (b^2 + a^2~{\rm
cot}^2~\theta_1)^{{m\o 2}}~{\rm cot}^n~\theta_1 ~d\theta_1} where $a$ and
$b$ are some constants on the sphere $S^3$, $n = 0, 1$ and 
$m = 1, -1$.\foot{In the notations of Appendix 1, the constants $a,b$ can be
extracted from $\langle \alpha \rangle_1^{\pm{1\o 2}}$ 
measured at a constant radius. 
There is also an
overall constant related to the expectation value $\langle \alpha \rangle$ that
we ignore here. Replacing $\theta_1$ by $\theta_2$ the constants $a,b$ 
should now be extracted from $\langle \alpha \rangle_2^{\pm{1\o 2}}$.}
It turns out that the shifts $\delta\psi_1, \delta x$ etc. can be
integrated to yield the following transformations on the
coordinates $\psi_1, x, \theta_1$: \eqn\coordinate{\eqalign{&
\psi_1 ~\to~ \psi_1 - {\left[1 - \left({a^2 - b^2 \o a^2}
\right)~{\rm sin}^2~\theta_1\right]^{m+1\o 4}
 \o \sqrt{\epsilon}~(a^{-1} {\rm sin}~\theta_1)^{m+1\o 2}} -
{\left(a^2 - b^2 \right)^{m\o 2} \o \sqrt{\epsilon}} 
~{\rm sin}^{-1}~\left[\left({a^2 -
b^2 \o a^2} \right)^{1\o 2}~ {\rm sin}~\theta_1 \right] \cr & x ~\to ~ x -
{\left(a^2 - b^2 \right)^{m\o 2} \o \sqrt{\epsilon}} 
{\rm ln}~{\left[ \left({a^2 - b^2 \o a^2}
\right)^{1\o 2} ~{\rm cos}~\theta_1 + \sqrt{1 - \left({a^2 - b^2 \o
a^2} \right)~{\rm sin}^2~\theta_1}
 \right]^{m+1 \o 2} \o \left[{\sqrt{1 - \left({a^2 - b^2 \o a^2}
\right)~
{\rm sin}^2~\theta_1} + \left({a^2 - b^2 \o a^2} \right)^{1-m \o 2}~
{\rm cos}~\theta_1 \o \sqrt{1 - \left({a^2 - b^2 \o a^2}
\right)~
{\rm sin}^2~\theta_1} - \left({a^2 - b^2 \o a^2} \right)^{1-m \o 2}
~{\rm cos}~\theta_1}\right]^{1\o (m+1)\sqrt{a^2-b^2}}}}}
with $\theta_1$ transforming as
$\theta_1 ~\to~ c\cdot \theta_1$ where $c$ is another constant. (Observe that these 
expressions have the required periodicity).
In the above transformations we have inherently assumed that $a >
b$. What happens for the case $a < b$?

This case turns out to be rather involved, but nevertheless
do-able. The transformation can again be integrated for both $m = \pm 1$
to give us
the following results (here we present the result only for $m = 1$):
\eqn\aleb{\eqalign{& \psi_1 ~\to~ \psi_1 +
\left(b^2 - a^2 \o a~\sqrt{\epsilon} \right)~{\rm ln}~{\left[\left(b^2 - a^2 \o
a^2 \right) ~{\rm sin}~\theta_1 + \sqrt{1 + \left(b^2 - a^2 \o
a^2 \right)~{\rm sin}^2~\theta_1}\right] \o {\rm exp}~\left[
{a^2~\sqrt{1 + \left(b^2 - a^2 \o a^2 \right)~{\rm
sin}^2~\theta_1} \o (b^2 - a^2)~ {\rm sin}~\theta_1}\right]}  \cr
& x ~\to ~ x - {a\o 2\sqrt{\epsilon}}~ {\rm ln}~ \left[{\sqrt{1 + \left(b^2 -
a^2 \o a^2 \right)~{\rm sin}^2~\theta_1}
 + {\rm cos}~\theta_1 \o \sqrt{1 + \left(b^2 -
a^2 \o a^2 \right)~{\rm sin}^2~\theta_1}
- {\rm cos}~\theta_1}\right] + \left(b^2 -
a^2 \o a~\sqrt{\epsilon} \right)~{\rm sin}^{-1}~ {\left(b^2 - a^2 \o a^2
\right)~{\rm cos}~\theta_1 \o \sqrt{1 + \left(b^2 - a^2 \o a^2
\right)^2}}}} with $\theta_1$ transformation remaining the same.
For the other $S^3$ the $y$ and $\theta_2$ transformation would
look similar to the above transformations on $x$ and $\theta_2$
transformations respectively. However $\psi_2$ transformation
will differ by relative signs. More details on the effect of these
transformations on the mirror metric is given in Appendix 1.

Taking the above transformations
 into account, and then performing the three T-duality
transformations we get the final mirror manifold (in the
delocalised limit) with the following form of the metric written
with $dx, dy, dz$ and $d\theta_i$:
\eqn\mirmanchange{\eqalign{ds^2 = &~~ g_1~\left[(dz -
b_{z\mu}~dx^\mu) + \hat\Delta_1~{\rm cot}~\hat\theta_1~ (dx -
b_{x\theta_1}~d\theta_1) + \hat\Delta_2~{\rm cot}~\hat\theta_2~(dy -
b_{y\theta_2}~d\theta_2)+ ..\right]^2 + \cr &~~~~~~~ + g_2~
[d\theta_1^2  + (dx - b_{x\theta_1}~d\theta_1)^2] +
g_3~[d\theta_2^2 + (dy - b_{y\theta_2}~d\theta_2)^2] +  \cr
& ~~~~~~~ + g_4~ [ d\theta_1~d\theta_2 - (dx - b_{x\theta_1}~d\theta_1)(dy -
b_{y\theta_2}~d\theta_2)]}} where we have used un-tilded
coordinates to avoid clutter (we will continue using this
coordinates in the rest of the paper unless mentioned otherwise).
The dotted part in the $dz$ fibration are the corrections to
$\theta_i$  terms from the scaling etc. We have written ${\rm
cot}~\hat\theta_i$ instead of ${\rm cot}~\theta_i$ to emphasize
the change in $\theta_i$. Its is interesting to note that (as we
saw earlier) this is the only change in $\theta$ because of
\scaletheta. All other changes due to scalings etc. have been
completely incorporated! Observe also that we now require only
four warp factors $g_1, g_2, g_3$ and $g_4$ instead of six that
we had earlier in \gis. The precise warp factors  can now be
written explicitly as: \eqn\gisnow{\eqalign{& g_1 = \alpha^{-1},
~~~g_2 = \alpha~j_{yy}, ~~~ g_3 = \alpha~j_{xx}, ~~~ g_4 =
2\alpha~j_{xy} \cr &~~~~~~~~~~~ \hat\Delta_1 = \sqrt{\gamma'\o
\gamma} ~r_0~\alpha, ~~~ \hat\Delta_2 = \sqrt{\gamma'\o \gamma + 4a^2}~r_0~\alpha}}
where $\alpha$ and $j$ are defined in \defalpha\ and \comedfi\ (with
${\rm cot}~\theta_i$ changed to ${\rm cot}~\hat\theta_i$),
respectively, and the $b$ fields have been rescaled according to
\tildeb. With these values the metric \mirmanchange\ can be
compared to \dsixdco.

\subsec{Physical meaning of $f_1$ and $f_2$}

The transformations that we performed in the previous subsection to 
bring the metric in the form \metresconichange\ using the functions
$f_1$ and $f_2$ can be given some physical meaning\foot{The discussion 
in this section is motivated from the conversations that one of us (R.T) 
had with the UPenn group, especially V. Braun and M. Cvetic.}. As discussed
earlier, the conversion from ($\phi_1, \theta_1$) and ($\phi_2, \theta_2$)
to ($x, \theta_1$) and ($y, \theta_2$) coordinates respectively is to 
write the metric as the metric of tori. Now the metric of the tori can be
generically written in terms of complex structures $\tau_i$ where $i = 1,2$
represent the two tori. We can define
\eqn\cstr{dz_1 = dx - \tau_1~d\theta_1, ~~~~~~~ dz_2 = dy - \tau_2~d\theta_2}
as the two coordinates of the two tori. The metric \lineel\ can therefore
be written in terms of $dz_i$ as:
\eqn\linaga{ds^2 = (dz + \Delta_1~{\rm cot}~\theta_1 ~dx + 
\Delta_2~{\rm cot}~\theta_2 ~dy)^2 + \vert dz_1 \vert^2 + \vert dz_2 \vert^2,}
with the complex structure not yet specified. The transformation that we
performed in the previous section, would therefore correspond to the 
following choice of the complex structure of the base tori:
\eqn\csbase{\tau_1 = f_1 + {i\o 2} \sqrt{\gamma \sqrt{h}}, ~~~~~~~~
\tau_2 = f_2 + {i\o 2} \sqrt{(\gamma + 4 a^2) \sqrt{h}}}
where we have already defined $h(r_0), \gamma(r_0), a$ in earlier sections. Observe that 
when $f_1 = 0 = f_2$ then the base is a torus with complex 
structure 
\eqn\basecs{\tau_1 = {i\o 2} \sqrt{\gamma \sqrt{h}},~~~~~~  
\tau_2 = {i\o 2} \sqrt{(\gamma + 4a^2) \sqrt{h}}} In this limit the 
metric has no cross terms. This is basically the metric that we started 
off with. The transformations in the previous subsection are therefore to 
convert $\tau_i \to \tau_i + f_i$ via $SL(2, R)$ transformations on the 
two tori\foot{For example using local $SL(2, R)$ 
matrices $\pmatrix{1&f_i\cr 0&1}$.}. 
In the limit when $f_i$ are very large, the base of the six 
dimensional manifold ($\theta_1, \theta_2, r = r_0$) is very large compared to the 
$T^3$ fiber ($x, y, z$). This situation is consistent with the fact 
that the generalised SYZ transformations require similar condition \syz\ for 
mirror rules to work properly (see also \louis). On the other hand, this limit is 
precisely the opposite to the one where the geometric transition takes place. This is one reason why we have 
to go through non-trivial manipulations to get to the final metric of the mirror manifold\foot{We thank the referee for pointing this 
out.}.

\subsec{$B$ Fields in the Mirror Setup}

%popo
There is another possibility that we haven't entertained yet.
This is to allow new $B$ field components in the resolved side.
Observe that the analysis presented above was done from the
resolved conifold setup when we only had $B_{NS}$ fields with
components $b_{x\theta_1}, b_{y\theta_2}$ and $b_{z\mu}$. What
happens if we switch on a cross component $b_{xy}$ that has legs
on both the spheres in the type IIB resolved conifold setup? Can
this generate a $d\theta_1 d\theta_2$ term? First, of course this
will not convert to a component of the metric under a mirror
transformation and will appear in the type IIA framework as a
$B$ field. This $B$ field will become a threeform field in
M-theory. We will discuss this later. Second, the mirror metric
will change. This change can be easily evaluated $-$ and we shall
do this below $-$ but before that lets see whether it is indeed
possible to switch on such component in the type IIB setup.

To analyze this, we go back to our fourfold scenario that we had
in section 3.  In the fourfold setup, no matter what choice we
make for the components of the $G$ fluxes, the metric will retain
its warped form and the only thing that could change will be the
exact value of the warp factor. Therefore we can choose an
additional component, say $G_{587a}$, and get the corresponding
$b_{xy}$ flux in type IIB.

Now under this choice of $B$ field, the metric will have the
additional  term $-{1\o G_{zz}}(G_{z\mu} dx^\mu + G_{zx} dx +
G_{zy} dy )^2$. One can show that the form of the $dz$
fibration structure remains the same, although the values will
differ by additional $b_{xy}$ terms. In the notations of the
earlier sections, let us assume the form of the metric to be:
\eqn\metfassu{\eqalign{ ds^2 = &~~(dz + {\rm fibration})^2 +
{\cal G}_{xx}~dx^2+ {\cal G}_{yy}~dy^2
 + {\cal G}_{\theta_1 \theta_1}~d\theta_1^2 +
 {\cal G}_{\theta_2 \theta_2}~d\theta_2^2 + 2{\cal G}_{xy}~dx dy + \cr
& ~~~~~~~ 2{\cal G}_{x\theta_1}~dx d\theta_1 + 2{\cal G}_{y
\theta_2}~dy d\theta_2 + 2{\cal G}_{x \theta_2}~dx d\theta_2 +
2{\cal G}_{y \theta_1} ~dy d\theta_1 + 2{\cal G}_{\theta_1
\theta_2}~d\theta_1 d\theta_2.}} Using the mirror rules given
earlier, one can work out all these     components. Since
$b_{xy}$ is non-zero, the analysis gets a little involved. If we
define again $\alpha^{-1} =j_{xx}j_{yy}-j_{xy}^2+b_{xy}^2$, this
time $b_{xy}$ being different from zero, we arrive exactly at the
form \mirman: \eqn\metnowinty{\eqalign{&{\cal G}_{xx}~dx^2+
2{\cal G}_{x\theta_1}~dx d\theta_1 + {\cal G}_{\theta_1
\theta_1}~d\theta_1^2 = \alpha ~j_{yy}~[d\theta_1^2 + ~(dx -
b_{x\theta_1}~d\theta_1)^2] \cr & {\cal G}_{yy}~dy^2 + 2{\cal
G}_{y \theta_2}~dy d\theta_2 +{\cal G}_{\theta_2 \theta_2}~
d\theta_2^2 = \alpha~j_{xx}~[d\theta_2^2 + (dy - b_{y\theta_2}~
d\theta_2)^2] \cr & {\cal G}_{xy}~dx~dy+ {\cal G}_{\theta_1
\theta_2}~d\theta_1 d\theta_2 + {\cal G}_{x \theta_2} ~dx
d\theta_2+ {\cal G}_{y \theta_1}~dy d\theta_1 = \cr &
~~~~~~~~~~~~~~-\alpha~j_{xy} (dx - b_{x\theta_1}~d\theta_1) (dy -
b_{y\theta_2}~d\theta_2)].}} Where we have given the precise warp
factors of every terms. Therefore, introducing a new component of
the $B$ field has not changed the form of the metric and has
failed to generate the $d\theta_1 d\theta_2$ term. What changes,
in the final mirror picture, is that we will now have a non zero
$B_{NS}$ flux in type IIA setup. We will comment on this later.

The above choice of $B$ fields in the resolved conifold side is
highly  unnatural (although it may be allowed from the
supergravity analysis). A more natural way to generate a $B$
field in the mirror side has already been taken into account when
we switched to the tilde-coordinates. In fact the cross terms in
the resolved conifold metric \metresconichange\ i.e. the $j_{x
\theta_1}$ and the $j_{y \theta_2}$ terms will be responsible to
give a non-zero $B$ field in the mirror picture. This way we can
give another physical meaning to the shifts that we performed in
\dxdychange. The background $B$ field in the type IIA background
can now be written in terms of the deformed conifold coordinates
as: \eqn\bintwoa{\eqalign{{\tilde B} = &~{2f_1\o
\sqrt{h^{1/2}\gamma}}~d\tilde x \wedge d{\tilde\theta_1} + {2f_2
\o \sqrt{h^{1/2}(\gamma + 4a^2)}}~d\tilde y \wedge
d{\tilde\theta_2}  \cr &~ + \left({2Af_1 \o
\sqrt{h^{1/2}\gamma}}~d\tilde\theta_1 + {2Bf_2 \o
\sqrt{h^{1/2}(\gamma + 4a^2)}}~d\tilde\theta_2\right) \wedge
d{\tilde z}}} with all other components vanishing in the limit
$\epsilon \to 0$. These $B$ fields are in general large, because
the transformations \dxdychange\ that we performed in the
resolved conifold setup is large. In the limit where we define
$\tilde B \equiv \epsilon^{-1/2} \hat B$, the finite part $\hat
B$ will be given by: \eqn\hatb{{\hat B \o \sqrt{\alpha}} = d{x}
\wedge d\theta_1 - d{y} \wedge d\theta_2 + (A~d\theta_1 -
B~d\theta_2) \wedge d{z}} modulo an overall sign if we employ the
opposite choice in \ftwos\ and \foneequa. Again, we have omitted
the tildes in the final expression. Notice also the following interesting
facts: 

\noindent $\bullet$ We can replace $dx$ and $dy$ by the corresponding one forms
$D\hat x$ and $D\hat y$ because of the wedge structure (at constant $b$). 
We will soon use this property (in the next sub-section) to get another
form of the $B$ fields that is more adapted to our mirror set-up.

\noindent $\bullet$ When we use an integrable complex structure for the two tori
in \csbase, i.e when we use $\langle\alpha\rangle_1, \langle\alpha\rangle_2$
instead of $\alpha$ (see Appendix 1), the $B_{NS}$ field takes the following form:
\eqn\bnow{{\hat B} = \sqrt{\langle\alpha\rangle_1} ~d{x} 
\wedge d\theta_1 - \sqrt{\langle\alpha\rangle_2}~ d{y} \wedge d\theta_2 + 
(A~\sqrt{\langle\alpha\rangle_1}~d\theta_1 -
B~\sqrt{\langle\alpha\rangle_2}~d\theta_2) \wedge d{z}.}
As one can easily see, this is a pure gauge! Thus even though we 
have large $B$ field in this scenario, the effect of this is nothing as it is
a gauge artifact. More on this will appear in  forthcoming papers \toappear, \bdkkt.

\subsec{The Mirror Manifold}

{}From the detailed analysis in the above two subsections,  we
can summarize the following: our metric of the mirror manifold
has strong resemblance to the metric of D6 wrapped on deformed
conifold, but they differ because of non-trivial B-dependent
fibration of some of the terms. As we see, this is the key
difference between the two metrics (apart from the non-trivial
warp factors). The manifold \mirmanchange\ is generically non-K\"ahler
whereas the metric \dsixdco\ could be K\"ahler in some limit. The
metric evaluated in \mirmanchange\ is actually after we perform a
coordinate transformation and
therefore we see no $\psi_0$ dependence in the final picture. In
the {\it usual} coordinate system, our ansatz therefore, for the
exact metric in type IIA will be to take the ``usual'' $D6$ brane
wrapped on the deformed conifold (i.e. eq. \dsixdco) and
replace the ($ dz, dx, dy, {\tilde g}$)
by \eqn\redpsietc{\eqalign{& dz ~\to ~ dz - {b}_{z\mu}~dx^\mu \cr
& dx ~\to ~ dx - b_{x\theta_1}~d\theta_1 \cr & dy ~\to ~ dy -
b_{y\theta_2}~d\theta_2 \cr & {\tilde g}_i(r_0, \theta_1, \theta_2)
~ \to ~ g_i(r_0, \theta_1, \theta_2)}} with the remaining terms
unchanged. Observe that before this replacement (i.e in the absence of fluxes) 
\dsixdco\ is exactly
\metcomptwo\ up to ${\rm cos}~\psi$ and ${\rm sin}~\psi$
dependences. We believe, as discussed above, this has to do with
the delocalization of the $\psi$ coordinate. In the presence of fluxes,
the final answer for
the type IIA metric therefore will be to convert 
\mirmanchange\ into\foot{\noindent This ansatz
can actually be given a little more rigorous derivation. To see
this from \mirmanchange, perform the following local
transformation $$\pmatrix{D{\hat y}\cr d\theta_2} ~\to
~\pmatrix{{\rm cos}~\psi_0 & -{\rm sin}~\psi_0 \cr {\rm sin}~\psi_0 &
{\rm cos}~\psi_0}~\pmatrix{D{\hat y} \cr d\theta_2}$$ \noindent
where $D{\hat y} \equiv dy + b_{y\theta_2}~d\theta_2$. The change
in the $dz$-fibration structure will be in such a way as to
restore back ${\rm cot}~\theta_2~dy$ via the {\it reverse}
transformation a-la \psichange.}: \eqn\fiiamet{\eqalign{ds_{IIA}^2
= &~~ g_1~\left[(dz - {b}_{z\mu}~dx^\mu) + \Delta_1~{\rm
cot}~\hat\theta_1~(dx - b_{x\theta_1}~d\theta_1) + \Delta_2~{\rm
cot}~ \hat\theta_2~(dy - b_{y\theta_2}~d\theta_2)+ ..\right]^2
\cr & ~~~~~~~~~~~~~~ +~ g_2~ {[} d\theta_1^2 + (dx -
b_{x\theta_1}~d\theta_1)^2] + g_3~[ d\theta_2^2 + (dy -
b_{y\theta_2}~d\theta_2)^2{]}  \cr & ~~~~~~~~~~~~~~ + ~ g_4~{\rm
sin}~\psi_0~{[}(dx - b_{x\theta_1}~d\theta_1)~d \theta_2 + (dy -
b_{y\theta_2}~d\theta_2)~d\theta_1 {]}\cr & ~~~~~~~~~~~~~~ +
~g_4~{\rm cos}~\psi_0~{[}d\theta_1 ~d\theta_2 - (dx -
b_{x\theta_1}~d\theta_1) (dy - b_{y\theta_2}~d\theta_2)],}} where
$g_i$ are again given by \gisnow. Similarly the finite part of
the background $B$ field can be transformed from \hatb\ to the
following form involving the ${\rm sin}~\psi_0$ and ${\rm cos}~\psi_0$
dependences as: \eqn\hatbnow{{\hat B \o \sqrt{\alpha}} = dx
\wedge d\theta_1 - dy \wedge d\theta_2 + A~d\theta_1\wedge dz -
B~({\rm sin}~\psi_0 ~dy - {\rm cos}~\psi_0~d\theta_2) \wedge dz.} The
type IIA coupling on the other hand can no longer be constant
even though in the type IIB side we start with a constant
coupling. The constant coupling on the type IIB side is
generically fixed by RR and NS fluxes via a superpotential
(though not always). If we start with a type IIB coupling $g_B$
(constant or non-constant), the type IIA theory is given by a
non-constant coupling $g_A$, that depend on the coordinates of
the internal space as \eqn\ccons{g_A = {g_B \o \sqrt{1-
{\epsilon\o \alpha}}}.} Observe that a small coupling in the type
IIB side implies a small coupling on the mirror manifold.
Therefore any perturbative calculation in type IIB side will have
a corresponding perturbative dual in the mirror side. This is
another advantage that we get from the mirror manifolds.

The above analysis more or less gives the complete background for
the mirror case (the RR background will be dealt with shortly).
For later comparison we would however need the three form NSNS
field strength defined as $H = d\hat B$. This is basically the
finite part of the three form, and is given by \foot{Written in terms of 
$\sqrt{\langle\alpha\rangle_1}$ and $\sqrt{\langle\alpha\rangle_2}$ this is exactly zero, 
and therefore serves as a gauge artifact.}:
\eqn\hfine{\eqalign{H & = -\sqrt{\alpha^3} A~(A~dA + B~dB) \wedge
d\theta_1 \wedge dz + \sqrt{\alpha^3} (A~dA + B~dB) \wedge dy
\wedge d\theta_2 \cr &  + \sqrt{\alpha}~dA \wedge d\theta_1
\wedge dz + \sqrt{\alpha^3}~B~(A~dA + B~dB) \wedge ({\rm
sin}~\psi_0 ~dy - {\rm cos}~\psi_0~d\theta_2) \wedge dz \cr &  -
\sqrt{\alpha^3}~(A~dA + B~dB)\wedge dx \wedge d\theta_1 ~
 -\sqrt{\alpha} ~dB \wedge ({\rm sin}~\psi_0 ~dy - {\rm cos}~\psi_0~d\theta_2) \wedge dz,}}

\noindent where we have used the following simplifying
definitions  in the above form of $H$: \eqn\sifoo{dA =
\del_{\theta_i}A ~d\theta_i + \del_r A~dr =\del_{\theta_i}A ~d\theta_i , ~~~~ d\sqrt{\alpha} =
-\sqrt{\alpha^3}(A~dA + B~dB)} with similar definition for $dB$.
Notice that $H$ involves all components of the mirror manifold
and therefore will be spread over the whole space.

\noindent We now need to show the following things:

\noindent (1) The manifold is explicitly non-K\"ahler i.e. $dJ \ne
0$, where $J$ is the fundamental two form. The manifold should
also be non-Ricci flat and have an $SU(3)$ structure. Recall that $SU(3)$ 
structure {\it doesn't} imply Ricci-flatness.

\noindent (2) The complex structure should in general be
non-integrable. Therefore the manifold should be non-complex and
non-K\"ahler. The properties of such manifolds have been
discussed earlier in \louis, \dal.

\noindent (3) We have to calculate the superpotential and show
that the holomorphic three form $\Omega$ is in general {\it not}
closed.

\noindent Some of these details will be addressed in later
sections of this paper. We will leave a more elaborate discussion
for part II. We now go to the M-Theory analysis.

\newsec{Chain 2: The M-theory Description of the Mirror}

Now that we have obtained the mirror metric, it is time to go to
the second chain of fig. 1 and lift the type IIA configuration to
M-theory. Initial studies have been done in \amv. We will use
their ideas to go to another $G_2$ holonomy manifold which is
related to the previous one by a flop. But before moving ahead,
we will require some geometric details of the background. These
geometric details will help us to formulate the background in a
way so that the procedure of flop will be simple to see.

\subsec{One Forms in M-theory}

We first need to define one forms in M-theory. These one forms are
different from the ones presented in say \amv, \brand, \cveticone\
as they
have contributions from the $B$ fields in the resolved conifold
side. These $B$ fields are in general periodic variables and
therefore let us denote them by angular coordinates $\lambda_1,
\lambda_2$ as ${\rm tan}~\lambda_1 \equiv a_1 b_{x\theta_1}, {\rm
tan}~\lambda_2 = a_2 b_{y\theta_2}$, with $a_1, a_2$ constants.
These one-forms are however only defined {\it locally} because
the $B$ fields that we will use in the definition are not
globally defined variables. The existence of these one forms can
be argued from the consistency of the metric\foot{Observe that
for $r_0 \approx 0$ the metric \fiiamet\ is exactly the metric of $D6$ branes
wrapping an $S^3$ of a deformed conifold (because ($b_{x\theta_1}, b_{y\theta_2}$)$\to ~0$ at IR), 
and therefore will have
similar one-forms as in \amv, \brand, \cveticone\
by which we can express
the metric.}. They are
given by: \eqn\oneformsM{\eqalign{& \sigma_1 = {\rm
sin}~\psi_1~dX + {\rm sec}~\lambda_1~{\rm cos}~(\psi_1 +
\lambda_1)~d\Theta_1 \cr & \sigma_2 = {\rm cos}~\psi_1~dX - {\rm
sec}~\lambda_1~{\rm sin}~(\psi_1 + \lambda_1)~d\Theta_1 \cr &
\sigma_3 = d\psi_1 + n_1~{\rm cot} ~\hat\Theta_1~dX - n_2~{\rm
tan}~\lambda_1~{\rm cot} ~\hat\Theta_1~d\Theta_1}} where we have
defined new coordinates $\psi_1, \psi_2, X, \Theta_1,
\hat\Theta_1$.
 Their relation to $z, x, \theta_1, \hat\theta_1$ will be determined
as we proceed with our calculation.
 We have
also included two functions $n_1$ and $n_2$ in the definition of
$\sigma_3$. These are functions of all the coordinates, and will
be analyzed later. Using these, we can define another set of one
forms with $X, \Theta_1$ etc. replaced by $Y, \Theta_2$ etc. as:
\eqn\seconefor{\eqalign{& \Sigma_1 = -{\rm sin}~\psi_2~dY +  {\rm
sec}~\lambda_2~{\rm cos}~(\psi_2 -
 \lambda_2)~d\Theta_2 \cr
& \Sigma_2 = -{\rm cos}~\psi_2~dY - {\rm sec}~\lambda_2~{\rm
sin}~(\psi_2 - \lambda_2)~d\Theta_2 \cr & \Sigma_3 = d\psi_2 -
n_3~{\rm cot} ~\hat\Theta_2~dY +
 n_4~{\rm tan}~\lambda_2~{\rm cot} ~\hat\Theta_2~d\Theta_2.}}
We have chosen the respective signs with the foresight of making
a simple identification with our original variables possible. The
above set of one forms will suffice to define the corresponding
seven dimensional manifolds in M-theory. Although, not important
for our work here, we can make some interesting simplifications.
The quantities $b_{x\theta_1}$ and $b_{y\theta_2}$, as discussed
above, are basically periodic variables, and for small
$\lambda_i$ we can define another angular coordinates $\beta_1$
and $\beta_2$ that modify the original $\psi_1, \psi_2$ as
\eqn\lambpsi{\beta_1 = \psi_1 - b_{x\theta_1}, ~~~~~~ \beta_2 =
\psi_2 - b_{y\theta_2}.} With these choices of angles, we can
define another set of one forms in M-theory in the following way:
\eqn\anosetone{\eqalign{& \tilde\sigma_1 = {\rm sin}~\psi_1~dX +
{\rm cos}~\beta_1~d\Theta_1  \qquad \tilde\Sigma_1 = {\rm
sin}~\psi_2~dY + {\rm cos}~\beta_2~d\Theta_2 \cr & \tilde\sigma_2
= {\rm cos}~\psi_1~dX + {\rm sin}~\beta_1~d\Theta_1 \qquad
\tilde\Sigma_2 = {\rm cos}~\psi_2~dY + {\rm
sin}~\beta_2~d\Theta_2,}} with $\tilde\sigma_3$ and
$\tilde\Sigma_3$ being identical to $\sigma_3$ and $\Sigma_3$,
respectively. This way of writing the one-forms helps us to
compare them to the one-forms given in \amv, \brand, and \cveticone. 
Observe
that for small background values of $b_{x\theta_1}$ and
$b_{y\theta_2}$ the above set \anosetone\ is the same as
\oneformsM\ and \seconefor. Furthermore, the field strength
vanishes locally, and these one forms satisfy the $SU(2)$
algebra. This is true because over a small patch the
$b_{x\theta_1}$ and $b_{y\theta_2}$ values are constants.
Therefore we can approximate \eqn\dxdxpatch{D\hat x \equiv dx -
b_{x\theta_1}~d\theta_1 = d(x - b_{x\theta_1}~\theta_1), ~~~~~
D\hat y \equiv dy - b_{y\theta_2}~d\theta_2 = d(y -
b_{y\theta_2}~\theta_2)} as exact one forms. In this way
\oneformsM\ and \seconefor\ appear like the usual one forms for
the $G_2$ manifold and satisfy an $SU(2) \times SU(2)$ symmetry.
Globally, there is no $SU(2) \times SU(2)$ symmetry because our
manifold is no longer a K\"ahler manifold. This is also clear
from the one-forms  \oneformsM\ and \seconefor. We will use this
identification many times to compare our results to the
ones from literature. Also, having an exact form for $D\hat x$
and $D\hat y$ locally means vanishing type IIB $B$ fields. In
this way we will be able to extend our results to the case with
torsion.

\subsec{M-theory Lift of the Mirror IIA Background}

To perform the M-theory lift, we need the field strength
$F_{mn}$ which comes from the mirror dual of the three-form
$H_{RR} \equiv {\cal H}$ and five-form $F_5$
 in type IIB theory with $D5$ on a resolved conifold. The five
form appears because a $D5$ wrapped on an $S^2$ with $B_{NS}$
fluxes gives rise to a $D3$ brane source as we discussed in the
beginning of this paper. The background value of the RR potential
is given in \pandoz\ as: \eqn\hrrbg{\eqalign{& {\cal H} = c_1~(dz
\wedge d\theta_2 \wedge dy - dz \wedge d\theta_1 \wedge dx) +
c_2~{\rm cot}~\hat\theta_1~dx \wedge d\theta_2 \wedge dy -
c_3~{\rm cot}~\hat\theta_2 ~dy \wedge d\theta_1 \wedge dx \cr &
F_5 = K(r)~(1 + \ast) ~dx \wedge dy \wedge dz \wedge d\theta_1
\wedge d\theta_2,}} with $c_i$ is a constant coefficient with
$K(r)$ being a function of the 
global transverse coordinate $r$ (we could as well define $F_5$ with 
$K(r_0)$ locally, as we did for ${\cal J}_i$ in \bfico). 
This means
that we have the following components of the RR three form:
${\cal H}_{\mu x z}, {\cal H}_{\mu y z}$ and ${\cal H}_{\mu x
y}$. The T-duality rules for the RR field strengths with
components along the T-dual directions are given in \fawad:
\eqn\fadhu{\eqalign{& {\tilde F}^{(n)}_{ijk....} = F^{(n+1)}_{x
ijk....} - n B_{x[i} F^{(n-1)}_{jk....]} + n(n-1)j_{xx}^{-1}
B_{x[i}j_{\vert x\vert j} F^{(n-1)}_{x\vert k....]} \cr & {\tilde
F}^{(n)}_{xij....} = F^{(n-1)}_{ij....} - (n-1)j_{xx}^{-1}
j_{x[i}F^{(n-1)}_{x\vert jk....]}}} where $n$ denote the rank of
the form, $x$ is the T-duality direction and $B$ is the NS field.
Notice also that in the above relation, the duality direction $x$
is inert under anti-symmetrization. Under a mirror
transformation, the RR three-form will give rise to the following
gauge potentials in type IIA theory: \eqn\gaugepot{\eqalign{&
F_{z\theta_1} = {\cal H}_{xy\theta_1}, ~~~~ F_{z\theta_1} = {\cal
H}_{xy\theta_2} \cr & F_{y\theta_1} = -{\cal H}_{xz\theta_1} +
{\cal H}_{xy\theta_1}\left[{j_{yz}j_{xx} - j_{xy}j_{xz} \o j_{yy}
j_{xx} - j_{xy}^2} - {\cal B}_{zy}\right]\cr & F_{x\theta_2}=
{\cal H}_{yz\theta_2} - {\cal H}_{xy\theta_2} \left[
{j_{xy}(j_{yz}j_{xx} - j_{xy}j_{xz}) \o j_{xx}(j_{yy} j_{xx} -
j_{xy}^2)}
 +{\cal B}_{zx} -{j_{xz}\o j_{xx}}\right]}}
which simplifies, after choosing the signs of ${\cal B}_{mn}$ in
a way that makes the fibration structure in the mirror
consistent, as: \eqn\simsol{\eqalign{& F_{z\theta_1} = -c_3~{\rm
cot}~\hat\theta_2, ~~~~~~~~~~~~~~ F_{z\theta_2} = -c_2{\rm
cot}~\hat\theta_1 \cr & F_{y\theta_1} = c_1 - 2c_3~\alpha~B~{\rm
cot}~{\hat\theta_2}, ~~ F_{x\theta_2} = c_1 - 2c_2~\alpha~A~{\rm
cot}~{\hat\theta_1}.}} {}From above we will eventually extract
gauge potentials $A_x$ and $A_y$ (the $A_z$ potential can be
absorbed in the definition of $dx_{11}$)\foot{There is a simple
reason for this. We had earlier defined $dz \equiv d\psi_1 -
d\psi_2$ and $dx_{11} \equiv d\psi_1 + d\psi_2$. Thus $dx_{11} =
d\psi_1 - d\psi_2 + 2 d\psi_2 = dz + 2d\psi_2$. Therefore, any
additional $dz$ dependent terms should be absorbed in $dx_{11}$
by changing the coefficient in front of $dz$ in $dx_{11}$. For
example $dx_{11} + A_z~dz = (1+A_z)~dz + 2 d\psi_2 = d\tilde
x_{11}$ when $A_z$ is a pure gauge.}.

The analysis done above only gave us some of the gauge fields in
type IIA theory. To get the other components, we need to get the
mirror dual of the five form $F_5$. One can easily show that the
$F_5$ part contributes to $F_{\theta_1 \theta_2}$. The precise
value turns out to be \eqn\fiveformv{\eqalign{F_{\theta_1
\theta_2} & =~ K(r)\vert_{r\to r_0} - b_{x\theta_1} {\cal H}_{yz\theta_2} -
b_{y\theta_2} {\cal H}_{xz\theta_1} - {\cal B}_{z\theta_2}~{\cal
H}_{xy\theta_1} - {\cal B}_{z\theta_1}~{\cal H}_{xy\theta_2} \cr
& = ~ K(r)\vert_{r\to r_0} - c_1~(b_{x\theta_1} - b_{y\theta_2}) -
2c_3~\alpha~B~b_{y\theta_2}~{\rm cot}~\hat \theta_2 +
2c_2~\alpha~A~b_{x\theta_1}~{\rm cot}~\hat\theta_1}} where again
the sign in ${\cal B}_{z\theta_1}$ and ${\cal B}_{z\theta_2}$ have
been chosen opposite to the definitions employed in sec. 4.2. The
above mentioned gauge potentials will eventually appear as metric
components in M-theory. As is well known, the M-theory metric
components will typically look like \eqn\mthmetlook{G^M_{\mu\nu}
= e^{-{2\phi \o 3}} g^{IIA}_{\mu\nu} - e^{4\phi\o 3} A_\mu A_\nu,
~~~~~ G^M_{\mu~11} = - e^{4\phi \o 3} A_\mu} where $A_\mu$ are
the gauge fields and $\phi$ is the type IIA dilaton. In fact, we
can use the above definitions to absorb some terms in $dx^{11}$
when we define the gauge fields. Up to some warp factors our
ans\"atze for the gauge fields will therefore be:
\eqn\gaugefields{\eqalign{& A_x = \Delta_3 ~{\rm
cot}~\hat\theta_1, ~~~~~~~~~A_{\theta_1} =
-\Delta_3~b_{x\theta_1}~{\rm cot}~\hat\theta_1 \cr & A_y = -
\Delta_4~{\rm cot}~\hat\theta_2, ~~~~~ A_{\theta_2} =
\Delta_4~b_{y\theta_2}~{\rm cot}~\hat\theta_2}} where
$\Delta_{3,4}$ are some specific functions of $\theta_i, x, y$
and $z$.   Now combining everything together we can write the
part of the M-theory metric originating from the gauge fields as:
\eqn\gaupot{A\cdot dX \equiv \Delta_3~{\rm cot}~\hat\theta_1~(dx
- b_{x\theta_1}~d\theta_1) - \Delta_4~{\rm cot}~\hat\theta_2~(dy
- b_{y\theta_2}~d\theta_2)} where we remove any $z$ dependences.
The above potentials \gaupot\ are basically the
wrapped $D6$ brane sources that have been converted to geometry
giving rise to the M-theory metric \eqn\mteorymet{\eqalign{ds^2
&= e^{-{2\phi \o 3}}
(h^{-1/2}~ds_{0123}^2+h^{1/2}\gamma'~dr^2)\cr & + e^{-{2\phi \o
3}}~ds^2_{IIA} + e^{4\phi \o 3}~(dx_{11} + \Delta_3~{\rm
cot}~\hat\theta_1~d\hat x - \Delta_4~{\rm cot}~\hat\theta_2~d\hat
y)^2}} with $x_{11}$ being the eleventh direction, and
$ds^2_{IIA}$ is the metric given in \fiiamet. We have also used
the definition of $\hat x$ and $\hat y$ (introduced in section 5)
to write the fibration structure in a compact form. Recall also
that we are using the un-tilded coordinates henceforth.

The metric \mteorymet\ is basically the M-theory metric that we
are  looking for. To see how the $G_2$ structure appears from
this, we need to write the metric using the one forms that we
gave in the previous section. This will also help us to perform a
flop in the metric.

\noindent {}From the M-theory metric \mteorymet\ we see that the
total fibration structure is \eqn\fibstr{ g_1~e^{-{2\phi \o 3}}
(dz + \Delta_1~{\rm cot}~\hat\theta_1~d\hat x + \Delta_2~{\rm
cot}~\hat\theta_2~d\hat y)^2 + e^{4\phi \o 3}~(dx_{11} +
\Delta_3~{\rm cot}~\hat\theta_1~d\hat x - \Delta_4~{\rm
cot}~\hat\theta_2~d\hat y)^2.} In writing this, the careful
reader might notice that we have used ${\rm cot}~\hat\theta_i$ in
\hrrbg. However, we have two options here: we can either absorb
the scaling etc. in the definition of warp factors or keep the
warp factors as they are and change $\theta$ in ${\rm cot}\theta$
to accommodate this. Furthermore, the scaling of $\theta$, which
only affects this term, is actually of ${\cal O}(1)$ and could be
ignored in the subsequent calculations. We will however not
assume any approximations and continue using $\hat\theta_i$ in
the fibration. Observe that all other one forms are defined wrt
$\theta_i$.

If we now identify $dz \equiv  d\psi_1 - d\psi_2$ and $dx_{11}
\equiv d\psi_1 + d\psi_2$, then one can easily see that the above
fibration can be written in terms of the one forms that we
devised earlier in \oneformsM\ and \seconefor, as
\eqn\ondev{\alpha_3^2~ (\sigma_3 + \Sigma_3)^2  + \alpha_4^2~
(\sigma_3 - \Sigma_3)^2} with $\alpha_i$ being the relevant warp
factors; and we also identify ($X, Y, \Theta_1, \Theta_2$) to
($x, y, \theta_1, \theta_2$). The coefficients $a_i$ are simply
the identity, so $\tan\lambda_1=b_{x\theta_1}$ and
$\tan\lambda_2=b_{y\theta_2}$. This way of writing the $z$-- and
$x_{11}$-- fibration also forces us to set $n_1=n_2=\Delta_1$ and
$n_3=n_4=\Delta_2$. \ondev\ is consistent with the expected form
for the M-theory lift of the $D6$ configuration. Let us now look
at the other possible combinations of the one forms.

The first combination is to consider the sum of the squares of
the difference between the one forms. In other words, we will
consider: \eqn\sumofdif{(\sigma_1 - \Sigma_1)^2 + (\sigma_2 -
\Sigma_2)^2.} This give rise to the following algebra:
\eqn\algeb{\eqalign{ & ({\rm sin}~\psi_1~dX + {\rm
sec}~\lambda_1~{\rm cos}~(\psi_1 + \lambda_1)~d\Theta_1 + {\rm
sin}~\psi_2~dY -
 {\rm sec}~\lambda_2~{\rm cos}~(\psi_2 - \lambda_2)~d\Theta_2)^2 + \cr
& ({\rm cos}~\psi_1~dX - {\rm sec}~\lambda_1~{\rm sin}~(\psi_1 +
\lambda_1)~d\Theta_1 + {\rm cos}~\psi_2~dY + {\rm
sec}~\lambda_2~{\rm sin}~(\psi_2 - \lambda_2)~d\Theta_2)^2  = \cr
&~~~ d\Theta_1^2 + d\Theta_2^2 + (dX - {\rm
tan}~\lambda_1~d\Theta_1)^2 + (dY - {\rm
tan}~\lambda_2~d\Theta_2)^2 + \cr & ~~-2~{\rm cos}~(\psi_1 -
\psi_2)~[d\Theta_1~d\Theta_2 - (dX - {\rm
tan}~\lambda_1~d\Theta_1)~(dY - {\rm tan}~\lambda_2~d\Theta_2)] +
\cr & ~~-2~ {\rm sin}~(\psi_1 - \psi_2)~[d\Theta_1~(dY - {\rm
tan}~\lambda_2~d\Theta_2) + d\Theta_2 ~ (dX - {\rm
tan}~\lambda_1~d\Theta_1)].}} This is more or less the expected
form, but differs from \fiiamet\ by some
 relative signs. To fix the signs we need to evaluate the other possible combination of
one forms, i.e. $(\sigma_1 + \Sigma_1)^2 + (\sigma_2 +
\Sigma_2)^2$.
This time the algebra will yield:
\eqn\algebyiel{\eqalign{& ({\rm sin}~\psi_1~dX + {\rm sec}~\lambda_1~{\rm cos}~(\psi_1 +
 \lambda_1)~d\Theta_1 + {\rm sin}~\psi_2~dY +
{\rm sec}~\lambda_2~{\rm cos}~(\psi_2 + \lambda_2)~d\Theta_2)^2 + \cr
& ({\rm cos}~\psi_1~dX - {\rm sec}~\lambda_1~{\rm sin}~(\psi_1 + \lambda_1)~d\Theta_1 +
{\rm cos}~\psi_2~dY - {\rm sec}~\lambda_2~{\rm sin}~(\psi_2 + \lambda_2)~d\Theta_2)^2  = \cr
& ~~~ d\Theta_1^2 + d\Theta_2^2 + (dX - {\rm tan}~\lambda_1~d\Theta_1)^2 +
(dY - {\rm tan}~\lambda_2~d\Theta_2)^2 + \cr
& ~~ + 2~{\rm cos}~(\psi_1 - \psi_2)~[d\Theta_1~d\Theta_2 +
(dX - {\rm tan}~\lambda_1~d\Theta_1)~(dY - {\rm tan}~\lambda_2~d\Theta_2)] + \cr
& ~~ + 2~{\rm sin}~(\psi_1 - \psi_2)~[- d\Theta_1~(dY - {\rm tan}~\lambda_2~d\Theta_2) +
 d\Theta_2 ~ (dX - {\rm tan}~\lambda_1~d\Theta_1)].}}
It differs from \algeb\ only by overall minus signs in the
$\cos\psi$ and $\sin\psi$ terms. To get the exact form of the
metric that we have in \mteorymet, we write
\eqn\mmetgeneric{ds^2 = {\alpha}_1^2 ~\sum_{a = 1}^2 (\sigma_a +
\xi\Sigma_a)^2 + {\alpha}_2^2 ~ \sum_{a = 1}^2 (\sigma_a - \xi
\Sigma_a)^2 + {\alpha}_3^2 ~(\sigma_3 + \Sigma_3)^2 +
{\alpha}_4^2 ~ (\sigma_3 - \Sigma_3)^2 + {\alpha}_5^2 ~dr^2,}
where we have introduced the factor $\xi$ for $\Sigma_1$ and
$\Sigma_2$ to account for the different warp factors that the
directions $(x,\theta_1)$ and $(y,\theta_2)$ have\foot{In the notations that
we used here, they are tori of course. But it would be easy to get to the sphere case once we know the global
type IIA metric. 
Observe also that when we switch off the $\lambda_i$ the metric reduces to the 
well known $G_2$ form.}. 
By comparison
with \fiiamet\ and \mteorymet\ we determine the warp factors
${\alpha}_i$ with $\xi=\sqrt{g_3/g_2}$ to be:
\eqn\warpyiden{\eqalign{{\alpha}_1 &= {1\o 2} e^{-{\phi\o
3}}\sqrt{\xi^{-1}g_4 + 2g_2}, ~~~ {\alpha}_2 = {1\o 2} e^{-{\phi
\o 3}}\sqrt{2g_2-\xi^{-1}g_4},\cr {\alpha}_3 &= e^{2\phi\o 3}, ~~~
{\alpha}_4 = e^{-{\phi\o 3}}\sqrt{g_1}, ~~~ {\alpha_5} =
e^{-{\phi\o 3}}\sqrt{\gamma'\sqrt{h}}.}} Observe also that,
writing the metric as \mmetgeneric, one can easily identify it to
the metric presented in \amv, \brand, \cveticone\
where the vielbeins in
terms of $\sigma_a$ and $\Sigma_a$ are written (see for example
eq. 2.7 of \brand). This will be useful in the following.

Before moving ahead let us pause for a second to reflect on the metrics 
\mmetgeneric\ and the type IIA metric \fiiamet. In \fiiamet\ we have kept the 
$dz$ fibration structure distinct from the $\psi$ rotation. In fact the rotation
generates constant warp factors of ${\rm sin}~\psi_0$ and ${\rm cos}~\psi_0$ 
in \fiiamet. But in \mmetgeneric\ we see that the one forms \oneformsM\ and 
\seconefor\ can in fact be used to write the M-theory metric with arbitrary $\psi$! 
Indeed now the relation between the non-constant $\psi$ and $z$ will become 
\eqn\depsi{z ~ = ~\psi_1 - \psi_2 ~=~ a_s \psi} where $a_s = a_s(r_0)$ is the 
constant of proportionality\foot{From \miapmet\ we see that $a^2_s ~ = ~ 
r_0^2 \sqrt{h}~{\del\gamma\o \del r^2}\vert_{r = r_0}$. Using now the small $r$ 
behavior of $\gamma$ \smlarbe, we see that ${\del\gamma\o \del r^2}\vert_{r = r_0} ~ = ~ {1\o \sqrt{6} a} + 
{\cal O}(r_0^2)$ and $h$ is of ${\cal O}(1)$ so that $a_s ~ \sim ~ r_0$. This will be useful later when we 
try to study the scaling behavior of our M-theory 
metric.}.  
With this in mind, now we see that the M-theory metric 
allows a type IIA solution which takes us away from the $\psi = \psi_0$ point and 
allow non constant ${\rm sin}~\psi$ and ${\rm cos}~\psi$ in \fiiamet. Thus the local
metric in type IIA is delocalised only along the $r$ direction (i.e defined at 
$r = r_0$) and is now expressed as
\eqn\filament{\eqalign{ds_{IIA}^2 
= &~~ ds^2_{0123} + g_1\left(D\hat z + \Delta_1~{\rm
cot}~\hat\theta_1~D \hat x  + \Delta_2~{\rm
cot}~ \hat\theta_2~D \hat y  + ..\right)^2 ~ +
\cr & ~~~~~ + g_0~dr^2 + g_2~ {(} d\theta_1^2 + D \hat x^2) 
 + g_3~(d\theta_2^2 + D \hat y^2)~ + \cr & ~~~~~ +  g_4~{\rm
sin}~(z a_s^{-1})~{}(D \hat x \cdot d \theta_2 + D \hat y \cdot d\theta_1) + 
~g_4~{\rm cos}~(z a_s^{-1})~(d\theta_1 \cdot d\theta_2 - D\hat x \cdot D\hat y)}}
where we have already defined $D\hat z, ~D\hat y$ and $D\hat x$ earlier. The 
effect of the additional six branes that are mirror dual to the type IIB seven branes
is already accounted for in the local picture above. The full global story is 
presented in \bdkkt\ where things get pretty involved because of the presence of 
orientifold six-planes, ungauged localised two form fluxes and non-trivial three form fluxes,
along with the additional D6 branes. We will not discuss these issues further here, and the 
readers can find more details in \bdkkt. 

To extend our earlier analysis, we will now require all the
components  of the seven dimensional metric. We denote the
M-theory metric as $ds^2 = {\cal G}_{mn}~dx^m~dx^n$, where $m,n =
x, y, z, \theta_1, \theta_2, r, a$ and $x^a \equiv x^{11}$. The
various components are now given by: \eqn\mecoone{\eqalign{&
(1)~~{\cal G}_{xx} = g_1~e^{-{2\phi\o 3}}~ \Delta_1^2~ {\rm
cot}^2~\hat\theta_1 + g_2~e^{-{2\phi\o 3}} + e^{4\phi \o 3}~
\Delta_3^2 ~ {\rm cot}^2~\hat\theta_1 \cr & (2)~~ {\cal G}_{yy} =
g_1~e^{-{2\phi\o 3}}~ \Delta_2^2~ {\rm cot}^2~\hat\theta_2 +
g_3~e^{-{2\phi\o 3}} + e^{4\phi \o 3}~ \Delta_4^2 ~{\rm
cot}^2~\hat\theta_2 \cr & (3)~~ {\cal G}_{\theta_1 \theta_1} =
e^{-{2\phi \o 3}}~(g_1~ \Delta_1^2~ {\rm cot}^2~\hat\theta_1 +
g_2)~b^2_{x\theta_1} + g_2~e^{-{2\phi \o 3}} +
b_{x\theta_1}^2~e^{4\phi \o 3}~\Delta_3^2~{\rm
cot}^2~\hat\theta_1 \cr & (4)~~ {\cal G}_{\theta_2 \theta_2} =
e^{-{2\phi \o 3}}~(g_1~ \Delta_2^2 ~{\rm cot}^2~\hat \theta_2 +
g_3)~b^2_{y\theta_2} + g_3~e^{-{2\phi \o 3}} +
b_{y\theta_2}^2~e^{4\phi \o 3}~\Delta_4^2~{\rm
cot}^2~\hat\theta_2 \cr & (5) ~~ {\cal G}_{x \theta_1} =-
e^{-{2\phi \o 3}}~(g_1~ \Delta_1^2~ {\rm cot}^2~\hat\theta_1 +
g_2)~b_{x\theta_1} - b_{x\theta_1}~e^{4\phi \o 3}~\Delta_3^2 ~
{\rm cot}^2~\hat\theta_1 \cr & (6)~~ {\cal G}_{y \theta_2} =-
e^{-{2\phi \o 3}}~(g_1~ \Delta_2^2~ {\rm cot}^2~\hat\theta_2 +
g_3)~b_{y\theta_2} -  b_{y\theta_2}~e^{4\phi \o 3}~\Delta_4^2 ~
{\rm cot}^2~\hat\theta_2 \cr & (7)~~ {\cal G}_{x \theta_2} =-
e^{-{2\phi \o 3}}~\left(g_1~\Delta_1~\Delta_2~{\rm cot}~\hat
\theta_1~{\rm cot}~\hat\theta_2 - {g_4 \o 2}~{\rm
cos}~\psi\right) ~b_{y\theta_2}  + {g_4 \o 2}~e^{-{2\phi \o
3}}~{\rm sin}~\psi + \cr & ~~~~~~~~~~~~~~~~~~ + e^{4\phi \o
3}~\Delta_3~\Delta_4~{\rm cot}~\hat\theta_1~{\rm
cot}~\hat\theta_2~b_{y\theta_2} \cr & (8)~~ {\cal G}_{y \theta_1}
= - e^{-{2\phi \o 3}}~\left(g_1~\Delta_1~\Delta_2~{\rm cot}~
\hat\theta_1~{\rm cot}~\hat\theta_2 - {g_4 \o 2}~{\rm
cos}~\psi\right) ~b_{x\theta_1}  + {g_4 \o 2}~e^{-{2\phi \o
3}}~{\rm sin}~\psi + \cr & ~~~~~~~~~~~~~~~~~~  + e^{4\phi \o
3}~\Delta_3~\Delta_4~{\rm cot}~\hat\theta_1~{\rm
cot}~\hat\theta_2~b_{x\theta_1} \cr & (9) ~~ {\cal
G}_{\theta_1\theta_2} = e^{-{2\phi \o
3}}~\left(g_1~\Delta_1~\Delta_2~{\rm cot}~\hat\theta_1~{\rm
cot}~\hat\theta_2 - {g_4 \o 2}~{\rm cos}~\psi\right)
~b_{x\theta_1}~b_{y\theta_2}  + {g_4 \o 2}~e^{-{2\phi \o 3}}~{\rm
cos}~\psi +\cr & ~~~~~~~~~~~~~~~~~~ - {g_4 \o 2}~e^{-{2\phi \o
3}}~{\rm sin}~\psi~(b_{x\theta_1} + b_{y\theta_2}) - e^{4\phi \o
3}~\Delta_3~\Delta_4~{\rm cot}~\hat\theta_1~{\rm
cot}~\hat\theta_2~b_{x\theta_1}~b_{y\theta_2} \cr & (10)~~ {\cal
G}_{xy} = e^{-{2\phi \o 3}}~\left(g_1~\Delta_1~\Delta_2~{\rm cot}~
\hat\theta_1~{\rm cot}~\hat\theta_2 - {g_4 \o 2}~{\rm
cos}~\psi\right) - e^{4\phi \o 3}~\Delta_3~\Delta_4~{\rm
cot}~\hat\theta_1~{\rm cot}~\hat\theta_2 \cr & (11) ~~ {\cal
G}_{zx} = g_1~e^{-{2\phi \o 3}}~\Delta_1~{\rm cot}~\hat\theta_1
~~(12) ~~{\cal G}_{zy} = g_1~e^{-{2\phi \o 3}}~\Delta_2~{\rm
cot}~\hat\theta_2 ~~ (13)~~{\cal G}_{zz} = g_1~e^{-{2\phi \o
3}}\cr & (14) ~~ {\cal G}_{z \theta_1} = -g_1~e^{-{2\phi \o
3}}~\Delta_1~{\rm cot}~\hat\theta_1~b_{x\theta_1} ~~ (15)~~ {\cal
G}_{z \theta_2} = -g_1~e^{-{2\phi \o 3}}~\Delta_2~{\rm
cot}~\hat\theta_2~b_{y\theta_2} \cr & (16)~~ {\cal G}_{rr} =
e^{-{2\phi \o 3}}~\gamma'~\sqrt{h} ~~~ (17) ~~{\cal G}_{ax} =
\Delta_3~e^{4\phi \o 3}~ {\rm cot}~\hat\theta_1 ~~~ (18) ~~{\cal
G}_{ay} = - \Delta_4~e^{4\phi \o 3}~ {\rm cot}~\hat\theta_2 \cr &
(19) ~~{\cal G}_{a\theta_1} = -\Delta_3~e^{4\phi \o 3}~ {\rm
cot}~\hat\theta_1~b_{x\theta_1} ~~~~~ (20)~~{\cal G}_{a\theta_2}
= \Delta_4~e^{4\phi \o 3}~ {\rm cot}~\hat\theta_2~b_{y\theta_2}
~~(21)~~{\cal G}_{aa} = e^{4\phi \o 3}}}
%where we have all the
%non zero components of the metric. Observe that we have kept both
%$g_2$ and $g_3$. This would therefore be the precise metric of the
%seven manifold.
The remaining seven components are all vanishing for this
specific case: \eqn\vancom{ {\cal G}_{rx} = {\cal G}_{ry} = {\cal
G}_{rz} = {\cal G}_{r\theta_1} = {\cal G}_{r\theta_2} = {\cal
G}_{ra} = {\cal G}_{az} = 0} with the spacetime metric
$e^{-{2\phi \o 3}}~h^{-{1\o 2}}$. Therefore \mecoone\ and
\vancom\ are basically the 28 components of our metric.

Having gotten the precise components and the one forms that
describe our  background, we can now get back to some of the
questions that we raised earlier in the type IIA section. Our
first question was to verify the non-K\"ahler nature of the type
IIA picture \filament. To do this we need the vielbeins. They can
be calculated from the one forms \oneformsM\ and \seconefor.
%We will also use the simplifying assumption: $g_2 = g_3$ for the
%warp factors. This simply means that the sphere metrics all come
%with the {\it same} warp factor. It is easy to go to a more
%generic description where we take $g_2 \ne g_3$, but we will not
%do so here. Our assumption therefore will be:
With the assumption \eqn\ourassum{n_1 = n_2 = \Delta_1 =
\Delta_3, ~~~~~~~ n_3 = n_4 = \Delta_2 = \Delta_4} the vielbeins
are now easy to determine. They can be extracted from the metric
components \mecoone, \vancom\ and the one forms \oneformsM,
\seconefor\ if we put $\phi = 0$ in \mecoone, and are defined as
\eqn\viel{e^a \equiv e^a_\mu ~dx^\mu = e^a_x~dx + e^a_y~dy +
e^a_z~dz + e^a_{\theta_i}~d\theta_i + e^a_r~dr,} where $a = 1,
..., 6$ and $e^a_\mu$ are given by (recall that
$\xi=\sqrt{g_3/g_2}$): \eqn\vielcomp{\eqalign{& {e^1_x \o \sqrt{2
g_2 - g_4/\xi}} = {1\o 2} {\rm sin}~\psi_1 = {e^3_x \o
\sqrt{g_4/\xi + 2 g_2}}, ~~~ {e^1_{\theta_1} \o \sqrt{2
g_2-g_4/\xi}} = {{\rm cos}~(\psi_1 + \lambda_1) \o 2~{\rm
cos}~\lambda_1} = {e^3_{\theta_1} \o \sqrt{g_4/\xi + 2 g_2}} \cr
& {e^1_y \o \sqrt{2 g_2- g_4/\xi}} = {1\o 2} {\rm sin}~\psi_2 =
{- e^3_y \o \sqrt{g_4/\xi + 2 g_2}}, ~~~
 {- e^1_{\theta_2} \o \sqrt{2 g_2 - g_4/\xi}} = {{\rm cos}~(\psi_2 -
 \lambda_2) \o 2~{\rm cos}~\lambda_2} =  {e^3_{\theta_2} \o \sqrt{g_4/\xi + 2 g_2}} \cr
& {e^2_x \o \sqrt{2 g_2 - g_4/\xi}} = {1\o 2} {\rm cos}~\psi_1 =
 {e^4_x \o \sqrt{g_4/\xi + 2 g_2}}, ~~~
{-e^2_{\theta_1} \o \sqrt{2 g_2 - g_4/\xi}} =
 {{\rm sin}~(\psi_1 + \lambda_1) \o 2~{\rm cos}~\lambda_1} =
  {-e^4_{\theta_1} \o \sqrt{g_4/\xi + 2 g_2}} \cr
& {e^2_y \o \sqrt{2 g_2 - g_4/\xi}} = {1\o 2} {\rm cos}~\psi_2 =
 {- e^4_y \o \sqrt{g_4/\xi + 2 g_2}}, ~~~
 {e^2_{\theta_2} \o \sqrt{2 g_2 - g_4/\xi}} =
 {{\rm sin}~(\psi_2 - \lambda_2) \o 2~{\rm cos}~\lambda_2} =
  {-e^4_{\theta_2} \o \sqrt{g_4/\xi + 2 g_2}}\cr
& e^5_x = \sqrt{g_1}~\Delta_1 ~{\rm cot}~\hat\theta_1, ~~e^5_y =
\sqrt{g_1}~\Delta_2 ~{\rm cot}~ \hat\theta_2, ~~ e^5_{\theta_i} =
 -\sqrt{g_1}~\Delta_i~{\rm tan}~\lambda_i ~{\rm cot}~\hat\theta_i,
~~e^5_z =\sqrt{g_1}}} with the remaining components all vanishing
except $e^6_r$ given by $e^6_r = \sqrt{\gamma'\sqrt{h}}$, where
we have defined $\gamma'$ in the beginning of section 5. These
are the local vielbeins for our case, and the metric \filament\
can also be written in terms of \vielcomp. In fact, as is well
known, both the metric and the fundamental two form $J$ for our
case can be written using the vielbeins \vielcomp\ as:
\eqn\metj{ds^2 = \sum_{a = 1}^6~ e^a \otimes e^a, ~~~~~~~~~ J =
\sum_{a,b} ~e^a \wedge e^b.} One may also define complex
vielbeins as $(e^1+ie^2), ~(e^3+ie^4)$ and $(e^5+ie^6)$, then we
get $J=(e^1\wedge e^2)+(e^3\wedge e^4)+(e^5\wedge e^6)$. This
reads in components as: \eqn\jcomponents{\eqalign{J &
=(\alpha_2^2-\alpha_1^2)~\sin\psi dx\wedge dy -
(\alpha_1^2+\alpha_2^2)~dx \wedge d\theta_1 +
(\alpha_1^2+\alpha_2^2)~dy \wedge d\theta_2 \cr & +
(\alpha_2^2-\alpha_1^2)~[\cos\psi -
\sin\psi~\tan\lambda_2]~dx\wedge d\theta_2 -
(\alpha_2^2-\alpha_1^2)~[\cos\psi -
\sin\psi~\tan\lambda_1]~dy\wedge d\theta_1 \cr & -
(\alpha_2^2-\alpha_1^2)~[\cos\psi~(\tan\lambda_1+\tan\lambda_2) +
\sin\psi~(1-\tan\lambda_1~\tan\lambda_2)]~d\theta_1\wedge
d\theta_2 \cr & +
\alpha_4\alpha_5\Delta_1\cot\hat\theta_1~dx\wedge dr +
\alpha_4\alpha_5\Delta_2\cot\hat\theta_2~dy\wedge dr +
\alpha_4\alpha_5~dz\wedge dr \cr &
-\alpha_4\alpha_5\Delta_1\cot\hat\theta_1\tan\lambda_1~d\theta_1\wedge
dr - \alpha_4\alpha_5\Delta_2\cot\hat\theta_2\tan\lambda_2
~d\theta_2\wedge dr.}} From here one can easily compute $dJ$ and
show that in general $dJ \ne 0$ because of terms like
$d\lambda_i$. This implies that the metric \filament\ is in
general not K\"ahler.

Having reached the explicit form of the background, it is now time
to pause again a little to discuss some of the expected mathematical
properties of these backgrounds. The six dimensional manifold
that we gave in \filament\ has an $SU(3)$ structure, is
non-K\"ahler and in general could be non-complex. Mathematical
properties of such manifolds have been discussed in some details
in \salamon, \gauntlett, \lust, \louis\ though an explicit
example have never been given before. To our knowledge, the
examples that we gave in this paper, are probably the first ones.

The holonomy of these manifolds are measured not wrt to the
usual  Riemannian connection, but wrt so-called torsional
connection. Earlier concrete examples of this were given in \sav,
\beckerD, \GP, \bbdg, \bbdgs. These examples dealt mostly with
heterotic theory and were non-K\"ahler but compact with integrable
complex structures. The examples that we gave here are mostly
non-compact, though compact examples could also be constructed
with some effort. Both the heterotic cases and the type II cases
are examples of torsional manifolds. As discussed in \salamon,
\louis\ and \lust, the torsional manifolds are classified by {\it
torsion-classes} ${\cal W}_i$, with $i = 1, 2, ...., 5$. In fact,
the torsion ${\cal T}$ belongs to the five classes:
\eqn\torso{{\cal T} \in {\cal W}_1 \oplus {\cal W}_2 \oplus {\cal
W}_3 \oplus  {\cal W}_4 \oplus {\cal W}_5} which in turn is
related to the $SU(3)$ irreducible representations. To determine
the torsion classes for our case, we need the ($3,0$) form
$\Omega$: \eqn\Omegad{\Omega \equiv \Omega_+ + i \Omega_- = (e^1
+ i e^2) \wedge (e^3 + i e^4) \wedge (e^5 + i e^6)} where $e^i$
are computed in \vielcomp. From the definition of the fundamental
two form $J$ in  \metj\ it can be observed that both
$\Omega_{\pm}$ are annihilated by $J$ and $\Omega_+ \wedge
\Omega_- = {2\o 3} J \wedge J \wedge J$. For our case,
\eqn\omepm{\eqalign{ & \Omega_+ = e^1 \wedge e^3 \wedge e^5 - e^2
\wedge e^4 \wedge e^5 -  e^1 \wedge e^4 \wedge e^6 - e^2 \wedge
e^3 \wedge e^6 \cr & \Omega_- = e^1 \wedge e^4 \wedge e^5 +  e^2
\wedge e^3 \wedge e^5 +  e^1 \wedge e^3 \wedge e^6 -  e^2 \wedge
e^4 \wedge e^6}} Using the three forms, and the fundamental two
form $J$ all the torsion classes can be constructed. The
fundamental two form $J$ and the holomorphic ($3,0$) form
$\Omega$ are not covariantly constant, as we had observed earlier
in \rstrom, \beckerD, \bbdg, \bbdgs, and they obey
\eqn\jandomega{\eqalign{& {\cal D}_m J_{np} = \nabla_m J_{np} -
{\cal T}_{mn}^{~~~r}J_{rp} - {\cal T}_{mp}^{~~~r}J_{nr} = 0 \cr &
{\cal D}_m \Omega_{nmp} = \nabla_m \Omega_{npq} - {\cal
T}_{mn}^{~~~r}\Omega_{rpq} - {\cal T}_{mp}^{~~~r}\Omega_{nrq} -
{\cal T}_{mq}^{~~~r}\Omega_{npr} = 0.}} The fact that the complex
structure is also not integrable might be argued from the
definition of the complex vielbeins: $e^i + i e^{i+1}$, that we
gave earlier. A more detailed analysis of the mathematical
discussion that we gave here will be relegated to part II. The
fact that the fundamental two form is not covariantly constant
implies that $dJ$ involves (3,0) and a (2,1) $+ {\rm c.c}$
pieces. The definition of $d$ becomes:
\eqn\defofd{d\omega^{(p,q)} = d\omega^{(p-1,q+2)} +
d\omega^{(p,q+1)} +  d\omega^{(p+1,q)} +  d\omega^{(p+2,q-1)}}
which, for an integrable complex structures will not have the
$d\omega^{(p-1,q+2)} +  d\omega^{(p+2,q-1)}$ pieces.

The lift of these six dimensional manifolds in M-theory gives rise
to $G_2$ manifolds in M-theory equipped with a $G_2$ structure.
The $G_2$ structure is specified by a three form $\tilde\Omega$
that could be easily evaluated from the vielbeins. We have
already defined six vielbeins earlier. Let us define $e^7 =
\sigma_3 + \Sigma_3$. Using this we can use the $SU(3)$ structure
to determine a $G_2$ structure on the seven manifold as:
\eqn\thrM{\eqalign{\tilde\Omega & = J \wedge e^7 + \Omega_+ \cr &
= e^1 \wedge e^3 \wedge e^5 - e^2 \wedge e^4 \wedge e^5 -  e^1
\wedge e^4 \wedge e^6 - e^2 \wedge e^3 \wedge e^6 \cr & ~~~~ + e^1
\wedge e^2 \wedge e^7 + e^3 \wedge e^4 \wedge e^7 +  e^5 \wedge
e^6 \wedge e^7.}} This determines the $G_2$ structure induced
from the $SU(3)$ structure of \filament. The $G$ fluxes will give
rise to a three form $G_3 = \ast_7 G$ on the seven manifold. This
$G_3$ is {\it not} related to the torsion, the torsion three form
being given by \eqn\torthrf{{\cal\tau} = -\ast d\tilde\Omega -
\ast\left[{1\o 3} \ast(\ast d\tilde\Omega \wedge \tilde\Omega)
\wedge \tilde\Omega \right].} To see how the connections in both
the theories behave, the reader may want to look into \ivan. The
equivalent to the torsion classes are now the four modules \gray:
\eqn\torMM{{\cal T} \in \chi_1 \oplus \chi_2 \oplus \chi_3 \oplus
\chi_4} where $\chi_i$ are the ${\bf 1, 14, 27, 7}$ of $SO(7)$.
For our case we do not have a closed three form and therefore the
manifold will have a $G_2$ structure. A $G_2$ structure of the
type $\chi_1 \oplus \chi_3 \oplus \chi_4$ is in general
integrable with a Dolbeault cohomology given in \graytwo. For
further details see Appendix 2.

There are many questions that arises now regarding the seven
dimensional manifold that we presented. They will be tackled in the
sequel to this paper. To continue,
we will assume that the manifold has an
explicit $G_2$ structure. It goes without saying, of course that,
since we followed strict mirror rules the background that we get
should always preserve the set of conditions required.

\newsec{Chain 3: M-theory Flop and Type IIA Reduction}

Having the $G_2$ manifold, we can perform
a  flop transition and reduce to the corresponding type IIA picture.
Before we perform the flop, we revisit the one-forms
of the previous section. We will continue following
the steps laid out in \brand, \cveticone, 
since the local behavior of our metric is
almost that of \brand, \cveticone.

To proceed further, first define a set of $2 \times 2$ matrices
$N_1$ and $N_2$ in the following way:
\eqn\nonemat{N^{\lambda_1}_1 \equiv \pmatrix{ \sigma_3 &
e^{i\psi_1} ~[e^{i\lambda_1}~{\rm sec}~\lambda_1~d\theta_1 - i
~dx] \cr e^{-i\psi_1}~[e^{-i\lambda_1}~{\rm
sec}~\lambda_1~d\theta_1 + i~ dx] & -\sigma_3}}
\eqn\ntwomat{N^{\lambda_2}_2 \equiv \pmatrix{ \Sigma_3 & \xi
e^{i\psi_2}~[e^{-i\lambda_2} ~{\rm sec}~\lambda_2~d\theta_2 + i~
dy] \cr \xi e^{-i \psi_2}~[e^{i\lambda_2}~ {\rm
sec}~\lambda_2~d\theta_2 - i~ dy] & -\Sigma_3}} where $\sigma_3$
and $\Sigma_3$ are the one form appearing in \oneformsM\ and
\seconefor, and we have again introduced the factor $\xi$ to
account for the asymmetry in the $\hat x$ and $\hat y$ terms. The
off--diagonal terms in the matrices are contributions from the
torsional part of the metric in the type IIA theory. Observe
that, in the absence of $B$ fields (in the original type IIB
theory), these matrices will take the following known form:
\eqn\matchange{\eqalign{N^0_1 & \equiv\pmatrix{ d\psi_1 +
\Delta_1~{\rm cot}~\hat\theta_1~dx & e^{i\psi_1} (d\theta_1 - i~
dx) \cr e^{-i \psi_1}(d\theta_1 + i~ dx) & -d\psi_1 -
\Delta_1~{\rm cot}~\hat\theta_1~dx} \cr N^0_2 & \equiv\pmatrix{
d\psi_2 - \Delta_2~{\rm cot}~\hat\theta_2~dy & \xi e^{i\psi_2}
(d\theta_2 + i~ dy) \cr \xi e^{-i \psi_2}(d\theta_2 - i~ dy) &
-d\psi_2 + \Delta_2~{\rm cot}~\hat\theta_2~dy}}}
which can be easily derived from the one forms given in \brand, \cveticone.

The matrices \nonemat\ and \ntwomat\ can be combined in various
ways to create new $2 \times 2$ matrices for our space. A generic
combination will be \eqn\gencov{N^{\lambda_1 \lambda_2}_{[a, b]}
= a~N^{\lambda_1}_1 - b~N^{\lambda_2}_2} with integral $a,b$.
Using various choices of $a,b$ we can express our eleven
dimensional metric. Locally however, one can show, that there are
two choices given by $N^{\lambda_1 \lambda_2}_{[1, 1]}$ and
$N^{\lambda_1 \lambda_2}_{[0, 1]}$ that are specifically useful
to write the M-theory metric. In the absence of $B$ fields (in
the original type IIB picture) these matrices are related to the
left invariant one forms $\omega$ and $\tilde\omega$ for the
$G_2$ spaces, i.e. \eqn\leftone{N^{00}_{[1, 1]}~ \to~
\tilde\omega, ~~~~~~~ N^{00}_{[0, 1]} ~ \to ~ \omega} as an
equality up to $SU(2)$ group elements. Of course, since for our
case there is no underlying $SU(2)$ symmetry globally (our
manifold being non-K\"ahler from the start i.e. directly in the
type IIA case) these relations are only in local sense. In the
presence of $B$ fields $-$ or non trivial fibrations in the
mirror/M-theory set-up $-$ the metric can still be written in
terms of $N^{\lambda_1 \lambda_2}_{[a, b]}$  even though they are
no longer related to $\omega$ and $\tilde\omega$. In fact, if we
remove the restriction on $a,b$ in \gencov\ as simple constants
and allow more generic values for them, we can express our
M-theory metric completely in terms of \gencov\ in the following
way: \eqn\mmetgec{ds^2 = - \left[{\rm det}~N^{\lambda_1
\lambda_2}_{[\beta, \beta]} - {\rm Tr}^2~(N^{\lambda_1
\lambda_2}_{[\gamma, \gamma]}\cdot \Gamma_3)\right] - \left[{\rm
det}~N^{\lambda_1 \lambda_2}_{[\delta, -\delta]} - {\rm
Tr}^2~(N^{\lambda_1 \lambda_2}_{[\epsilon, -\epsilon]}\cdot
\Gamma_3)\right]} where $\Gamma_3$ is the third Pauli matrix, and
$\beta, \gamma, \delta$ and $\epsilon$ can be extracted from
\warpyiden\ as: \eqn\bacvalve{\eqalign{& \beta = {e^{-{\phi \o
3}}\o {2}}~\sqrt{2g_2 - g_4/\xi}, ~~~~~\gamma = {e^{-{\phi \o
3}}\o {4}}~\sqrt{4g_1 - 2g_2 + g_4/\xi}\cr & \delta = {e^{-{\phi
\o 3}}\o {2}}~\sqrt{2g_2+g_4/\xi}, ~~~~~ \epsilon = {e^{-{\phi \o
3}}\o {4}}~\sqrt{4 e^{2\phi} - 2g_2 -g_4/\xi}.}}
In the absence of fluxes
i.e. $\lambda_i = 0$, on the other hand, it is conjectured that
the M-theory metric can
be written in terms of $N^{\lambda_1 \lambda_2}_{[a, b]}$ as \amv
\eqn\mmefg{ds^2 = - {\rm det}~N^{00}_{[A, A]} - {\rm
det}~N^{00}_{[B, -B]}} where the values of $A$ and $B$ are
defined at the radial distances $r= r_0$. We see that
this type of metric is not realized in out set-up
because \mmetgec\ do not reduce to \mmefg\ by making
$\lambda_i = 0$.
To compare our result \mmetgec\ to the one without fluxes, we
need the metric of the $G_2$ manifold with terms of the form
${\rm det}~N^{00} - {\rm Tr}^2~(N^{00}\cdot \Gamma_3)$. Taking
this limit is of course possible and an example of this has been
given in \brand, \cveticone. Our manifold would therefore resemble this
scenario, although one has to be careful here. The identification
is only local where the $\lambda_i$ values are approximately
constant. The matrices $N_i$ in the presence and in the absence
of fluxes differ by $\lambda_i$ terms, giving rise to one forms
that do not have any underlying $SU(2)$ symmetry. To summarize
the situation, our metric \mmetgec\ in the presence of fluxes
gives rise to new $G_2$ holonomy manifolds that have not been
studied before. In the absence of fluxes, \mmetgec\ gives rise to
one of the examples studied in \brand\ (see e.g. equation (3.5),
and discussion) which in our notation would look like:
\eqn\gukbran{ds^2 = - \left[{\rm det}~N^{00}_{[a, a]} - {\rm
Tr}^2~(N^{00}_{[b, b]}\cdot \Gamma_3)\right] - \left[{\rm
det}~N^{00}_{[c, -c]} - {\rm Tr}^2~(N^{00}_{[d, -d]}\cdot
\Gamma_3)\right]} with $a, b, c$ and $d$ are defined at the radial
coordinate $r = r_0$ as: \eqn\abcdvalue{\eqalign{& a = \sqrt{{4 r_0^2 +
12 r_0 - 27 \o 48}}, ~~~ b = {\sqrt{{36 r_0 - 4 r_0^2 - 81}}
 \o 24}, ~~~ c = \sqrt{4 r_0^2 - 12 r_0 - 27  \o
48} \cr
& ~~~~~~~~~~~~~~~~~~~~ d = \sqrt{{16 r_0^4 -
48  r_0^3 - 148 r_0^2 + 108 r_0 + 324 \o 48(4r_0^2 - 9)}}}}
As mentioned in \brand, \cveticone, there are alternative expressions for the
values of $a, b, c$ and $d$ obtained by scaling the metric. Then
the scale factor will appear in the set of relations \abcdvalue.
This will be useful for us to get rid of $\epsilon^{-1}$ factors
in the type IIA three form $d\tilde B$ with $\tilde B$ given earlier
in \bintwoa. 
We will discuss this soon. For more
details the readers can see sec. 3 of \brand. This is the point
where our conclusions would differ from the results of \amv\  where
\mmefg\ is presented as the M-theory lift of the type IIA configuration.

To proceed further, we need to perform the flop transition and
go  to the corresponding type IIA picture. On the other hand, if
we followed \amv, we
would see from the one forms \oneformsM\ and \seconefor\ that we
now have two possible directions along which we can compactify
and come down to type IIA: $dx$ and $d\theta_1$ (or
correspondingly $dy$ and $d\theta_2$). But the $\theta_1$
direction is not globally defined as it comes with the
corresponding type IIA $b_{x\theta_1}$ field. So the
compactification to type IIA should rather be performed along
$dx$. To
see this, let us first write the M-theory metric in the following
suggestive way: \eqn\metsugges{\eqalign{ds^2 = & ~ e^{-{2\phi \o
3}} g_1 (dz + \gamma_1~d\theta_1 + \gamma_2~D\hat y)^2 + e^{4\phi
\o 3} (dx_{11} +\gamma_3~d\theta_1 + \gamma_4~D\hat y)^2 + \cr &
~+~e^{-{2\phi \o 3}}~g_3~(d\theta_2^2 + D\hat y^2) + e^{-{2\phi
\o 3}}~g_2~{\rm cot}^2~\lambda_1~d\theta_1^2 + g_5~(dx + {\cal
A})^2 - g_5~{\cal A}^2 + \cr & ~+~e^{-{2\phi \o 3}}~ g_4~{\rm
cos}~(\psi + \lambda_1)~{\rm sec}~\lambda_1~d\theta_1~d\theta_2 +
e^{-{2\phi \o 3}}~g_4~{\rm sin}~(\psi + \lambda_1)~{\rm
sec}~\lambda_1~D\hat y ~d\theta_1}} where we have already defined
$D\hat y \equiv dy - {\rm tan}~\lambda_2~d\theta_2$ earlier, and
the $dx$ fibration structure is represented as $dx + {\cal A}$,
${\cal A}$ being the corresponding one form that will appear in
type IIA as gauge fields. The other variables appearing in
\metsugges\ can be defined as follows:
\eqn\defmet{\eqalign{\gamma_1 & = -\Delta_1 ~{\rm
cot}~\hat\theta_1~{\rm tan}~\lambda_1, ~~~~~~~ \gamma_2 =\Delta_2
~{\rm cot}~\hat\theta_2, ~~~~~~~
 \gamma_3 = -\Delta_3 ~{\rm cot}~\hat\theta_1~{\rm tan}~\lambda_1 \cr
\gamma_4 & =-\Delta_4 ~{\rm cot}~\hat\theta_2, ~~~~ g_5 \equiv
e^{4\Phi \o 3} = (g_1~\Delta_1^2 ~{\rm cot}^2~\hat\theta_1 +
g_2)~e^{-{2\phi \o 3}} + e^{4\phi \o 3} \Delta_3^2~{\rm
cot}^2~\hat\theta_1 \cr {\cal A}& = A_1 ~dz + A_2 ~d\theta_1 +
A_3 ~d\theta_2 + A_4 ~D\hat y + A_5 ~dx_{11}}} where $\Phi$ is the
type IIA dilaton and $A_i$ are the components of the gauge fields
defined in the following way: \eqn\comgaud{\eqalign{& A_1 =
{e^{-{2\phi \o 3}}~g_1~\Delta_1~{\rm cot}~\hat\theta_1 \o (g_2 +
g_1~\Delta_1^2 ~{\rm cot}^2~\hat\theta_1)~e^{-{2\phi \o 3}} +
e^{4\phi \o 3} \Delta_3^2~{\rm cot}^2~\hat\theta_1}, ~~~ A_2 = -
{\rm tan}~\lambda_1 \cr & A_3 = {{1\o 2} e^{-{2\phi \o 3}}~g_4~
{\rm sin}~\psi\o (g_2 + g_1~\Delta_1^2 ~{\rm
cot}^2~\hat\theta_1)~e^{-{2\phi \o 3}}
 + e^{4\phi \o 3} \Delta_3^2~{\rm cot}^2~\hat\theta_1} \cr
& A_5 = { e^{4\phi \o 3} ~\Delta_3~ {\rm cot}~ \hat\theta_1 \o
(g_2 + g_1~\Delta_1^2 ~{\rm cot}^2~\hat\theta_1)~e^{-{2\phi \o
3}} + e^{4\phi \o 3} \Delta_3^2~{\rm cot}^2~\hat\theta_1}\cr &
A_4 = {{\rm cot}~\hat\theta_1{\rm cot}~\hat\theta_2~(e^{-{2\phi
\o 3}}~g_1~ \Delta_1~\Delta_2 -  e^{4\phi \o
3}~\Delta_3~\Delta_4) - {1\o 2} e^{-{2\phi \o 3}}~g_4~{\rm
cos}~\psi \o (g_2 + g_1~\Delta_1^2 ~ {\rm
cot}^2~\hat\theta_1)~e^{-{2\phi \o 3}} + e^{4\phi \o 3}
\Delta_3^2~ {\rm cot}^2~\hat\theta_1}}}
The metric \defmet\ would
basically be our answer, but if we look closely  we see that
\defmet\ do not resemble the expected resolved conifold metric in
the absence of $b_{x\theta_1}, b_{y\theta_2}$. 
%This is somewhat
%expected because we have to do a coordinate transformation to go
%to the right metric after dimensional reduction. In other words
Therefore instead of reducing along $dx$, which would not give us the
expected resolved conifold metric in the absence of
$b_{x\theta_1}, b_{y\theta_2}$ without further coordinate
transformation, we will consider performing a flop in M-Theory
and then reduce along $dx_{11}$.
To consider the flop and the subsequent change in the metric we
have to do a transformation to our M-theory metric \mmetgeneric.
Before moving ahead, let us clarify one minor thing regarding the
scalings of the metric \mmetgeneric. As discussed in \brand, \cveticone,
scaling the metric from ${\cal G} ~\to ~ a^2 ~{\cal G}$ keeps the
$G_2$ structure intact. We can use this freedom of rescaling the
metric to remove the $\epsilon^{-{1\o 2}}$ factor in \bintwoa. The
coordinates of the $G_2$ manifold are given in terms of $x, y,
z, \theta_1, \theta_2$ and $x_{11}$ at a point $r = r_0$. 
Let us scale the radial
coordinate $r_0$ as $r_0 ~\to ~ \epsilon^{1\o 6}~r_0$. We also want to
scale $dz$ as $\epsilon^{1\o 6}~dz$ so that
$d\psi$ would not scale. This is important, since now
the ${\rm sin}~\psi$ and ${\rm cos}~\psi$ in the metric \filament\
will remain unchanged. If we now rescale $$x, y, z, \theta_i,
x_{11}  ~\to ~ \epsilon^{1\o 6}~x, \epsilon^{1\o 6}~y,
\epsilon^{1\o 6}~z, \epsilon^{1\o 6}~\theta_i, \epsilon^{1\o
6}~x_{11}$$ \noindent the metric \filament\ or the complete
M-theory metric \mmetgeneric\ will have an overall scale of
$\epsilon^{1\o 3}$ (provided, of course, that we also scale the 
$B$ field ${\tilde b}_{mn}$ in the fibration accordingly). 
This metric preserves $G_2$ structure as
before, but now the three form flux $C_3$ coming from the two
form $B_{NS}$  in \hatbnow\ as $C_3 = \hat B \wedge dx_{11}$ will
be finite. Thus, we can make the background finite using the
rescaling freedom, and therefore this gives us confidence in
considering only the finite part of the background \hatbnow\ and
the metric \filament\ without any $\epsilon$ dependences anywhere.

After this detour, it is now time to consider the issue of flop
on  the M-theory metric \mmetgeneric. This way we will be able to
connect the final answer, after dimensional reduction, to the
type IIA metric implied above in \metsugges. A simple way to
guess the answer would be to restore the case without any
torsion. In the absence of torsion the type IIA reduction should
be a resolved conifold with fluxes and no $D6$ branes. This would
imply that the metric looks like two tori. 
If we now use our one forms \oneformsM\ and \seconefor,
one way to generate this would be if we consider the
transformation on $N^{\lambda_1 \lambda_2}_{[a,b]}$ as:
\eqn\floptra{N^{\lambda_1 \lambda_2}_{[1,0]} ~~\to ~~N^{\lambda_1
\lambda_2}_{[{1+f \o 2}, {1\o 2}]}, ~~~~~~~ N^{\lambda_1 \lambda_2}_{[0,
-1]} ~~\to ~~ N^{\lambda_1 \lambda_2}_{[{1-f \o 2}, {1\o 2}]}} upto
possible conjugations. Locally, the above relation will imply a
similar relation in the absence of type IIB fluxes. We have also
kept a parameter $f$ in \floptra. Therefore,
the transformation \floptra\ will convert our case \mmetgec\ to:
\eqn\afterflop{ds^2_{\rm Flop} = -\left[{\rm det}~N^{\lambda_1
\lambda_2}_{[\beta, 0]} - {\rm Tr}^{2}~(N^{\lambda_1
\lambda_2}_{[\gamma, 0]}\cdot \Gamma_3)\right] - \left[{\rm
det}~N^{\lambda_1 \lambda_2}_{[f\delta, \delta]} - {\rm
Tr}^{2}~(N^{\lambda_1 \lambda_2}_{[f \epsilon, \epsilon]}\cdot
\Gamma_3)\right]} upto possible rescaling, and $\beta, \gamma,
\delta$ and $\epsilon$ have already been given in \bacvalve.

In the limit $f \to 0$ the above metric gives the right tori
parts but fails to give the fibration structure correctly. This
implies that a global definition {\it a-la} \floptra\ may not be
possible here. What went wrong? A careful study of \afterflop\
reveals that in the summation of the one forms $\Sigma_a -
f~\sigma_a, ~a = 1, 2,3$ we had used the same $f$ for all the
three terms. In the limit where $f \to 0$ this gives the right
torus metric but wrong fibration. A way out of this can be
immediately guessed by having a different factor in the third
term, i.e. having $\Sigma_3 - g~\sigma_3$, and the rest with $f$.
Locally, this is exactly the one predicted by \brandtwo, and
therefore, using out patch argument, we can extend this to all
other patches. This implies the following metric after we make a
flop in M-theory: \eqn\afterflopagain{\eqalign{ds^2_{\rm Flop} = &
-\left[{\rm det}~N^{\lambda_1 \lambda_2}_{[\beta, 0]} - {\rm
Tr}^{2}~(N^{\lambda_1 \lambda_2}_{[\gamma, 0]}\cdot
\Gamma_3)\right] \cr & - \left[{\rm det}~N^{\lambda_1
\lambda_2}_{[f\delta, \delta]} + {\rm Tr}^{2}~(N^{\lambda_1
\lambda_2}_{[{f \delta \o 2}, {\delta \o 2}]}\cdot \Gamma_3) -
{\rm Tr}^{2}~(N^{\lambda_1 \lambda_2}_{[{g \alpha_3 \o 2},
{\alpha_3 \o 2}]}\cdot \Gamma_3) \right]^{g \to 1}_{f \to 0}}}
where the variables have been defined in \bacvalve\ and
\warpyiden. Thus, before flop the metric is given by \mmetgec\
and after flop it is given by \afterflopagain.

\subsec{The Type IIA Background}

To obtain the type IIA theory we can reduce either  via $dz$ or
via $dx_{11}$. This will not lead us back to the type IIA theory
we started with because of the change induced in the metric by
the flop. To have a one to one correspondence with the type IIA
picture before flop, let us reduce along direction $dx_{11}$. The
new $G_2$ metric can now be written in the following suggestive
way: \eqn\gtwosuggest{\eqalign{ds^2 & =  e^{4\phi \o 3}\left[dz +
\Delta_1~{\rm cot}~\hat\theta_1~(dx - b_{x\theta_1}~d\theta_1) +
\Delta_2 ~{\rm cot}~\hat\theta_2~(dy -
b_{y\theta_2}~d\theta_2)\right]^2  \cr +& e^{-{2\phi \o
3}}\left({g_2\o 2} - {g_4 \o 4 \xi}\right)\left[d\theta_1^2 + (dx
- b_{x\theta_1}~d\theta_1)^2\right]  + e^{-{2\phi \o
3}}\left({g_2\o 2} + {g_4 \o 4\xi}\right)\left[d\theta_2^2 + (dy -
b_{y\theta_2}~d\theta_2)^2\right] \cr & ~~~~~ + {1\o 4}
e^{-{2\phi \o 3}}~g_1~\left[dx_{11} + 2\Delta_1~ {\rm
cot}~\hat\theta_1~(dx - b_{x\theta_1}~d\theta_1)\right]^2}} which
clearly shows that the base is locally a resolved conifold.
Note, that we have again used the freedom to absorb $A_z$ into $dx_{11}$.
The
$dx_{11}$ term in \gtwosuggest\ is basically the fibration over
which we have to reduce to get to type IIA theory. As mentioned
above, we can also reduce along $dz$, as the $dx_{11}$ and $dz$
directions can be easily exchanged among each other. The metric
\gtwosuggest\ is thus the right $G_2$ metric after flop and could
be compared to \metsugges. A redefinition of the coordinates of
\metsugges\ and some coordinate transformation would relate
\metsugges\ to \gtwosuggest\ and would also simplify the form of
the gauge potential given earlier in \comgaud. The final type IIA
metric after a dimensional reduction turns out to be:
\eqn\iiafinal{\eqalign{ds^2 = & {1\o 4}\left(2g_2 - {g_4\o
\xi}\right)\left[d\theta_1^2 + (dx - b_{x\theta_1}~d\theta_1)^2
\right]  + {1\o 4} \left(2g_2 + {g_4\o
\xi}\right)\left[d\theta_2^2 + (dy -
b_{y\theta_2}~d\theta_2)^2\right]  \cr & + e^{2\phi}\left[dz +
\Delta_1~{\rm cot}~\hat\theta_1~(dx - b_{x\theta_1}~d\theta_1) +
\Delta_2 ~{\rm cot}~\hat\theta_2~(dy -
b_{y\theta_2}~d\theta_2)\right]^2}} which is precisely the metric
of a resolved conifold when we switch off $b_{x\theta_1}$ and
$b_{y\theta_2}$ (or consider it locally over a patch where
$b_{x\theta_1}$ and $b_{y\theta_2}$ are constants). In the
presence of $b_{x\theta_1}$ and $b_{y\theta_2}$ we get the
``usual'' metric but shifted by the generic ansatz that we
proposed in \redpsietc. Therefore, we can now make a precise
statement: {\it the metric before geometric transition is given by
\filament, and after the transition is given by \iiafinal}. The
type IIA metric has two tori whose radii are proportional to
\eqn\radoft{r_1 = {1\o 2} \sqrt{2g_2 - g_4\sqrt{g_3 g^{-1}_2}}, ~~~~
r_2 = {1\o 2}
\sqrt{2g_2 +g_4\sqrt{g_3 g^{-1}_2}}.}
One of them would shrink to zero
size while the other doesn't when we approach the origin. The
type IIA coupling is now given by \eqn\iicofinal{ g_A = 2^{-{3\o
2}} e^{-{\phi \o 2}} \alpha^{-{3\o 4}}} where $\alpha$ is defined
in \defalpha. Observe that the coupling is not a constant but is
a function of the internal coordinates, and it doesn't blow up
anywhere in the internal space. This background is the expected
background after we perform a geometric transition on \filament.
This means that the $D6$ branes in \filament\ should completely
disappear and should be replaced by fluxes in the type IIA
picture. From the $G_2$ manifold that we had in \gtwosuggest, we
see that this is indeed the case, and the gauge fluxes are given
by: \eqn\gaufinalii{{\cal A}\cdot dX =  2\Delta_1~{\rm
cot}~\hat\theta_1~(dx - b_{x\theta_1}~d\theta_1)} which, as one
can easily check, looks like the remnant of $D6$ brane sources
modified appropriately by our ans\"atze \redpsietc. There are also
$B_{NS}$ fields that originate from the dimensional reduction of
the three form fields in M-theory. Since we are reducing along the
direction $dx_{11}$ they would be the same $\hat B$ field that we
had in \hatbnow. The only difference will be that the finite part
(which is of course $\hat B$ itself) is now the exact solution as
we had removed the $\epsilon^{-1/2}$ dependence by scaling our
$G_2$ manifold before flop. The $B_{NS}$ can be written down
directly from \hatbnow\ as:
 \eqn\hatbnowfinal{{B \o \sqrt{\alpha}} =
 dx \wedge d\theta_1 - dy \wedge d\theta_2 +
 A~d\theta_1\wedge dz - B~({\rm sin}~\psi ~dy -
{\rm cos}~\psi~d\theta_2) \wedge dz,} which will again be a pure gauge artifact.
Combining \iiafinal,
\iicofinal, \gaufinalii\ and \hatbnowfinal, we recover the
precise background after geometric transition in type IIA picture.

\subsec{Analysis of Type IIA Background and Superpotential}

In this section we will try to verify the non-K\"ahler nature of
our background and the corresponding superpotential. Other detail
aspects, for example non integrability of complex structure,
torsion classes etc., will be left for part II of this paper. To
check the non-K\"ahlerity of this background we will have to
determine the corresponding vielbeins. They can be easily
extracted from \iiafinal, and are given by:
\eqn\vielsfinal{\eqalign{e & = \pmatrix{e^1_x & e^1_y & e^1_z &
e^1_{\theta_1}& e^1_{\theta_2}& e^1_r \cr \noalign{\vskip -0.20
cm}  \cr e^2_x & e^2_y & e^2_z & e^2_{\theta_1}& e^2_{\theta_2}&
e^2_r \cr \noalign{\vskip -0.20 cm}  \cr e^3_x & e^3_y & e^3_z &
e^3_{\theta_1}& e^3_{\theta_2}& e^3_r \cr \noalign{\vskip -0.20
cm}  \cr e^4_x & e^4_y & e^4_z & e^4_{\theta_1}& e^4_{\theta_2}&
e^4_r \cr \noalign{\vskip -0.20 cm}  \cr e^5_x & e^5_y & e^5_z &
e^5_{\theta_1}& e^5_{\theta_2}& e^5_r \cr \noalign{\vskip -0.20
cm}  \cr e^6_x & e^6_y & e^6_z & e^6_{\theta_1}& e^6_{\theta_2}&
e^6_r} \cr \noalign{\vskip -0.20 cm}  \cr & = \pmatrix{0 & 0 & 0
& r_1 & 0 & 0 \cr \noalign{\vskip -0.20 cm}  \cr
 r_1 & 0 & 0 & -r_1~b_{x\theta_1} & 0 & 0 \cr
\noalign{\vskip -0.20 cm}  \cr 0 & 0 & 0 & 0 & r_2 & 0 \cr
\noalign{\vskip -0.20 cm}  \cr 0 &  r_2 & 0 & 0 &
-r_2~b_{y\theta_2} & 0 \cr \noalign{\vskip -0.20 cm}  \cr e^\phi
\Delta_1~ {\rm cot}~\hat\theta_1 & e^\phi \Delta_2~ {\rm
cot}~\hat\theta_2  & e^\phi & - e^\phi \Delta_1~ {\rm
cot}~\hat\theta_1 b_{x\theta_1}  & - e^\phi \Delta_2~{\rm
cot}~\hat\theta_2 b_{y\theta_2} & 0 \cr \noalign{\vskip -0.20
cm}  \cr
 0 & 0 & 0 & 0 & 0 & e^6_r}}}
where $r_1$ and $r_2$ are the radii of the two tori as defined
earlier. We have kept $e^6_r$ undefined here. But this can also
be easily seen to be the usual vielbein for the resolved conifold
case in the type IIB picture. Now to check the non-K\"ahlerity we
have to construct the fundamental two form ${\cal J}$ using these
vielbeins. Before evaluating this, observe that in the absence of
$b_{x\theta_1}$ and $b_{y\theta_2}$ the manifold should be
K\"ahler with a K\"ahler form $J$. In the presence of
$b_{x\theta_1}$ and $b_{y\theta_2}$ the fundamental form ${\cal
J}$ can be written as a linear combination of the usual K\"ahler
form $J$ and additional  $b_{x\theta_1}$ and $b_{y\theta_2}$
dependent terms, as \eqn\twoform{{\cal J} = J +
e^{\phi}~(\Delta_1~{\rm cot}~\hat\theta_1~b_{x\theta_1}~e^6_r
\wedge d\theta_1 +
 \Delta_2~{\rm cot}~\hat\theta_2~b_{y\theta_2}~e^6_r \wedge d\theta_2)}
where $dJ = 0$. From above it is easy to see that $d{\cal J} \ne
0$ in general  because of non- zero $db_{x\theta_1}$ and
$db_{y\theta_2}$. Therefore the manifold \iiafinal\ is a
non-K\"ahler manifold\foot{As this point one might wonder about the global 
behavior of \iiafinal. Before geometric transition the global type IIA picture 
had extra six branes and other defects. After geometric transition we would still 
expect some of the six branes and possibly other defects to reappear. One has to 
carefully do the flop operation in the presence of these objects to see how 
many of them would survive in the type IIA side. More details on this will be presented 
elsewhere.}.  
For completeness, let us also write down
all the components of the type IIA metric:
\eqn\twoacomp{\eqalign{g & = \pmatrix{g_{xx} & g_{xy} & g_{xz} &
g_{x\theta_1} & g_{x\theta_2} \cr \noalign{\vskip -0.20 cm}  \cr
g_{xy} & g_{yy} & g_{yz} & g_{y\theta_1} & g_{y\theta_2} \cr
\noalign{\vskip -0.20 cm}  \cr g_{xz} & g_{yz} & g_{zz} &
g_{z\theta_1} & g_{z\theta_2} \cr \noalign{\vskip -0.20 cm}  \cr
g_{x\theta_1} & g_{y\theta_1} & g_{z\theta_1} &
g_{\theta_1\theta_1}& g_{\theta_1\theta_2} \cr \noalign{\vskip
-0.20 cm}  \cr g_{x\theta_2} & g_{y\theta_2} & g_{z\theta_2} &
g_{\theta_1\theta_2} & g_{\theta_2\theta_2}} \cr \noalign{\vskip
-0.25 cm}  \cr & = \pmatrix{C_1 & e^{2\phi} AB & e^{2\phi} A &
-b_{x\theta_1} C_1 & -e^{2\phi} b_{y\theta_2} AB \cr
\noalign{\vskip -0.20 cm}  \cr
 e^{2\phi} AB & D_1 & e^{2\phi} B & -e^{2\phi} b_{x\theta_1} AB
 & -b_{y\theta_2} D_1 \cr
\noalign{\vskip -0.20 cm}  \cr e^{2\phi} A & e^{2\phi} B  &
e^{2\phi} &  -e^{2\phi} b_{x\theta_1} A & -e^{2\phi}
b_{y\theta_2} B \cr \noalign{\vskip -0.20 cm}  \cr -b_{x\theta_1}
C_1 &  -e^{2\phi} b_{y\theta_2} AB & -e^{2\phi} b_{x\theta_1} A &
C + b^2_{x\theta_1}C_1 & e^{2\phi} AB b_{x\theta_1} b_{y\theta_2}
\cr \noalign{\vskip -0.20 cm}  \cr
 -e^{2\phi} b_{y\theta_2} AB & -b_{y\theta_2} D_1  & -e^{2\phi}
 b_{y\theta_2} B  & e^{2\phi} AB b_{x\theta_1} b_{y\theta_2}  &
D + b^2_{y\theta_2}D_1}}} where $A$ and $B$ have been defined
earlier in \defAandB. The other
 variables appearing in \twoacomp\ can be defined as follows:
\eqn\defiiac{C_1 = C + A^2~e^{2\phi}, ~~ D_1 = D + B^2~e^{2\phi},
~~ C = {g_2\o 2} - {g_4 \o 4\xi}, ~~ D = {g_2\o 2} + {g_4 \o
4\xi}} Thus, \twoacomp\ is the final answer for the type IIA
background without any $D6$ branes and with two-- and three--form
field strengths. There are many questions that arise from the
explicit background that we have in \iiafinal\ and \twoacomp. Let
us elaborate them:

\noindent $\bullet$ The first issue is related to the choice
of complex structure for our manifold. The complex structure is
written in terms of the fermions, and therefore we have to see
how the fermions transform under three T-dualities. From the
generic analysis of \kachruone, we see that the T-dual fermions
give rise to a complex structures that is in general not integrable
(in other words, the Nijenhaus tensor does not vanish). Therefore we
will get a non-complex manifold.

\noindent $\bullet$ The next issue is related to the
non-K\"ahlerity of our manifold. The naive expectation
(also from the results of \louis) would be that the manifold
we get in type IIA will be
{\it half-flat}. This comes from the fact that \iiafinal\ is
non-K\"ahler and also non-complex. Half-flat manifolds are
classified by torsion classes. For our case all these can be
explicitly derived from the metric. Below we will show that the
naive expectation is {\it not} realized in string theory and
our manifold will be more general than a half-flat manifold.

\noindent $\bullet$ The third issue is the asymptotic behavior of 
our metric. We haven't yet checked whether the metric that we derived above is 
non-degenerate and non-singular. Although unrelated, a similar 
metric with identical $B$ dependent fibration structure found in \sav,\bbdg,\bbdgs,
showed a good asymptotic behavior and was non-degenerate and non-singular. For the present case
however, we don't know the full global metric as our type IIB starting point was the local 
metric that ignored the seven branes. Once the full global story becomes clear, we should study the 
singularity behavior of the metric. 

\noindent $\bullet$ Last but not the least, we need to determine
the superpotential that governs our type IIA background. Before
doing so, let us start by recalling which the fields are present in
M--theory. The NS field \hatbnow\ is lifted to a three-form
$C$ by adding a leg in the $x^{11}$ direction. Its derivative is
$G = d C$. For the compactification on a $G_2$ manifold X with
the invariant 3-form $\tilde\Omega$ defined earlier in \thrM, 
the form of the superpotential was
first proposed in \gukov\  as: \eqn\potu{W = \int_{X} \tilde\Omega \wedge
G.} This form of the superpotential has been corrected in \ach,
\bw\ in order to make the right hand side a complex quantity. The
superpotential becomes\foot{Of course, the superpotential is 
for a {\it generic} $G_2$ structure. If the structure group is a 
subgroup of $G_2$ then the superpotential will be a truncation of the one 
that we mention here. These details have been addressed recently in \bertwo.
We thank K. Behrndt for correspondence on this issue.} 
\eqn\potc{W = \int_{X} ( \tilde\Omega + i C) \wedge
G.} Then, when reducing from 11 dimensions to 10 dimensions as in
\amv, \brandtwo, instead of just obtaining the volume of the
resolved conifold $J$, we get the complexified volume $J + i B$,
and this is the quantity that enters in the 10 dimensional potential
to give
\eqn\potzu{\tilde{W}_{1} = \int_{X_6}  (J + i B) \wedge d H_{3}.}
This is true because $\tilde\Omega$ descends to $J$ as it loses one leg
but $G$ descends as an RR 4-form. The RR 4-form should originate from
D4 branes. Since we know that our brane configurations did not
contain any D4 branes, there is no contribution from
$dH_{3}$ in the superpotential.

But this is not the full story because of the properties of the
compactification manifold. By considering the dimensional
reduction on the manifold \iiafinal, we have to use the fact that
the manifold does not have a closed (3,0) form. Let us first
recall the results for the case without torsion \tp, \pandoz. In
that case the condition that the (3,0) form is closed\foot{This
statement is equivalent to saying that the manifold is Ricci
flat.} is related to a  differential equation for a function
depending on the radial coordinate. The result was a one
parameter family of Calabi-Yau metrics on the resolved conifold.

In our case, the situation is different. We consider equation
\iiafinal\ and we read off the vielbeins from \vielsfinal.
The holomorphic 3-form is built as before as: \eqn\holof{\Omega =
(e_1 + i e_2) \wedge (e_3 + i e_4)  \wedge (e_5 + i e_6).} {}From
\holof\ we see that the condition for the (3,0) form to be closed
implies a differential equation which involves the functions
$b_{x \theta_{1}}, b_{y \theta_2}$, as they appear in
$\hat{x},~\hat{y},~\hat{z}$. As the functions  $b_{x \theta_{1}},
b_{y \theta_2}$ are arbitrary, the differential equation will not
have a solution for generic values of $b_{x \theta_{1}}, b_{y
\theta_2}$, so our situation is different from the one of \tp. It
also differs from the situation of \pandoz\ in the sense that our
manifold is not Ricci flat because the same differential equation
does not have a solution for generic values of $b_{x \theta_{1}},
b_{y \theta_2}$. For our case one can explicitly evaluate the
three-forms. Using the definitions of $D\hat x$ and $D\hat y$ we can
express our result as:
\eqn\ugapm{\eqalign{& {\Omega_+ \o \sqrt{4g_2^2 - {g_4^2 ~\xi^{-2}}}} =
e^6_r~d\theta_1 \wedge d\theta_2 \wedge dr - e^\phi~d \theta_1 \wedge
D\hat y \wedge (dz + \Delta_1~{\rm cot}~\hat\theta_1~D\hat x)~ + \cr
& ~~~~~~~~~~~~~~~~~ - e^6_r~D\hat x \wedge D\hat y\wedge dr + e^\phi~
d\theta_2 \wedge D\hat x \wedge (dz +
\Delta_2~{\rm cot}~\hat\theta_2~D\hat y)\cr
& {\Omega_- \o \sqrt{4g_2^2 - {g_4^2~\xi^{-2}}}} = e^\phi~d\theta_1 \wedge
d\theta_2 \wedge (dz + \Delta_1~{\rm cot}~\hat\theta_1~D\hat x +
\Delta_2~{\rm cot}~\hat\theta_2~D\hat y)~+ \cr
& ~~~~~~~~~~~~~~~~~ + e^6_r~(d\theta_1 \wedge D\hat y \wedge dr +
D\hat x \wedge d\theta_2 \wedge dr) - e^\phi~D\hat x \wedge D\hat y \wedge
dz}}
where $e^6_r$ is the associated vielbein for the $r$ direction.
{}From above we see that both $d\Omega_+$ and $d\Omega_-$ will not vanish
for the background that we have.
The existence of  $b_{x \theta_{1}}, b_{y \theta_2}$
implies the  non-closeness of the holomorphic 3-form $\Omega$.
Therefore our manifold is a specific non-complex, non-K\"ahler
manifold that is not half-flat.
Manifolds with an $\Omega$ which is not closed have been studied
in \louis\ where four forms $F^{2,2} \propto (d \Omega)^{2,2}$
correspond to harmonic forms measuring flux, and they can be
expanded in some basis\foot{For integrable complex structures one
could expand in $h^{1,1}$ basis, although if the manifold is
simultaneously non-K\"ahler this would be tricky. Recall also that we are
using $d\Omega \equiv d\left[\Omega^{(3,0)}\right] = 
d\Omega^{(2,2)} + d\Omega^{(3,1)} +
d\Omega^{(4,0)}$ for the non-complex manifold.}. 
The four
forms can then be combined with the independent holomorphic 2
forms to give contributions to the superpotential as
\eqn\extrasup{(J + i B) \wedge d \Omega.} This way we
encounter a first concrete example where the superpotential gets
an extra piece from the non  closed holomorphic 3-form. The case in
\louis\ involved an $\Omega$ with only the real part non--
closed and the manifold was a half flat manifold. Our case is
more general, as both $d \Omega_{+}$ and $d \Omega_{-}$ can be
non zero as functions of $b_{x \theta_{1}}, b_{y \theta_2}$.

\newsec{Discussion and future directions}

The subject of the present work was to clarify issues concerning
NS fluxes in geometric transitions and the non-K\"ahler geometries
arising in
the mirror pictures. Our starting point was the observation of
Vafa \vafai\ that the closed string dual to D6 branes wrapped
on a deformed conifold
is not a K\"ahler geometry. To obtain this mysterious departure from
K\"ahlerity, we started\foot{In terms of the dual ${\cal N} = 1$ gauge theory 
this is the IR of the gauge theory. In terms of geometry this is the region
where $r$, the radial parameter, is small.} 
from a IIB picture with D5 branes wrapped on a
$P^1$ cycle inside the resolved conifold and went to the mirror
picture by performing three T-dualities on the fiber $T^3$. The
result was a
non-K\"ahler geometry whose metric could be given precisely as a
non-K\"ahler deformation of a deformed conifold.
We then lifted this to M theory where the result was a new $G_2$ manifold with
torsion. A flop inside the $G_2$ manifold and a reduction to type IIA
brought us to the closed IIA picture with a non-K\"ahler deformation of the
resolved conifold. The latter non-K\"ahlerity can then be traced back
to the existence of the NS flux in the initial type IIB picture.
Our final result is not quite a half-flat manifold as anticipated
earlier \louis, but
is actually more general because the imaginary part of the
holomorphic 3-form is also not closed.

On the way we also solved some puzzles regarding the
T-duality between branes wrapped on the deformed and resolved conifold.
Previous attempts started from the deformed
conifold and the problems
encountered were related to the fact that this is not a toric variety.
We started with the resolved conifold which is a toric variety and
identified the $T^3$ fibration.

\subsec{Future directions}

There are many unanswered questions that we left for future work. A sample of
them are as follows:

\noindent $\bullet$ In the
figure we have drawn, there is an extra step which should be covered,
the mirror symmetry
that goes from the closed IIA to closed type IIB and it would be
interesting to see whether this would give rise to
 a K\"ahler geometry with
NS flux or to a non-K\"ahler geometry. Unfortunately, there is an
immediate problem that one would face while doing the mirror transformation.
The background that we have has lost the isometry along the $z$ direction.
Recall that the type IIB background that we started with in the beginning
of the duality chain had complete isometries along the $x, y$ and $z$
directions. In the final type IIA picture the metric does surprisingly have
all the isometries, but the $B$ field breaks it. Maybe a transformation
of the form \tranthe\ could be used here to get the mirror metric.

\noindent $\bullet$ To get the cross terms in the type IIA
mirror metric (on the
$D6$ brane side) we had used only a set of restricted coordinate
transformations which only lie on the two $S^3$ directions of the
corresponding type IIB picture. It will now be interesting to
see whether this could be generalized for the case where we could consider
$\delta r$ variations. In particular, for a particular $S^3$ parametrized
by ($\psi_1, \theta_1, x$), one should now consider both $\delta r$ and
$\delta \theta_2$ variations for a given variation of $x, \theta_1$ and
$\psi_1$. Similar discussion should be done for the other $S^3$ parametrized
by ($\psi_2, \theta_2, y$).

\noindent $\bullet$ In the type IIA mirror background we have $B$ fields
both before and after geometric transitions. On the $D6$ brane side, the
presence of a $B$ field amounts to having non-commutativity on the world
volume of $D6$ branes. However, this $B$ field is in general not a
constant, and therefore may not have such a simple interpretation. This is
somewhat related to a discussion on $C$-deformation in \ooguri. The
$C$-deformations in general violates Lorentz invariance. It will be
interesting to see if there is any connection to our result, or if the
precise background that we propose does indeed realize the Lorentz violation 
and $C$-deformations.

\noindent $\bullet$  As discussed above, the manifold that we have
in 11 dimensions has a $G_2$
structure. However we haven't evaluated the holonomy of the
manifold and it should be
interesting to do so in order to check that the supersymmetry is preserved.
One immediate thing to check would be whether the manifold could
become complex. In other words whether the complex structure is
integrable or not.
This is an interesting question and can only be answered after we trace the
behavior of fermions when we do the mirror transformation.
Generic studies done
earlier have shown that in general these mirror manifolds {\it do not} have
an integrable complex structure. In addition to this, there is also the
question of the {\it choice} of the complex structure. Recall that in the
type IIB theory which we started out with,
% the complex structure is in general {\it fixed} by the choice of background
%fluxes.
the background fluxes generate a superpotential that {\it fixes} the complex
structure. This would
imply that the K\"ahler structures are all fixed in the type IIA picture.
On the other hand, if we start with a type IIB framework with, say,
$h^{1,1} =1$
one might also be able to fix the complex structure in the mirror  just
by fixing the K\"ahler structure in the type IIB side via (non-perturbative)
corrections to the type IIB superpotential. Thus the choice of
complex structure in type IIA will be uniquely fixed. It will be
important to see if the choice of complex structure that we made here is
consistent with the value fixed by the superpotential.

\noindent $\bullet$ There is another important aspect of geometric
transition that one needs to carefully verify. This has to do with
the disappearance of $D6$ branes when we perform the
geometric transition. Since $D6$ branes support gauge fluxes, the
disappearance of $D6$ branes would imply that after geometric transition
there cannot be any localized gauge fluxes. To show this aspect, the
M-theory lift will be very useful. Recall that the 
world volume couplings (and interactions) of
$D6$ branes can be extracted from M-theory lagrangian using the normalizable
harmonic (1,1) form of the corresponding Taub-NUT space \imamura.
Now that we have fluxes and also a background non-K\"ahler geometry in
type IIA theory (or in other words a torsional $G_2$ manifold in M-theory)
the analysis of the normalizable harmonic form is much more complicated.
In the presence of fluxes this has been considered in \robbins. It was found
there that the harmonic forms themselves change by the backreaction
of the fluxes on geometry. It will now be important to evaluate this
harmonic form and show that it is normalizable.
This would allow a localized gauge
flux to appear in the type IIA scenario, proving that the $G_2$
metric is indeed the lift of the $D6$ brane on a non-K\"ahler geometry.
This harmonic
form should either vanish or become non normalizable {\it after}
we do a flop. This will show that we do not expect a localized
gauge flux in type IIA theory and therefore the $D6$ branes have
completely disappeared! This will confirm Vafa's scenario. However as
discussed in \brand, the issue of normalisable form is subtle here. In the
absence of torsion, the usual form is badly divergent even before the
transition. In the presence of torsion (or fluxes in the type IIB
theory) the back reaction of fluxes on geometry might make this
norm well behaved, as has been observed in \robbins\ in a different context.
This will be useful for comparison.

\noindent $\bullet$ The $G_2$ manifold is explicitly non-K\"ahler (because it
is odd dimensional), and appears from a fibration over another
six-dimensional non-K\"ahler manifold. As we saw earlier, this manifold
is also neither complex nor half-flat.
Thus we have a new $G_2$ manifold whose slice is a specific six dimensional
non-K\"ahler space. Therefore one should be able to use
Hitchin flow equations \hitchin\
to construct the $G_2$ manifold from the six dimensional non-K\"ahler
space. When the base is a half-flat manifold, this has already been done
in the work of Hitchin \hitchin. The flow equations take the following
form:
\eqn\flowe{ dJ = {\del \Omega_+ \o \del t}, ~~~~~ d\Omega_- =
- J \wedge {\del J \o \del t}}
where $t$ is a real parameter that determines the $SU(3)$ structure
of the base and is related to the vielben $e_7$. 
For our case, since we know almost every detail of the
metric, it will be interesting to check whether the $G_2$ manifold
follows from the flow equations\foot{For some more details on Hitchin's flow
equations related to a generic lift of six-manifolds to seven-manifolds, the
readers may want to consult \dal. We thank G. Dall'Agata for correspondence
on this issue.}.

\noindent $\bullet$ The non-K\"ahler manifold that we get in type IIA theory
both before and after geometric transtions should fit in the
classification of torsion classes \salamon. Since they are more generic than
the half-flat manifolds, the torsion classes will be less constrained. It will
also be interesting to find the full $G_2$ structure for the M-theory
manifolds.

\noindent $\bullet$ The analysis that we performed in this paper starting with
$D5$ wrapped on resolution $P^1$ cycle of a resolved conifold  is only the
IR description of the corresponding gauge theory. The full analysis should 
involve additional $D3$ branes in the type IIB picture, so that the cascading 
behavior can be captured. However the cascade being an infinite sequence of 
flop transitions, makes simple supergravity description of the full theory
a little more involved. It will be interesting to pursue this 
direction to see how the type IIA mirror phenomena works. We hope to address
this issue in near future.

\vskip.2in

\centerline{\bf Acknowledgments}

Its our pleasure to thank Lilia Anguelova, Klaus Behrndt, Volker Braun,
Mirjam Cvetic,
Gianguido Dall' Agata,
Katrin Becker, Shamit Kachru, Amir-Kian Kashani-Poor, Dieter Luest,
Leopoldo Pando-Zayas and Cumrun Vafa 
for many interesting discussions and
useful correspondences. The work of M.B. is supported by NSF grant
PHY-01-5-23911 and an Alfred Sloan Fellowship. The work of K.D.
is supported in part by a Lucile and Packard Foundation
Fellowship 2000-13856. A.K. would like to acknowledge support
from the German Academic Exchange Service (DAAD) and the
University of Maryland.~R.T. is supported by DOE Contract
DE-AC03-76SF0098 and NSF grant PHY-0098840.

\vskip.2in

\newsec{\bf Appendix 1: Algebra of $\alpha$}

In the earlier sections we have defined $\alpha$ as
$\alpha = {1\o 1 + A^2 + B^2}$ where $A,B$ are given in \defAandB. This
quantity $\alpha$ is a very crucial quantity as the finite transformation
depends on it. While integrating the finite shifts we saw that we need to
approximate the transformation on a particular sphere, parametrised by
($\psi_i, x_i, \theta_i$), as though the other sphere components are
constants. In other words, to study the transformation on, say, sphere 1
we take the $\theta_2$ terms as a constant (or fixed at an average
value). In other words, any generic $\theta_i$ will be denoted as
\eqn\genthtea{\theta_1 = \langle\theta_1\rangle
+ \vartheta_1, ~~~~ \theta_2 =
\langle\theta_2\rangle + \vartheta_2}
where $\langle\theta_{1,2}\rangle$ 
are the average values of the $\theta$ coordinates.
To see how the $\alpha$ factor responds to this, let us first
define three useful quantities:
\eqn\thrusefi{\eqalign{&\langle \alpha \rangle_1 = 
{1 \o 1 + \Delta^2_1~{\rm cot}^2~\theta_1 +
\Delta^2_2~{\rm cot}^2~\langle\theta_2\rangle}, ~~
\langle \alpha \rangle_2 = {1 \o 1 + \Delta^2_1~{\rm cot}^2~
\langle\theta_1\rangle +
\Delta^2_2~{\rm cot}^2~\theta_2}\cr
& ~~~~~~~~~~~~~~~~~ \langle\alpha\rangle~ = {1 \o 1 +
\Delta^2_1~{\rm cot}^2~\langle\theta_1\rangle +
\Delta^2_2~{\rm cot}^2~\langle\theta_2\rangle}}}
Using the above definitions, one can easily show that $\sqrt{\alpha}$ has the
following expansion:
\eqn\alpexpa{\sqrt{\alpha} = \sqrt{\langle \alpha \rangle_1} \left( 1 +
{\rm x}~\Delta_2^2~\langle \alpha \rangle_1 ~
{\rm cot}^2~\langle\theta_2\rangle \right)^{-{1\o 2}}}
where we have kept the radial variations as constant as before, and the
quantity x appearing above being given by the following exact expression:
\eqn\valx{{\rm x} = -{4~ {\rm cosec}~2
\langle\theta_2\rangle~{\rm tan}~\vartheta_2~( 1
+ {\rm cot}~2\langle\theta_2\rangle~
{\rm tan}~\vartheta_2) \o 1 + {\rm cot}~\langle\theta_2\rangle
~{\rm tan}~\vartheta_2~ (2 + {\rm cot}~
\langle\theta_2\rangle~{\rm tan}~\vartheta_2)}.}
If the warp factor $\Delta_2$ is chosen in such a way that in \alpexpa\ the
quantity in the bracket is always small, then $\alpha$ will have the
following expansions at all points in the internal space:
\eqn\alepdg{\sqrt{\alpha} = \sqrt{\langle \alpha \rangle_1} - {1\o 2} ~{\rm x}~
\Delta_2^2~{\rm cot}^2~
\langle\theta_2\rangle~\langle \alpha \rangle_1^{3/ 2} + ....}
and therefore could be approximated 
simply as $\sqrt{\langle \alpha \rangle_1}$. Similar
argument will go through for $\langle \alpha \rangle_2$ for the other sphere.
Another
alternative way to write $\alpha$ is
\eqn\altalp{\sqrt{\alpha} = {\langle\alpha\rangle \o \sqrt{\langle \alpha
\rangle_1}} + ... = {\langle\alpha\rangle \o \sqrt{\langle \alpha \rangle_2}}
+ ...}
where again the dotted terms can be easily determined for different spheres.
The above two pair of expressions: 
$\sqrt{\alpha} = \sqrt{{\langle\alpha\rangle}_{1,2}}$ and
$\sqrt{\alpha} = 
{\langle\alpha\rangle \o \sqrt{{\langle\alpha\rangle}_{1,2}}}$ 
are responsible for
the two different set of coordinate transformations in \coordinate\ and
\aleb\ with $m = \pm 1$. 
Observe also that under the above approximations, some of the
components of the deformed conifold metric will now look like:
\eqn\defcnop{\eqalign{ds^2_{\theta_1 \theta_2} &= -2 \sqrt{\langle \alpha
\rangle_1 \langle \alpha \rangle_2}~j_{xy}~d\theta_1~d\theta_2,\cr
ds^2_{\theta_2 \theta_2} &= \langle \alpha \rangle_2 (1 + A_1^2)~d\theta_2^2,
\cr
ds^2_{\theta_1 \theta_1} &= \langle \alpha \rangle_1 (1 + B_1^2)~d\theta_1^2}}
where $A_1$ and $B_1$ are the values of $A$ and $B$ at the average values of
$\theta_1$ and $\theta_2$ respectively.

\vskip.2in

\newsec{\bf Appendix 2: Details on $G_2$ structures}

The threeform $\Omega$ that we described earlier in the context
of type IIA manifold can be fixed by a subgroup 
$SU(3)$ of $SO(6)$. In fact both $J$,
the fundamental form and $\Omega$ can be fixed simultaneously by $SU(3)$. The
torsion classes that we mentioned earlier are basically the measure of the 
non-closedness of $\nabla J$ and $\nabla \Omega$. Detailed discussions on this 
are in \grayone, \salamon. As we saw earlier, the type IIA manifold is neither
K\"ahler nor complex. Therfore the torsion is generic. When the complex 
structure becomes integrable the torsional connection is known as Bismut 
connection \bismut. 

Similarly the threeform $\tilde\Omega$ can be fixed by a
subgroup of $GL_7$. This is the exceptional Lie group $G_2$ which is a
compact simple Lie subgroup of $SO(7)$ of dimension 14. The existence of a
$G_2$ structure is equivalent to the existence of the fundamental three form
$\tilde\Omega$. The following interesting cases have been studied in the
literature for the torsion free $G_2$ case (for a more detailed review on this
the reader may look into the last reference of \ivan. A short selection 
on $G_2$ manifolds are in \salamontwo, \monar, \kath, \joyce):

\noindent $\bullet$ When
$\nabla \tilde\Omega = 0$, then holonomy is contained in
$G_2$ with a Ricci flat $G_2$ metric \bonan.

\noindent $\bullet$  When $d\tilde\Omega = d \ast \tilde\Omega = 0$ then the
fundamental form is harmonic and the corresponding $G_2$ manifold is
called parallel. First examples were constructed in \salamontwo. Later
on, first compact examples were given in \joyce, \kovalev.

\noindent $\bullet$ When
$\varphi \equiv \ast(\tilde\Omega \wedge \ast d\tilde\Omega) =0$
then the $G_2$ structure is known to be balanced, where $\varphi$ is known as
the Lee form. If the Lee form is closed, then the $G_2$ structure is locally 
conformally equivalent to a balanced one.
When $d\ast \tilde\Omega = \varphi \wedge \tilde\Omega$ then
the $G_2$ structure is called integrable.

In the presence of torsion we have already described the changes that one
would expect from the above mentioned conditions. In terms of the torsion
three form $\tau$ given in \torthrf,
the connection shifts from $\nabla$ mentioned above
to the usual expected form $\nabla + {1\o 2} \tau$. Now the manifold is no
longer Ricci flat and the curvature tensor takes the following form:
\eqn\curva{{\cal R}_{ijkl} = R_{ijkl} - {1\o 2} \tau_{ijm}\tau_{klm}
-{1\o 4} \tau_{jkm}\tau_{ilm} - {1\o 4} \tau_{kim} \tau_{jlm}}
where $R$ measures the curvature wrt the Riemannian connection. For the
case when the base is half-flat, a construction of $G_2$ manifolds satisfying
some of the features mentioned above is given in \ivan. For our case we
do not have a half-flat base, and therefore the manifold that we presented in
sections 6 and 7 are new examples of $G_2$ manifolds with torsion that
satisfy all the string equations of motion.

\listrefs

\bye